\documentclass[fleqn,usenatbib]{mnras}

\usepackage{newtxtext,newtxmath}

\usepackage[T1]{fontenc}
\usepackage{ae,aecompl}

\usepackage{graphicx}	
\usepackage{amsmath}	
\usepackage{xspace}
\usepackage{amsfonts,textcomp}
\usepackage[usenames]{color}
\usepackage{times}
\usepackage{hyperref}
\usepackage{ulem}
\usepackage{enumitem}
\usepackage[compatibility=false]{caption}
\usepackage{subcaption}
\usepackage{multirow}
\usepackage{tabulary}
\usepackage[para]{threeparttable}
\usepackage{array,booktabs,longtable,tabularx}
\newcolumntype{L}{>{\raggedright\arraybackslash}X}
\usepackage{ltablex}
\usepackage{comment}

\usepackage{etoolbox}

\renewlist{tablenotes}{enumerate}{1}
\makeatletter
\setlist[tablenotes]{label=\tnote{\alph*},ref=\alph*,itemsep=\z@,topsep=\z@skip,partopsep=\z@skip,parsep=\z@,itemindent=\z@,labelindent=\tabcolsep,labelsep=.2em,leftmargin=*,align=left,before={\footnotesize}}
\makeatother

\graphicspath{{../},{./}}

\newcommand{\lt}{\mbox{$L_{\rm T}$}\xspace}
\newcommand{\noise}{\mbox{$\mathcal{N}$}\xspace}
\newcommand{\lya}{\mbox{Ly-$\alpha$}\xspace}
\newcommand{\sir}{\mbox{$\mathcal{R}_{I,r}$}\xspace}
\newcommand{\siu}{\mbox{$\mathcal{R}_{I,U}$}\xspace}
\newcommand{\svar}{\mbox{$\mathcal{R}_{\rm WQ}$}\xspace}

\newcommand{\rmax}{\mbox{$r_{\mathrm{max},ij}$}\xspace}
\newcommand{\hmax}{\mbox{$h_{\mathrm{max},ij}$}\xspace}
\newcommand{\rexc}{\mbox{$r_{\mathrm{ex},ij}$}\xspace}

\newcommand{\hi}{{\sc H\,i}\xspace}

\definecolor{Orange}{rgb}{1.0,0.5,0.15}
\definecolor{Blue}{rgb}{0,0.08,0.65}
\definecolor{Blue2}{rgb}{0,0.4,0.6}

\definecolor{Red}{rgb}{0.65,0.08,0.05}
\definecolor{Green}{rgb}{0.15,0.45,0.25}
\definecolor{Pink}{rgb}{1.0,0.05,0.5}
\definecolor{Purple}{rgb}{0.3,0.,0.5}

\title[Cosmic Web mapping with  Lyman-alpha tomography: critical point clustering \& connectivity. ]{Forecasts for WEAVE-QSO: 3D clustering and connectivity  of critical points with  Lyman-$\alpha$ tomography}

\author[K. Kraljic, C. Laigle, C. Pichon, S. Peirani et al.]{\parbox[t]{\textwidth}{
K.~Kraljic$^{1}$\thanks{E-mail: katarina.kraljic@lam.fr}, C.~Laigle$^{2}$, C.~Pichon$^{2,3}$, S.~Peirani$^{4,2}$,
S.~Codis$^{5}$, J.~Shim$^6$,\\ C.~Cadiou$^7$,
D.~Pogosyan$^8$, S.~Arnouts$^{1}$, M.~Pieri$^{1}$, V. Ir\v si\v c$^{9,10}$, \\ 
S. S. Morrison$^{1,11}$, J. O\~norbe$^{12}$, I. P\'erez-R\`afols$^{13}$, G. Dalton$^{14,15}$
}
\\
\\
$^{1}$ Aix Marseille Univ, CNRS, CNES, LAM, Marseille, France\\
$^{2}$ CNRS and Sorbonne Universit\'e, UMR 7095, Institut d'Astrophysique de Paris, 98 bis, Boulevard Arago, F-75014 Paris, France\\
$^{3}$ IPhT, DRF-INP, UMR 3680, CEA, L'Orme des Merisiers, B\^at 774, 91191 Gif-sur-Yvette, France\\
$^{4}$ Observatoire de la C\^ote d'Azur, CNRS, Laboratoire Lagrange, Bd de l'Observatoire, CS 34229, F-06304 Nice Cedex 4, France\\
$^{5}$ AIM, CEA, CNRS, Universit\'e Paris-Saclay, Universit\'e Paris Diderot, Sorbonne Paris Cit\'e, F-91191 Gif-sur-Yvette, France\\
$^{6}$ Korea Institute for Advanced Study,  85 Hoegiro, Dongdaemun-gu, Seoul, 02455, Republic of Korea\\
$^{7}$ Department of Physics and Astronomy, University College London, Gower Street, London WC1E 6BT, United-Kingdom \\
$^{8}$ Department of Physics, University of Alberta, 412 Avadh Bhatia Physics Laboratory, Edmonton, Alberta, T6G 2J1, Canada\\
$^{9}$ Kavli Institute for Cosmology, University of Cambridge, Madingley Road, Cambridge CB3 0HA, UK\\
$^{10}$ Cavendish Laboratory, University of Cambridge, 19 J. J. Thomson Ave., Cambridge CB3 0HE, UK\\
$^{11}$  Department of Astronomy, University of Illinois at Urbana-Champaign, Urbana, IL 61801, USA\\
$^{12}$ Facultad de F\'isica, Universidad de Sevilla, Avda. Reina Mercedes s/n. Campus de Reina Mercedes, 41012 Sevilla, Spain \\
$^{13}$ Sorbonne Universit\'e, Universit\'e Paris Diderot, CNRS/IN2P3, LPNHE, 4 Place Jussieu, F-75252 Paris, France \\
$^{14}$ Department of Physics, University of Oxford, Parks Rd, Oxford, OX1 3RH, UK\\
$^{15}$ TFC RALSpace, UKRI, Harwell Campus, OX110QX, UK\\
}

\date{Accepted XXX. Received YYY; in original form ZZZ}

\pubyear{2019}

\begin{document}
\label{firstpage}
\pagerange{\pageref{firstpage}--\pageref{lastpage}}
\maketitle

\begin{abstract}
The upcoming WEAVE-QSO survey will target a high density of quasars over a large area, enabling the reconstruction of the 3D density field through Lyman-$\alpha$ tomography over unprecedented volumes smoothed on intermediate scales ({$\approx$~16~Mpc/h}). We produce mocks of the Lyman-$\alpha$ forest using LyMAS, and reconstruct the 3D density field between sightlines through Wiener filtering in a configuration compatible with the future WEAVE-QSO observations. The fidelity of the reconstruction is assessed by measuring one- and two-point statistics from the distribution of critical points in the cosmic web. In addition, initial Lagrangian statistics are predicted from first principles, and measurements of the connectivity of the cosmic web are performed. The reconstruction captures well the expected features in the auto- and cross-correlations of the critical points. This remains true after a realistic noise is added to the synthetic spectra, even though sparsity of sightlines introduces systematics, especially in the cross-correlations of points with mixed signature. Specifically, for walls and filaments, the most striking clustering features could be measured with up to 4 sigma of significance with a WEAVE-QSO-like survey. Moreover, the connectivity of each peak identified in the reconstructed field is globally consistent with its counterpart in the original field, indicating that the reconstruction preserves the geometry of the density field not only statistically, but also locally. Hence the critical points relative positions within the tomographic reconstruction could be used as standard rulers for dark energy by WEAVE-QSO and similar surveys.
\end{abstract}
\begin{keywords}
Galaxy formation -- Large-scale structure -- Surveys -- Topology
\end{keywords}



\section{Introduction}

The geometry and cosmic evolution {of large-scale structure} are our best probes to make sense of 
the accelerated expansion of the Universe. 
At $z>2$, the Lyman$-\alpha$ (\lya) forest absorption towards bright background sources is observable from ground-based optical instruments and can be used at intermediate {({$\sim$~1~Mpc/$h$}) to large ({$\sim$~200~Mpc/$h$})} scales as a tracer of the underlying density field. 
The prospect of using tomography of the \lya forest for 
reconstructing the cosmic web \citep{bondetal1996}  
has a long history \citep[see e.g.][]{Pichon2001,tomography,Caucci2008,Gallerani2011,2012MNRAS.420...61K,2014MNRAS.440.2599C,Ozbek2016,2019A&A...632A..94J,Horowitz2019,Horowitz2021}  
and is now within reach from current (e.g. CLAMATO: \citealp{Lee_2016,2018ApJ...861...60K,Lee2018, Newman2020}, eBOSS-Stripe 82 \citealp{eBOSS16-2020,Ravoux2020}) and upcoming quasar or star-forming galaxy surveys (\citealp[e.g. WEAVE-QSO:][]{Pieri2016}, Pieri et al. in prep.,  PFS: \citealp{PFS2014} or DESI: \citealp{2016arXiv161100036D}).
Such reconstruction represents an unparalleled opportunity,  as it gives us  access to many large and intermediate scales
\citep{Bernardeau2002}. Its success  
relies on the  orders-of-magnitude better sensitivity of detection of neutral hydrogen in absorption (when compared to 
emission), along Gpc-long lines-of-sight \citep[][]{Petitjean1995,Rauch1998}. 
Hence \lya tomography  provides means to characterise the expansion-driven geometry of the cosmic web
in the lead up to the epoch of dark energy domination.
\begin{figure}
\includegraphics[width=\columnwidth]{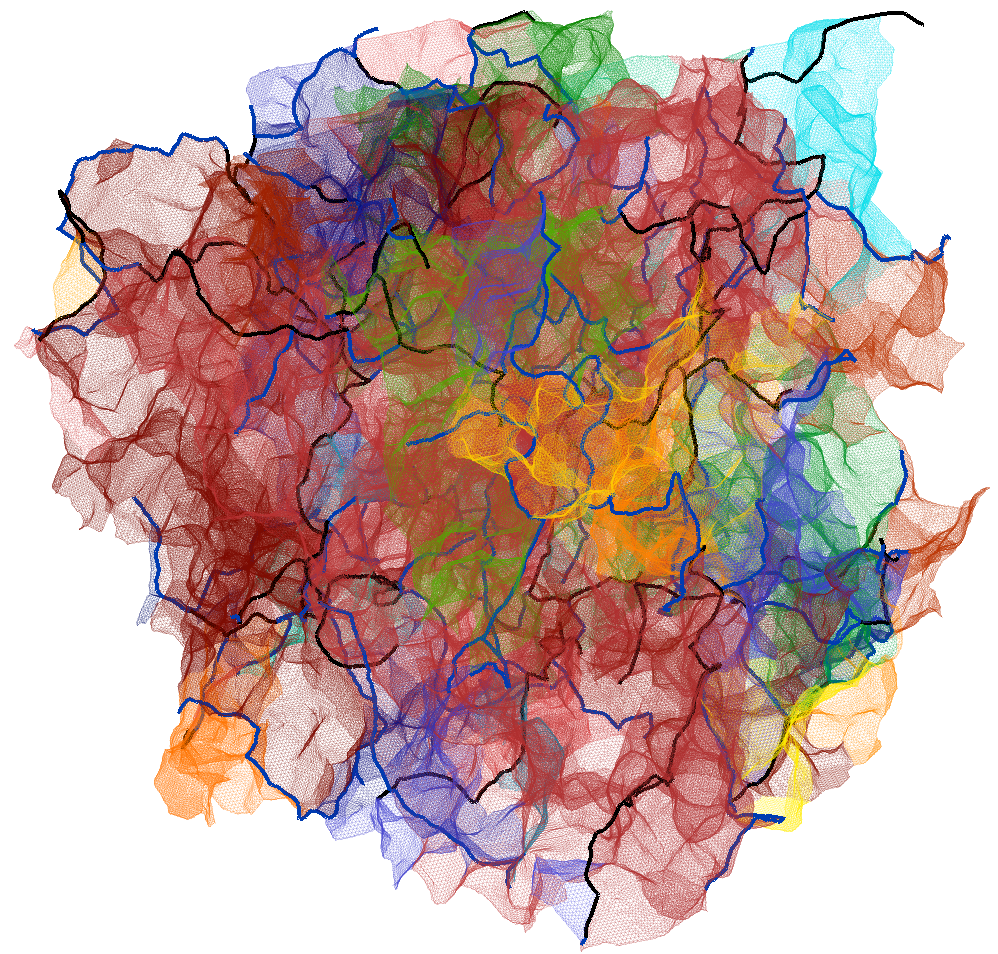}
\caption{The walls (colour coded randomly) and filaments (dark colour for all filaments, light colour for filaments of higher persistence) extracted from the DM density field of one of the mocks. 
The purpose of the reconstruction performed in this study is to recover as accurately as possible the geometry of this cosmic web, since it defines 
the metric in which we can constrain dark energy.
In order to assess this accuracy, we focus on the number counts and  clustering properties of 
the critical points associated with  peaks, voids, filaments and walls of the cosmic web. We also compute the connectivity of its nodes.
}
\label{fig:nice}
\end{figure}

Depending on the design of the surveys (sampling of background sources, availability of quasar and/or galaxy spectra, spectral resolution and signal-to-noise ratio), different scales and volumes will be accessible, making the tomographic reconstruction either more suitable for studies focused on co-evolution of galaxies and the intergalactic-medium (if filaments can be reconstructed at the Mpc-scale) or for cosmological analysis (if large volumes are available). 
Using idealized mocks, the pioneering work of \cite{Caucci2008} demonstrated that the topology of the cosmic web traced either by Minkowski functionals, such as the genus \citep{genus-paper}, or the skeleton \citep{sousbie112} could be {well-recovered} with this method. In the same vein, \cite{Horowitz2019} showed that cosmic web {structure classification} from
eigenvalues and eigenvectors of the pseudo-deformation tensor could be accurately performed, while \cite{Horowitz2021} focused on proto-cluster identification \citep[see also e.g. ][for complementary mock-based analyses of the quality of the reconstruction]{Cisewski14,Ozbek2016,2019A&A...632A..94J}. In particular, several reconstruction methods have been presented in the literature. Wiener filtering is the 
classical approach, but different procedures have also been proposed, either involving a sophistication of the standard Wiener Filter \citep[e.g.][]{Zihao2021}, a forward modelling approach \citep[e.g.][]{Porqueres2020, Horowitz2021}, or convolutional neural networks \citep{Harrington2021}.

Encouraged by these theoretical pursuits, 
three-dimensional reconstruction of the density field from the \lya forest has already been successfully performed in observational surveys, notably with the CLAMATO program \citep[see][for the latest release]{Clamato2ndrelease} and eBOSS-Stripe~82 \citep{Ravoux2020}.
The \lya forest has proven to be a powerful tracer of the density field, particularly sensitive to intermediate densities:
therefore tomographic reconstruction should allow us to characterise the geometry of the weakly {over- and under-dense regions} of the Universe, i.e. specifically the walls and filaments of the cosmic web (see Fig.~\ref{fig:nice} for an illustration). 

The clustering properties of maxima of 3D density fields were  predicted for  Gaussian random field by \cite{Regos1995} and revisited more recently by \cite{Baldauf2021}. Such predictions provide insight on  their dependency over cosmological parameters. 
More recently, \cite{Shim2021} systematically investigated the statistical properties of all critical points (i.e. the loci of zero gradient) of the cosmic  field of $\Lambda$CDM simulations, and in particular  the number counts and 
clustering properties of wall-like and filament-like saddle points\footnote{Recall that a saddle point is a point where the gradient is null, but that is neither a minimum nor a maximum.}. As they trace the relative  position of 
 walls and filaments (beyond the more standard peaks and voids), these saddle points help characterise 
 the global geometry and evolution of the  cosmic web \citep{Cadiou2020}.
They define the underlying topology, which makes them  robust to most systematics (e.g. biasing). Wall-saddle clustering encodes the typical size of voids \citep[][]{Stark2015}, while the cross-correlation of peaks and filament-type saddles is sensitive to the typical length of filaments.
 These sets of points probe less biased regions than galaxy surveys \citep{Desjacques2018},
hence their dynamics  are   better captured by perturbation theory \citep{Gay2012}. \cite{Shim2021} showed that the statistical properties of the set of critical points  
such as the size of the exclusion zones are weakly dependent on redshift,  hence could in principle be used as  standard rulers
\citep{Lazkoz2008,Appleby2021} to constrain alternative cosmology models \citep[e.g.][]{Bamba2012}.

The WEAVE-QSO survey, as part of the wider WHT Extended Aperture Velocity Explorer (WEAVE, \citealp{weave}) survey,  is potentially well-suited to deriving cosmologically meaningful statistics with critical points. Its large volume will make it possible to probe the large-scale structure over several  {thousands} of square degrees allowing us to characterise
the geometry of weakly over- and under-dense regions of the Universe, while its high density will allow to reach scales as small as $\sim 16~{\rm Mpc}/h$.
Could the  \lya  tomography reconstruction performed on surveys such as WEAVE-QSO be precise enough to measure the clustering of critical points and distinguish between different cosmological models?

In this paper, we investigate to what extent the cosmic web is correctly recovered with this technique by focusing on the clustering of critical points of the density field and their connectivity. We model the WEAVE-QSO survey and constrain the impact of its specificity on our ability to extract cosmological information from the clustering of  critical points. In particular, we will investigate what sets of critical pairs are least impacted by systematics and uncertainties inherited from the reconstruction.
We will also measure the individual connectivity of peaks identified in the reconstructed field and compare it to their counterparts in the original field, in order to verify that the geometry of the reconstructed field is robust  not only statistically, but also locally.

The paper is organised as follows:  in \S\ref{sec:data} we describe the method used to produce the mocks
and the corresponding estimators, in
\S\ref{sec:reconstruction}  we assess the  quality  of the reconstruction using one and two point statistics of critical points, and their connectivity. We discuss our results 
in \S\ref{sec:discussion-conclusions} and  conclude in \S\ref{sec:conclusions}.

Finally, Appendix~\ref{appendix:quality} investigates the expected accuracy on the reconstruction for future surveys (different configurations of sampling and noise), 
Appendix~\ref{appendix:GRF} sketches the steps involved in the computation of the Lagrangian two-point functions for the dark matter $\Lambda$CDM model smoothed over the relevant scales, 
Appendix~\ref{appendix:tables} provides a measurements of summary statistics, 
while Appendix~\ref{appendix:rarity} looks into the impact of the choice of 
rarity and smoothing length.

\section{Mock data and estimation methods}
\begin{figure}
\includegraphics[scale=0.5]{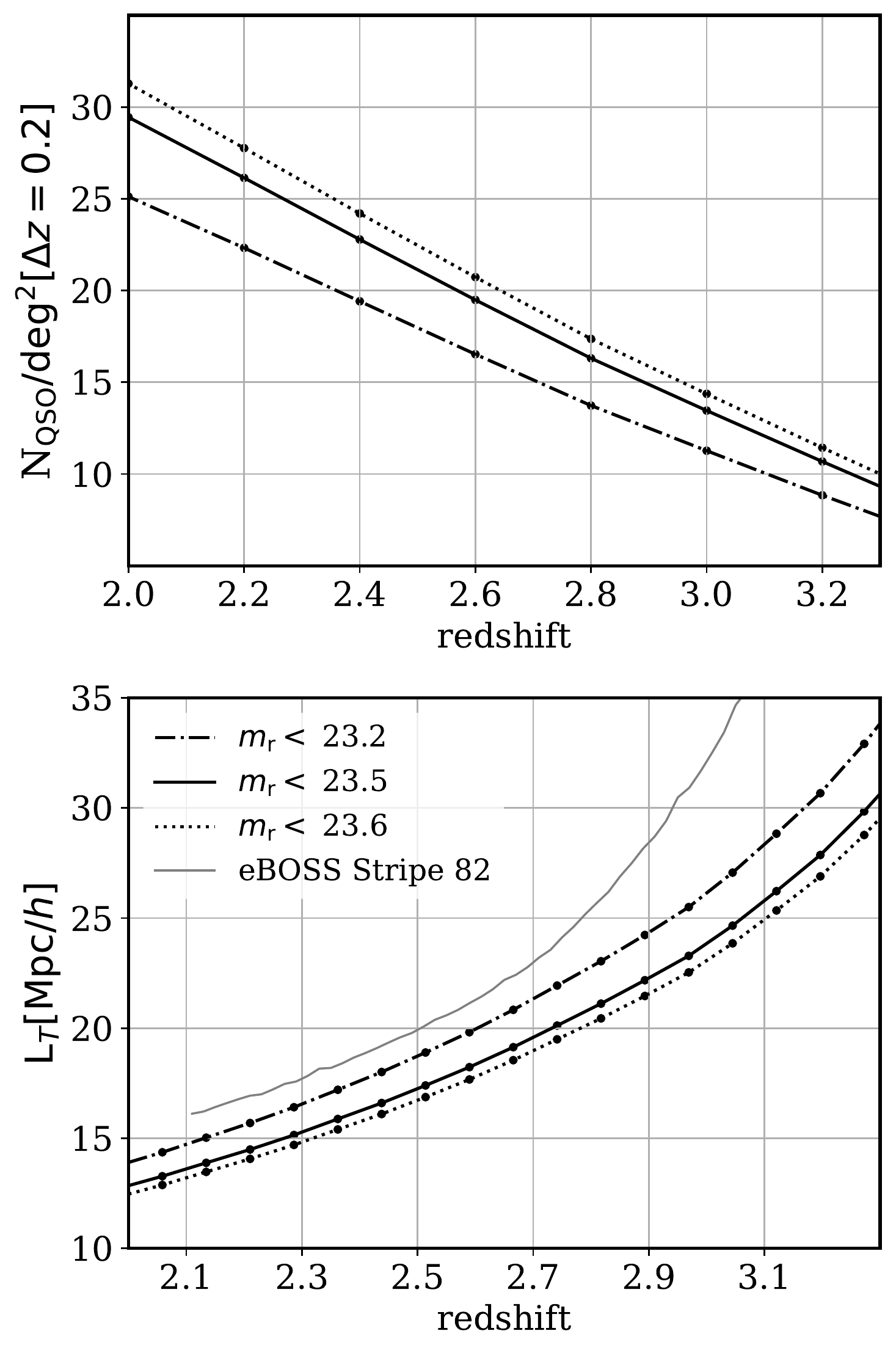}
\caption{\textit{Top:} Number of quasars per deg$^2$ per bin of $\Delta z=0.2$ as a function of the limiting magnitude. These counts follow from \protect\cite{PalanqueDelabrouille2016,PalanqueDelabrouilleCorr2016}. The configuration investigated in this study (corresponding to the HIGHDENS footprint, $m_r<23.5$) is represented by the solid black line. \textit{Bottom:} Mean separation between sightlines as a function of redshift, for the same limiting magnitude thresholds as in the \textit{top} panel, assuming that the full length between the \lya and Ly-$\beta$ wavelengths is usable. In grey, we have over-plotted for comparison the mean separation \lt in eBOSS-Stripe~82. 
}
\label{fig:quasmag}
\end{figure}
\begin{figure}
\includegraphics[scale=0.5]{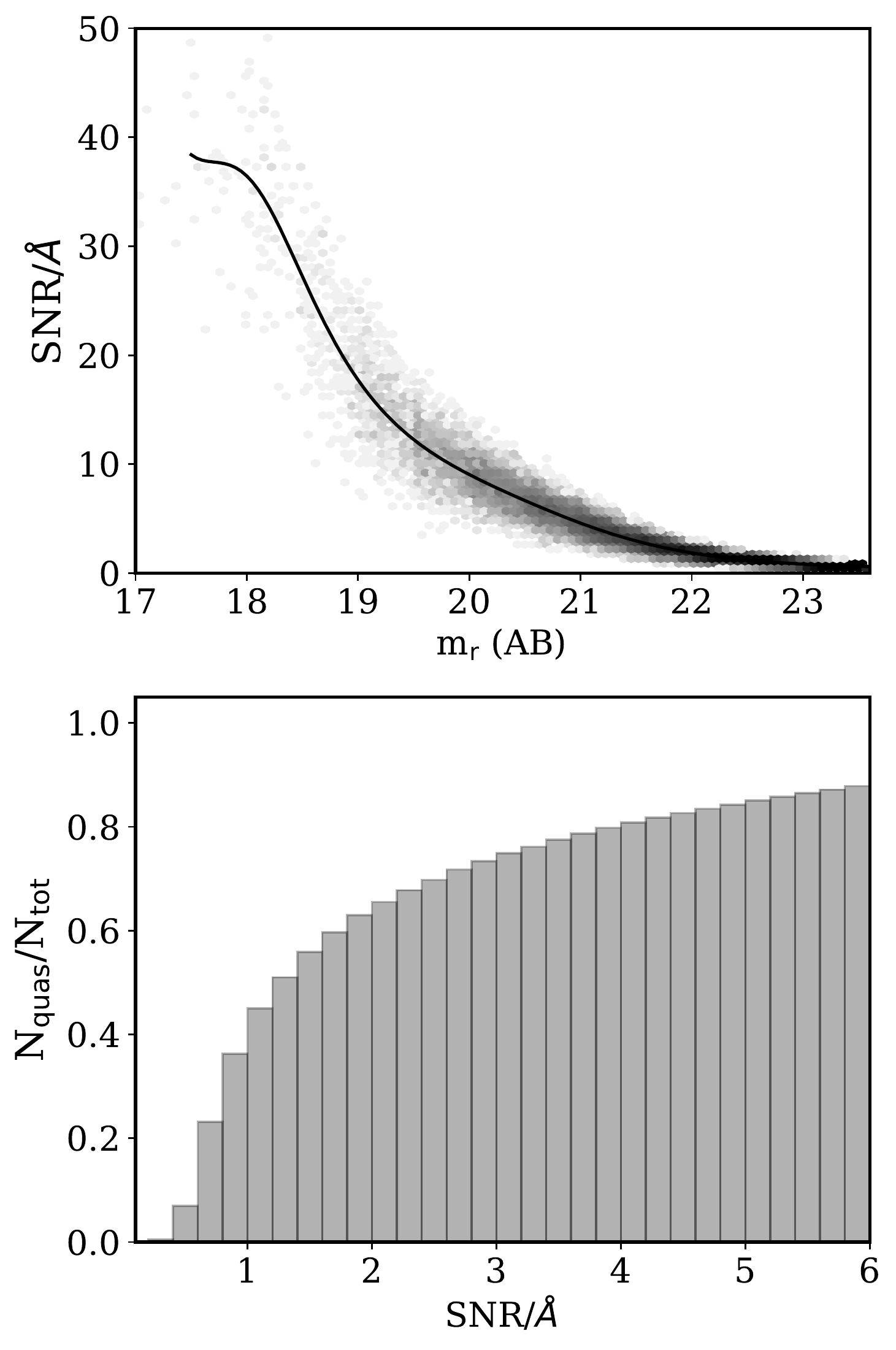}
\caption{\textit{Top}: SNR/\AA ~(defined in the continuum of the \lya forest) as a function of magnitude in the $r$-band as adopted in this study. \textit{Bottom}: corresponding cumulative count of quasars as a function of SNR/\AA~. These distributions correspond to reasonable forecasts for the WEAVE-QSO survey (\citealp{Morrison2019}, Jin et al in prep.). We expect about 50\% of quasars to have a SNR/\AA~larger than 1.4. 
}
\label{fig:SNRdistrib}
\end{figure}

\label{sec:data} 

Let us first describe briefly our mocks,
reconstruction method, 
critical point extraction and clustering estimators.

\subsection{Modelling the Lyman-$\alpha$ forest}
\label{sec:simus}
We rely on 5 $\Lambda$CDM simulation snapshots of (1Gpc$/h$)$^{3}$ at  $z=2.5$, run with \textit{WMAP7} cosmological parameters \citep[][]{Komatsu11}:  $\Omega_m=0.272$, $\Omega_\lambda=0.728$, $\sigma_8=0.81$, $n_s=0.967$. Each N-body simulation follows the time evolution of 2048$^{3}$ dark matter (DM) particles using {\sc Gadget2} \citep[][]{Springel2005}. Fig.~\ref{fig:nice} shows the  cosmic web of 
one of these simulations.

An estimator for the \lya forest flux from the DM  has been synthesised using  the hydrodynamical {\sc Horizon-AGN} simulation \citep{Dubois2014}, with LyMAS2,
an improved version of the \lya Mass Association Scheme \citep[LyMAS,][]{peirani14}. In brief, the optical depth of \lya absorption is  calculated along about one million sightlines extracted from {\sc Horizon-AGN}, based on the neutral hydrogen (\hi) density, whose evolution and distribution in the simulation is impacted by metal-dependent cooling, photoionization and heating from the UV background, feedback and metal enrichment. We work in the distant observer limit, assuming the sightlines are all parallel to one side of the box. 
LyMAS2 assigns a \lya flux in redshift space to each pixel in the N-body simulation assuming that the 3d flux correlations are
mainly driven by the correlations of the underlying DM (over)density and velocity dispersion fields.

To produce each large \lya forest mock catalogue, we have then derived first the DM density and the velocity dispersion fields
(on regular grids of 4096$^{3}$) from each gadget simulation at $z=2.5$. 
To save computational time throughout this work, the initial grids of each simulated volume are resampled on a Cartesian grid of $512^{3}$ (we kept
1 pixel of 8), and therefore the resulting resolution of the \lya spectra is 1.95~Mpc$/h$ per pixel, which corresponds to a resolution of about $ $2.6~\AA~per pixel in the \lya forest at $z=2.3$, or equivalently $\sim$R1560. We note that our simulated spectra are at the BOSS resolution
and are less resolved than the spectra which will be obtained with WEAVE (R$5000$), but this difference should not impair our analysis, since this latter relies on smoothed fields\footnote{Given that what primarily limits the resolution of the reconstructed map is the transverse separation between sightlines, the spectra are not exploited at the WEAVE fiducial spectral resolution. Indeed, several spectral resolution elements contribute to a given volume element in the reconstructed density field.}.
More details on the LyMAS2 implementation and \lya mock production are given in  \cite{peirani21}.  As a reminder, \lya flux is equal to $\exp(-\tau)$ where $\tau$ is the \lya optical depth, which at first order scales like the \hi density. In the following,  the  \hi field is estimated as simply being $-\log ({\rm flux})$. 

Note that redshift space distortions are included in the DM and flux (and consequently \hi) simulated fields 
but not in the Gaussian random field (GRF) predictions: this should be kept in mind when interpreting our results, e.g. in Fig.~\ref{fig:auto_dm_rarity10_16} below. We checked though that when smoothing the fields at the scales considered in this work (12-16 Mpc/$h$), redshift space distortions do not alter our results at the level of the expected accuracy.
Note that $\sigma_\mathrm{DM}\sim 0.1$ when smoothed over 16 Mpc/$h$ at redshift 2.5, so we are probing the regime accessible to perturbation theory \citep[][]{Gay2012}.

\subsection{Mimicking the distribution of sightlines}

\subsubsection{Quasar counts and separation between sightlines}
The number of observed quasar spectra defines the overall achievable transverse resolution of the reconstructed map. 
Fig.~\ref{fig:quasmag} presents the expected separation between sightlines as a function of redshift for different magnitude cuts,  where ${\rm m}_r$ is the magnitude in the $r$ filter passband. 
To compute it, we use the \cite{PalanqueDelabrouille2016,PalanqueDelabrouilleCorr2016} counts (their Table~7, model PLE), which provides the number of quasars per 100 deg$^2$ per bins of redshift ($\Delta(z)=1$) and per bins of magnitudes  in the $r$-band ($\Delta(m)=0.5$). The top panel of Fig.~\ref{fig:quasmag} displays these counts as a function of magnitude cuts, after having interpolated them on a finer redshift grid ($\Delta(z)=0.2$). Throughout this work we use only the portion of the rest-frame spectra between the \lya (1216\AA) and the Ly-$\beta$ (1026\AA) wavelength, to avoid contamination of the \lya forest by Ly-$\beta$ absorption lines\footnote{In practice, we should also exclude from the analysis the fraction of the sightlines in the direct vicinity of the quasar (over $\sim 30~{\rm Mpc}/h$) to mitigate the proximity effect (where the \hi distribution is affected by the ionizing UV flux from the quasar).}. For example, at $z\sim 2.3$, lines of sight are therefore usable along at most $\sim 165$~Mpc$/h$. This allows us to estimate the mean separation between sightlines, which is presented on the bottom panel of Fig.~\ref{fig:quasmag}. We also show, for comparison, the mean separation between sightlines reached in eBOSS-Stripe 82 \citep{Ravoux2020}\footnote{For eBOSS-Stripe 82, ${\rm L}_T$ has been estimated using the pixel map made publicly available at \href{https://zenodo.org/record/3737781\#.YUJbnn069h}{https://zenodo.org/record/3737781\#.YUJbnn069h}}.  

We emphasise that we choose the mean separation between sightlines for the correlation length used in the reconstruction (Section~\ref{sec:reconstruction}).
\cite{Ravoux2020}
made a different choice  on eBOSS-Stripe~82, using instead  the mean distance to the closest sightline, which returns a smaller value (about 10 Mpc$/h$ instead of $\sim 20~{\rm Mpc}/h$ at $z=2.5$). While this choice allows them to reach smaller scales in regions where sightlines are well clustered, the quality of their reconstruction is degraded in other places where sightlines are sparser (which they subsequently masked in the reconstructed map). Our choice of taking the mean separation between sightlines is  more conservative and allows us to obtain a more homogeneous quality of the reconstruction everywhere, which is required to derive robust statistics on the clustering of critical points.  

\subsubsection{WEAVE-QSO survey specifications}
The WEAVE-QSO survey is expected to begin in 2022 and will observe in various modes and configurations  towards a variety of survey goals (Jin et al, in prep, Pieri et al in prep). Two samples with particularly high density are of interest for the present work. We stress that the precise details of the survey plan may evolve over the coming years.

\paragraph*{The ``WIDE'' footprint:}  Over the 6000~deg$^2$ of the WIDE footprint,
a near-complete sample of quasars at $2.5<z<3$ with  ${\rm m}_r<23.5$ will be observed by WEAVE-QSO with a spectral resolution R5000, corresponding to $\sim 48$~sightlines per deg$^2$.  
This equates to a comoving volume of $\sim $13.6 (Gpc$/h$)$^3$. The target selection for this footprint is provided by the J-PAS survey (\citealp{Benitezetal2014}).

\paragraph*{The ``HIGHDENS'' footprint:} Over the 418~deg$^2$ of the HIGHDENS footprint, all quasars with $m_r<23.5$ and $z>2.15$ will be observed with WEAVE at R5000, corresponding to $\sim 111$~sightlines per deg$^2$. This configuration therefore reaches higher density compared to the WIDE footprint through its extension to $2.15<z<2.5$ (see Figure~\ref{fig:quasmag}). The volume covered by this area over this redshift range is however only about~0.73 (Gpc$/h$)$^{3}$. This footprint is also targeted by J-PAS and is placed within the WIDE footprint above. Its boundary is defined by the HETDEX main ``spring'' field  \citep{Gebhardtetal2021}. 

In this work, we adopt a simulated distribution of quasars which matches the redshift and the density expected on the HIGHDENS footprint, but we note that the WIDE footprint provides an equivalent distribution (in terms of limiting magnitude) over its higher redshift range.

\paragraph*{Signal-to-noise ratio distribution:} The \textit{top} panel of Fig.~\ref{fig:SNRdistrib} presents the signal-to-noise ratio (SNR) distribution as a function of magnitude which is expected to be compatible with the WEAVE-QSO main sample (Jin et al. in prep.). These values correspond to the SNR/\AA~ of the continuum in the \lya forest. The cumulative number of quasars as a function of SNR$/$\AA~is shown in the {bottom} panel. We expect about 50\% quasars to have a SNR/\AA~larger than~1.4.

\subsubsection{Adopted configurations}
We describe below the different sets of spectra used in our study to mimic WEAVE-QSO observations (\svar) and to test the effect of the different sources of noise: sparsity of the sightline distribution, irregularity of their spatial distribution, noise on spectra ($\mathbf{\mathcal{R_{\rm I,r}}}$, $\mathbf{\mathcal{R_{\rm I,U}}}$ and $\mathbf{\mathcal{N}}$). 
\label{sec:configurations}
\paragraph*{Fiducial realisation, \svar:}
Without attempting to perfectly reproduce the WEAVE-QSO selection function in one of the two above-mentioned footprints, we simply define a density of sightlines which is compatible with the specifications in the HIGHDENS footprint.
At $z\sim 2.3$ and for $m_r<23.5$, we populate each of our five (1 Gpc$/h$)$^3$ simulated boxes with about $\sim 30$ quasars per (100 Mpc$/h$)$^3$, their positions being drawn from a uniform distribution and quasars being assigned randomly a magnitude so that we reproduce the counts presented in the {top} panel of Fig.~\ref{fig:quasmag}. The flux along the sightlines is perturbed with white Gaussian noise, with a magnitude-dependent SNR, the resulting distribution of which is shown in Fig.~\ref{fig:SNRdistrib}. Note that because quasar positions are scattered randomly, clustering  will inevitably occur (``Poisson clumping"), however the projected small-scale clustering of our quasar distribution will be reduced with respect to what is measured in the real Universe. In particular, the fact that bright quasars (hence with high SNR) are expected to be more clustered  \citep[e.g.][]{Shenetal2009} could degrade a bit the homogeneity of the quality of our reconstructed map in real data (boosting the accuracy in patches where quasars are well clustered, and conversely degrading it elsewhere) and lead to results slightly worse than those obtain for \svar at the same quasar density.

The mean sightline separation, estimated as the square root of the area of the field divided by the number of sightlines in thin redshift slices, is about 15 Mpc$/h$. We note though that over $2.15<z<2.5$, i.e over the redshift range covered by the HIGHDENS footprint, this separation is expected to vary between 14 Mpc$/h$ and 17 Mpc$/h$. 

Finally, we note that our simulation set covers a volume of more than $6\times$ the expected volume of the HIGHDENS footprint. This should be factored in when discussing the statistical significance of our results in the context of the WEAVE-QSO survey. At the same time, the WIDE footprint is about $2.7\times$ larger than our simulated set, but covering a higher redshift range ($2.5<z<3$), which would lead to a larger separation between sightlines (as can be inferred from Fig.~\ref{fig:quasmag}).We discuss in section~\ref{sec:errorbudget} how the statistical significance of our results would change depending on the footprint. 

\paragraph*{Realisation without noise on spectra, $\mathbf{\mathcal{R_{\rm I,r}}}$:}
For testing the impact of noise on spectra, we also produce a realisation identical to \svar, but without perturbing the \lya flux to mimic noise on spectra. Comparing $\mathcal{R_{\rm I,r}}$ to $\mathcal{R_{\rm WQ}}$ allows to test the impact of noise in the Ly-$\alpha$ forest on the quality of the reconstruction. 

\paragraph*{Realisation with regular distribution of sightlines, $\mathbf{\mathcal{R_{\rm I,U}}}$:}
For convergence study, we produce a realisation with no noise on spectra and the same density of sightlines but with quasars regularly distributed across the box (so that we have exactly one sightline every $\sim 15 {\rm Mpc}/h$). Comparing $\mathcal{R_{\rm I,U}}$ to $\mathcal{R_{\rm I,r}}$ allows to test the impact of shot noise, i.e non-regularity in the distribution of sightlines, on the quality of the reconstruction.

\paragraph*{Realisation with noise only, $\mathbf{\mathcal{N}}$:}
In order to test the significance of our measurement, we also perform the tomographic reconstruction on 10 sets of sightlines (covering 1 (${\rm Gpc}/h$)$^3$ each, as for the other realisations) containing only white Gaussian noise (with an rms equal to the rms of the large-scale structure fluctuations in the simulated \lya forest, but without correlations along and between sightlines), with the same parametrization as for the other reconstructed sets. In the following, we will quantify the deviation from the signal produced by this ``noise-only" realisation.

\begin{figure*}
\includegraphics[width=0.99\textwidth]{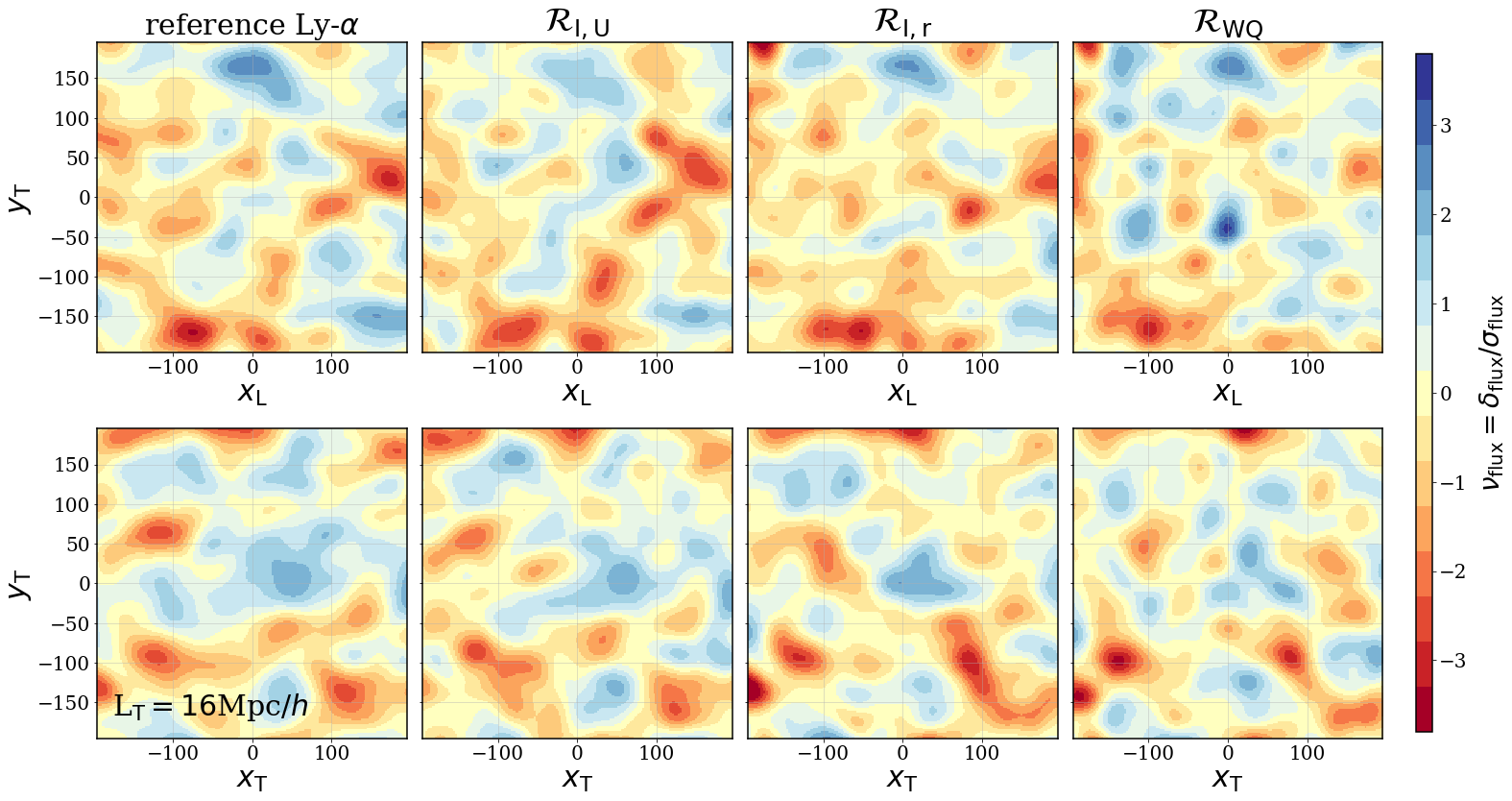}
\caption{A visualisation of 32 ${\rm Mpc/h}$ thick slices (parallel to the line-of-sight) in the flux contrast (in units of rms fluctuations) of the reference Ly-$\alpha$ field (\textit{left}, smoothed at 16 Mpc$/h$) and the reconstructed fields ($L_{\rm T}=16$~Mpc$/h$) when increasing sparsity/noise 
in the input dataset: a sparse but regular distribution of sightlines without noise on spectra ($\mathcal{R}_{\rm I,U}$, \textit{middle left}), a random distribution of sightlines without noise 
{added to} spectra  ($\mathcal{R}_{\rm I,r}$, \textit{middle right}) and a  random distribution of sightlines with a realistic SNR distribution (\textit{right}). As sparsity and noise increase, structures tend to be more disconnected (which will create more critical points, in particular wall-like and filament-like saddle points).
\textit{The bottom panel} shows transverse slices (i.e. perpendicular to the line-of-sight) through the same fields. The smoothing scale is identical in the \textit{top} and \textit{bottom} rows. 
}
\label{fig:Reconstruction}
\end{figure*}

\subsection{Inversion of the Lyman-$\alpha$ forest}
\label{sec:reconstruction}
\subsubsection{Reconstruction method}
The three-dimensional distribution of the \lya flux contrast $\delta$ is reconstructed by interpolating between the lines-of-sight using Wiener filtering in comoving space \citep[see][]{Pichon2001,Caucci2008,Lee2018}\footnote{See also, e.g. \cite{Zihao2021} for an alternative reconstruction method.}. 

Let	$\mathbf{D}$ be the 1-dimensional array representing the dataset (all sightlines placed end to end), and $\mathbf{M}$ is the 3-dimensional array  of the field estimated from the data.  Maximizing the penalized likelihood of the data given an assumed (zero mean Gaussian) prior for the flux contrast field yields
	\begin{equation}
	\textbf{M} = \mathbf{C}_{\delta^{3d} \!\delta } (\mathbf{C_{\delta \!\delta}} +\mathbf{N})^{-1} \mathbf{D}\,,
	\end{equation}
where $\mathbf{C}_{\delta^{3d} \!\delta}$ is the mixed parameter-data covariance matrix, and $\mathbf{C_{\delta \!\delta}}$ is the data covariance matrix. We assume that the noise is uncorrelated, therefore the noise covariance matrix can be expressed as $\mathbf{N}=n^{2}\mathbf{I}$. In addition, for  simplicity (and as commonly assumed in the literature), we assume a  normal covariance matrix  prior:
\begin{equation}
\mathbf{C_{\delta\! \delta}}(x_{1},x_{2},\mathbf{x}_{1\rm T},\mathbf{x}_{2\rm T})={\sigma^{2}_{\delta \!\delta}}\,e^{-\dfrac{\vert x_{1}-x_{2} \vert ^{2}}{2 L_{x}^2}}\!\!e^{-\dfrac{\vert \mathbf{x}_{1\rm T}-\mathbf{x}_{2\rm T} \vert^{2}}{2 L_{\rm T}^{2}}}\,,
\end{equation}
where ($x_{i}$,$\mathbf{x}_{i\rm T}$) are the coordinates of  points along and perpendicular to the line-of-sight. The mixed parameter-data covariance matrix $\mathbf{C}_{\delta^{3d} \!\delta}$ is taken of the same form. In principle one could get directly a better estimate of the data-data covariance matrix from the simulation. \cite{Ozbek2016} investigated how the reconstruction depends on the form used for the covariance matrix, and concluded that, at the scales they considered ($\sim 30~{\rm Mpc}/h$), the precise form of the covariance matrix has little impact.  However, at the scales of few ${\rm Mpc}/h$ probed in our study, the covariance matrix of the underlying flux density contrast field is expected to deviate more strongly from this normal form. Because generating covariance matrices is a computationally intensive process, investigating how the normal approximation impacts the quality of the reconstruction is beyond the scope of the present work. 

The reconstruction depends on the  normalisation,  $n^2/\sigma^{2}_{\delta \!\delta}$, involving the ratio of the noise matrix amplitude  (used for stabilizing the reconstruction) and the data-data prior covariance amplitude, and also on the correlation lengths  $L_{x}$  and $L_{\rm T}$, along and perpendicular to the line-of-sight respectively. 

\subsubsection{Specific settings toward WEAVE-QSO}
\label{sec:setWEAVEQSO}
For the  \svar mocks, $n^2$ is determined from the noise  on each sightline, so that the contribution of noisy sightlines to the reconstructed map is filtered. However, we set a cap to ${\rm SNR}=16$ to avoid the reconstruction being dominated by a few sightlines with very high SNR. 
The variance $\sigma_{\delta \delta}^2=0.06$ was directly estimated from the variance of the \lya flux on the noiseless simulated spectra.
For the data covariance matrix (encoding the correlation in the input simulation, the resolution of which is $\sim 2~{\rm Mpc}/h$, see Sec.~\ref{sec:simus}), we use a correlation length of $2~{\rm Mpc}/h$ in the three directions, while for the parameter-data covariance matrix we adopt  $L_{x}=2~{\rm Mpc}/h$ (which corresponds to our spectral resolution) and $L_{\rm T}=16~{\rm Mpc}/h$. This value for the transverse correlation length $L_{\rm T}$ is chosen because we cannot hope to reconstruct structures at a smaller scale than roughly the mean distance between sightlines.
Appendix~\ref{appendix:quality} explores how the reconstruction degrades when decreasing \lt, while Appendix~\ref{appendix:rarity} shows its impact on the clustering of critical points. 

In order to obtain an isotropic field, which is necessary in our analysis to investigate the clustering of critical points, the reconstructed three dimensional map is subsequently smoothed with an anisotropic Gaussian kernel of standard deviation 2 Mpc$/h$ in the transverse direction and 16 Mpc$/h$ in the longitudinal direction, which ensures a globally isotropic reconstruction at a scale of $\sqrt{16^2+2^2} \sim 16.1~{\rm Mpc}/h$. 

Because of the noise on spectra, some pixels on the input dataset can exhibit (non-physical) flux values larger than 1 or smaller than 0. Before performing the reconstruction, we cap these values to 1 and 0 respectively. 
Finally, to save time, the reconstruction is performed in parallel on overlapping boxes with a larger buffer regions of width $2.5\times {L}_{\rm T}$ (we checked that decreasing the width of the buffer region leads to spurious critical points).

After having performed the reconstruction on all set of simulated spectra (\siu, \sir, \svar and \noise), the flux contrast in the reconstructed map is converted into a pseudo \hi density, using the following transformation: $f:\delta \mapsto -\log ((\delta+1)\times \langle F \rangle)$, where $\langle F \rangle=0.795$ is the mean \lya flux in the simulation. In practice, given that the fluctuations in the flux contrast are of small amplitude (because the field is smoothed at such large scales), $f(\delta)\sim -\delta + \log(\langle F \rangle)$. The same transformation is applied to the  simulated Ly-$\alpha$ reference field after smoothing with an isotropic Gaussian kernel of standard deviation 16 Mpc$/h$.

Throughout this paper, these smoothed reconstructed maps converted into the pseudo \hi density are also compared with the original DM and {\sc Hi} smoothed with an isotropic Gaussian kernel of standard deviation 16 Mpc$/h$. We note that the only difference between the Ly-$\alpha$ reference field and the \hi reference field is that for the former smoothing has been applied on the flux before converting it into pseudo \hi density, while for the latter the order of the transformations is reversed.
Note finally that in what follows all statistics are computed in units of the root-mean square (rms) fluctuations of the field. 

\subsubsection{Visualisation of the reconstruction}

Fig.~\ref{fig:Reconstruction} shows the reconstruction in the various configurations in units of the rms fluctuations of each field. Projection in slices of thickness 32 Mpc/$h$ (twice the smoothing scale) parallel and perpendicular to the line-of-sight are plotted in the {top} and {bottom} panels respectively, 
for the Ly-$\alpha$ reference field ({extreme left}), the reconstructed field with a regular distribution of sightlines \siu ({middle left}), the reconstructed field with a random distribution of sightlines \sir ({middle right}) and the reconstructed field with a realistic noise on spectra \svar ({extreme right}). Adding noise on spectra and increasing sparsity creates more structures, which will be reflected in change in  the topology of the excursion set and therefore in the number and clustering of critical points.

\begin{figure}
\includegraphics[width=0.99\columnwidth]{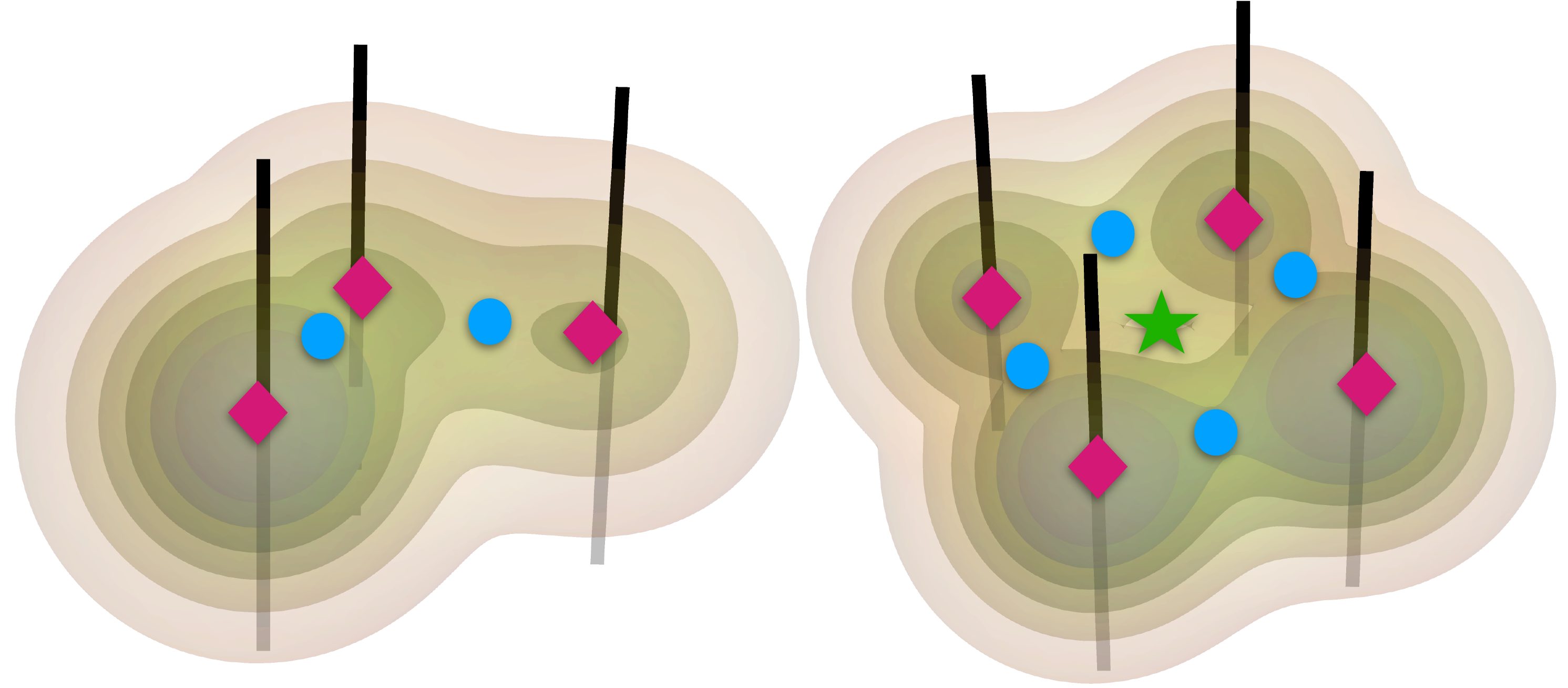}
\caption{The impact of adding noise along the line-of-sight on the topology of the  reconstructed field hence the number of critical points. 
For example, on the left panel, positive  noise along three (vertical) LOS (separated by more than the smoothing scale) are sufficient to generate two 
 spurious filament-like saddle points (blue circle), on top of the three generated maxima (magenta diamond).
On the right panel, positive noise along four  LOS   generates an extra spurious wall-like (green star) saddle point in between the line-of-sights.  Illustratively, uncorrelated noise could lead to such configurations resp. 1/4 and $1/8^\mathrm{th}$ of the time.
}
\label{fig:Noise-generate-extra}
\end{figure}

Fig.~\ref{fig:Noise-generate-extra} describes qualitatively how noise along the line-of-sight
may induce the appearance of extra critical points in their vicinity.
Assuming that the noise dominates the large-scale structure density within that region, the reconstruction of  spurious over-densities  will bridge the field in between sightlines. Such bridges will
contain extra filament-like and possibly wall-like saddle points (shown using the same colour coding as the previous figures). 
This effect would mostly impact sightline separations larger than the smoothing scale.

\subsection{Critical point statistics estimators} 
The critical points of a field are the points where the gradient vanishes \citep{Milnor1963}. They are classified by the sign of the eigenvalues of the matrix of second derivatives of the field: the signature of a critical point is the number of negative eigenvalues, from three for peaks to zero for voids.
For their identification, we implement a local quadratic estimator  based on a second-order Taylor-expansion of the density field \citep[see ][Appendix~G]{Gay2012}. 

In the following, the critical points are extracted from the \textit{smoothed density} fields. 
In the main text, the adopted smoothing scale is 16 Mpc/$h$. For DM and \hi, we first smooth the density fields with an isotropic 3D Gaussian kernel (of standard deviation 16~Mpc/$h$) before extracting the critical points.  For both  the \lya reference field (smoothed at 16 Mpc/$h$) and the reconstructed fields \siu, \sir, 
\svar and $\mathbf{\mathcal{N}}$, we applied first the transformation to pseudo \hi density (described at the end of section~\ref{sec:setWEAVEQSO}) before extracting the critical points. 

Following \cite{Shim2021}, we rely on the so-called Davis-Peebles estimator \citep{1983ApJ...267..465D} for the (cross) correlations, $\xi_{ij}$,
which is given by
\begin{equation}
1+\xi_{{ij}}(r) = \frac{\langle C_{i}C_{j}\rangle}{\sqrt{\langle C_{i}R_{j}\rangle \langle C_{j}R_{i}\rangle }}
\sqrt{\frac{N_{R_i} N_{R_j}}{N_{C_i} N_{C_j}}}.
\label{eq:xi}    
\end{equation}
Here $C_{i}$ stands for a catalogue of critical points $i\in\{{\cal P},{\cal F},{\cal W },{\cal V}\}$, while $R_i$ is a  catalogue with randomly distributed points following a uniform probability distribution in the same volume.  
The expectation,  $\langle XY \rangle$, measures the  number count of  pairs of critical points $X$ and $Y$  separated by $r$. The sample size, $N_{R_i}$, of the random catalogue is a factor of $100$ larger than the corresponding size of the simulated set $N_{C_i}$.

To avoid edge effects, we choose to discard from the statistics the critical points which are closer than $\approx 9.8$ Mpc/$h$ (5 grid cells) from the side of the box\footnote{Note that the simulation boxes are periodic by construction, however, this periodicity is not fully preserved by our implementation of the reconstruction.}. 

\section{Statistics of  critical points}
\label{sec:reconstruction}

Let us now quantify our ability to recover the statistics of critical points from WEAVE-QSO-like survey.
Since we follow closely the choices (estimator, rarity) made in \cite{Shim2021}, which strongly impacts the shape of the correlation functions, we will mostly focus our discussion  on the relative difference between the input and recovered relations, given that the origin of the features have been addressed in that paper.


\subsection{Critical points total number  counts}
\label{sec:counts}

Table~\ref{tab:crits_pts} shows the numbers of critical points of each type (peaks, filaments, walls and voids) in the flux contrast of the \lya and three reconstructed fields. For completeness, we also provide  the corresponding numbers for the DM and \hi density fields, as well as for the noise-only field $\mathbf{\mathcal{N}}$ that will serve as a reference for the quantification of the reconstructed signal.
For the DM density field, the number of peaks is higher than the number of voids and the number of filaments is higher than the number of walls. This deviates, as expected, from Gaussian random field (GRF) predictions, for which, due to the symmetry, the number of extrema (peaks and voids) is predicted to be equal, and similarly for  saddle points (filaments and walls), as is the case of the noise-only field \noise. In addition, for GRF, the ratio of the number of filaments to the number of peaks (or walls to voids) is predicted to be exactly $(29\sqrt{15}+18\sqrt{10})/(29\sqrt{15}-18\sqrt{10})\approx 3.05$, see Appendix~\ref{appendix:GRF}.

It is also expected \citep[see][]{Gay2012}, that at first non-Gaussian perturbative order, the total number of extrema (voids and peaks) and the total number of saddles (filaments and walls) is preserved, and therefore also their ratio.
Given the smoothing scales involved, we should not be far from this regime.
Indeed, for the DM density field, filaments and walls are found to be about 3 times (3.1 and the ratio is $\sim$ 2.8 after the borders removal) more abundant than peaks and voids.
Similarly, for the \hi density field derived from the DM density, the ratio between the number of saddles (filaments together with walls) and the number of extrema (peaks together with voids) is close to 3 (3.1 and $\sim$ 2.8 after the borders removal), and the number of identified peaks is larger than the number of voids, see Table~\ref{tab:crits_ratio_pts}. 
This ratio is  half the mean connectivity of the cosmic web \citep[][see \S~\ref{sub:connect}]{codisetal2018}. As such its robustness 
is not unexpected.

In contrast, filament-type saddles are found to be less numerous than walls. This could be caused by the effect of reduced periodicity of the \hi field as suggested by the fact that the trend reverses after the removal of the critical points near the borders. In practice, we remove five pixels (corresponding to $\approx 9.8$ Mpc/$h$) from each side of the box along each direction  to ensure consistency 
w.r.t the reconstructed fields without  periodicity.

For the \lya reference field the number of voids is higher than the number of peaks and the number of walls is higher than the number of filaments even after  boundary trimming. However, the ratio of saddles over extrema is preserved ($\sim$ 3.1 and 2.8 after the boundary removal, see Table~\ref{tab:crits_ratio_pts}).

For all types of reconstruction, the number of peaks is larger than the number of voids and the number of filaments is larger than the number of walls and the ratio of saddles over extrema is also close to 3.
However, the number of critical points in any reconstructed field is higher than the number of critical points of the original field. This fraction is lowest for \siu (1.15 at \lt = 16 Mpc/$h$) and highest for \svar (1.35 at \lt = 16 Mpc/$h$). This is expected, since \siu corresponds to the less noisy reconstruction (regular distribution of sightlines and no noise on spectra). Overall, all the different sources of noise (sparsity of the sightline distribution, irregularity of their spatial distribution, noise on spectra) result in an increase of the number of critical points.
For all types of reconstruction, this fraction is slightly higher for peaks than for voids. 

Nonetheless, as expected for sufficiently large volumes \citep[][]{Shim2021}, the ratio between the number of peaks and walls over filaments and voids remains close to one for all fields and reconstructions.

\begin{table}
\centering
\caption{Mean number of peaks ($\mathcal{P}$), filaments ($\mathcal{F}$), walls ($\mathcal{W}$), and voids
($\mathcal{V}$) for the  DM density, the \hi density, the Ly-$\alpha$ reference field and the three types of reconstruction used in this work at smoothing scale 16 Mpc/$h$. The second line reports the number of critical points after the borders removal. The errors represent the standard deviations.} 
\label{tab:crits_pts}
\begin{tabular*}{\columnwidth}{@{\extracolsep{\fill}}lcccc}
\hline
\hline
 & $\mathcal{P}$  & $\mathcal{F}$ & $\mathcal{W}$ & $\mathcal{V}$\\
\hline
\multirow{2}{*}{DM (density)} & 2045$\pm{38}$ & 6264$\pm{62}$ & 6170$\pm{58}$ & 1958$\pm{17}$ \\
& 1908$\pm{39}$ & 5273$\pm{59}$ & 5236$\pm{37}$ & 1874$\pm{17}$ \\
\hline
\multirow{2}{*}{\hi{} (density)} & 2161$\pm{27}$ & 6599$\pm{72}$ & 6736$\pm{68}$ & 2080$\pm{26}$ \\
& 2013$\pm{25}$ & 5587$\pm{63}$ & 5544$\pm{46}$ & 1976$\pm{23}$ \\
\hline
\multirow{2}{*}{Ly-$\alpha$} & 2132$\pm{27}$ & 6627$\pm{64}$ & 6644$\pm{59}$ & 2156$\pm{24}$ \\
& 2002$\pm{27}$ & 5604$\pm{56}$ & 5633$\pm{48}$ & 2031$\pm{26}$ \\
\hline
\multirow{2}{*}{\siu} & 2582$\pm{24}$ & 7678$\pm{61}$ & 7531$\pm{82}$  & 2409$\pm{38}$  \\
& 2376$\pm{25}$ & 6565$\pm{51}$ & 6410$\pm{69}$ & 2242$\pm{33}$ \\
\hline
\multirow{2}{*}{\sir} & 2697$\pm{34}$ & 7858$\pm{115}$ & 7742$\pm{86}$  & 2578$\pm{41}$\\
& 2463$\pm{22}$ & 6717$\pm{30}$ & 6606$\pm{50}$ & 2375$\pm{37}$ \\
\hline
\multirow{2}{*}{\svar} & 3003$\pm{34}$ & 8883$\pm{143}$ & 8856$\pm{192}$  & 2964$\pm{28}$  \\
& 2718$\pm{25}$ & 7627$\pm{45}$ & 7603$\pm{73}$ & 2714$\pm{28}$ \\
\hline
\multirow{2}{*}{\noise} & 3202$\pm{49}$ & 9576$\pm{150}$ & 9583$\pm{146}$  & 3191$\pm{39}$  \\
& 2894$\pm{41}$ & 8196$\pm{105}$ & 8200$\pm{97}$ & 2911$\pm{36}$ \\
\hline
\end{tabular*}
\end{table}

\begin{table}
\centering
\caption{Fraction of the mean number of filaments over peaks ($\mathcal{F}/\mathcal{P}$), filaments over walls ($\mathcal{F}/\mathcal{W}$) and voids over peaks ($\mathcal{V}/\mathcal{P}$) for the \lya reference field, the three types of reconstruction used in this work and noise-only field at smoothing scale 16 Mpc/$h$.} 
\label{tab:crits_ratio_pts}
\begin{tabular*}{\columnwidth}{@{\extracolsep{\fill}}lcccc}
\hline
\hline
 & $\mathcal{F}/\mathcal{P}$  & $\mathcal{F}/\mathcal{W}$ & $\mathcal{V}/\mathcal{P}$ & $\mathcal{W}/\mathcal{V}$\\
\hline
\lya & 3.11$\pm{0.02}$ & 0.99$\pm{0.01}$ & 1.01$\pm{0.01}$  & 3.08$\pm{0.02}$\\
\siu & 2.97$\pm{0.02}$ & 1.02$\pm{0.01}$ & 0.93$\pm{0.01}$ & 3.08$\pm{0.03}$\\
\sir & 2.91$\pm{0.03}$ & 1.02$\pm{0.01}$ & 0.96$\pm{0.01}$ & 3.0$\pm{0.03}$\\
\svar & 2.96$\pm{0.03}$ & 1.0$\pm{0.01}$ & 0.98$\pm{0.01}$ & 2.98$\pm{0.03}$\\
\noise & 2.99$\pm{0.02}$ & 0.99$\pm{0.01}$ & 0.99$\pm{0.01}$ & 3.0$\pm{0.02}$\\
\hline
\end{tabular*}
\end{table}

\subsection{One point function of critical points}
\label{sec:onept}

\begin{figure*}
\includegraphics[width=0.48\textwidth]{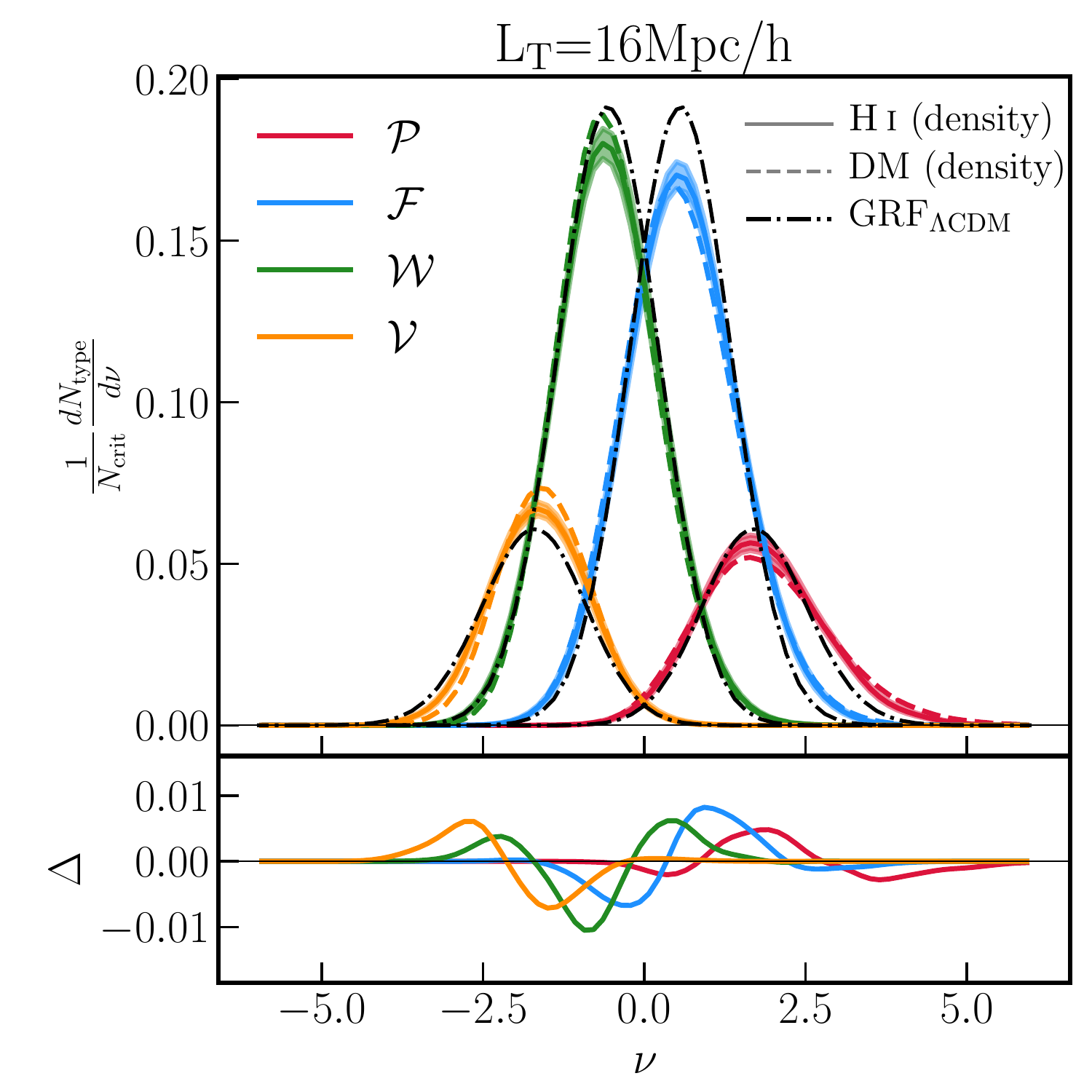}
\includegraphics[width=0.48\textwidth]{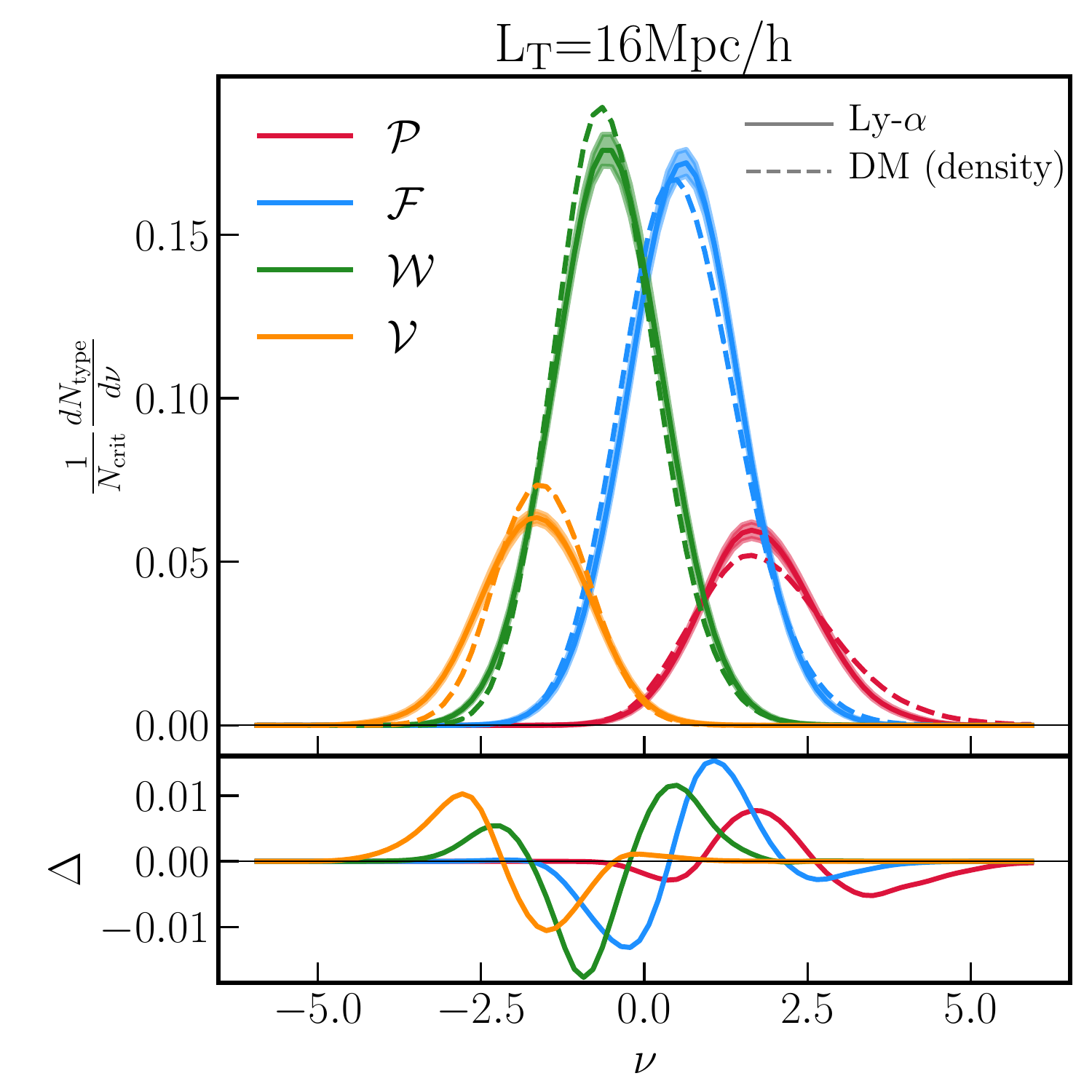}
\caption{\textbf{Effect of bias.} \textit{Top:} Relative number counts of critical points ($\mathcal{P}$: peaks, $\mathcal{F}$: filaments, $\mathcal{W}$: walls, and $\mathcal{V}$: voids) as functions of their rarity $\nu$ for the \hi{} density and \lya (solid lines, left and right panels, respectively), compared to dark matter density fields (dashed lines) and Gaussian random field with $\Lambda$CDM power spectrum (dash-dotted lines) at the smoothing scale of 16 Mpc/$h$. Shaded area corresponds to the standard deviation across five mocks, shown for clarity only for the \hi{} field.
\textit{Bottom:} Difference of the relative number counts of critical points in the \hi{} and dark matter density (left), and \lya and dark matter density (right).
}
\label{fig:1pt_stats_HI_dm_16LT}
\end{figure*}

Let us now study the distribution  of the critical points number counts as a function of  rarity in both the original and reconstructed fields. 
The rarity of the critical point is defined as $\nu \equiv \delta/\sigma$, with the over-density contrast of the smoothed density field $\delta \equiv \rho/\overline{\rho} - 1$ and $\sigma$ the rms fluctuation of the field $\sigma^2 \equiv \left<\delta^2 \right>$. Our purpose in choosing rarity is to  sample  populations that represent the same abundance for a given type of critical points. This allows us to limit the number of configurations we investigate. The lack of overlap in rarity values induces exclusion zones in correlation functions \citep[see \S\ref{sec:auto} below and Appendix~B of ][for a more extended discussion]{Shim2021}.  

To quantify the effect of bias, Fig.~\ref{fig:1pt_stats_HI_dm_16LT} shows the relative number counts of critical points (top panel) and their difference (bottom panel) as function of rarity for the DM and \hi density fields (left) and for DM density and \lya reference fields (right). For each type of critical point the number counts are normalised by the total number of critical points. Let us also stress that the rms fluctuation of each field used to define $\nu$ is computed independently for each of them.
In particular, this means the linear bias is factored in \citep[see e.g.][for the \hi linear bias]{2010MNRAS.407..567B}.
The bottom panel of the figure thus allows us to probe the non-linear bias of different critical points.
The maximum amplitude of the rarity distribution is higher for voids (walls) than for peaks (filaments) with the effect being stronger for DM density field. The  distributions clearly show a positively-skewed rarity  driven by gravitational clustering (as expected, see \cite{Gay2012}). 
At low rarity, more voids are identified in the \hi density fields compared to the DM density field, while at rarity corresponding to the maximum of the rarity distributions the trend is reversed. Similar behaviour is found for rarity distribution of walls, such that there are more walls at low rarity in the \hi density field compared to DM density field, while at intermediate rarity the number of walls is higher in the DM field. On the other hand, the trends are reversed for filaments and peaks. At their respective intermediate rarity (near the maxima of rarity distributions), more filaments (peaks) are identified in the \hi density field than in the DM density field, while at highest rarity the number of filaments (peaks) is higher for the DM. 
Interestingly, the differences of the relative number counts (bottom panel) for peaks and voids are symmetric w.r.t. $\nu = 0$ (similarly for filaments and walls), which is likely to reflect the fact that these critical points are oppositely biased tracers.
Qualitatively, similar trends are found when comparing \lya fluxes with the DM density field, with enhanced relative differences highlighting the observational  bias associated with measuring the fluxes. This is reflected by the differences in  the tails of the PDFs, capturing  rare events such as the flux saturation  of the densest peaks. 

\begin{figure}
\includegraphics[width=0.48\textwidth]{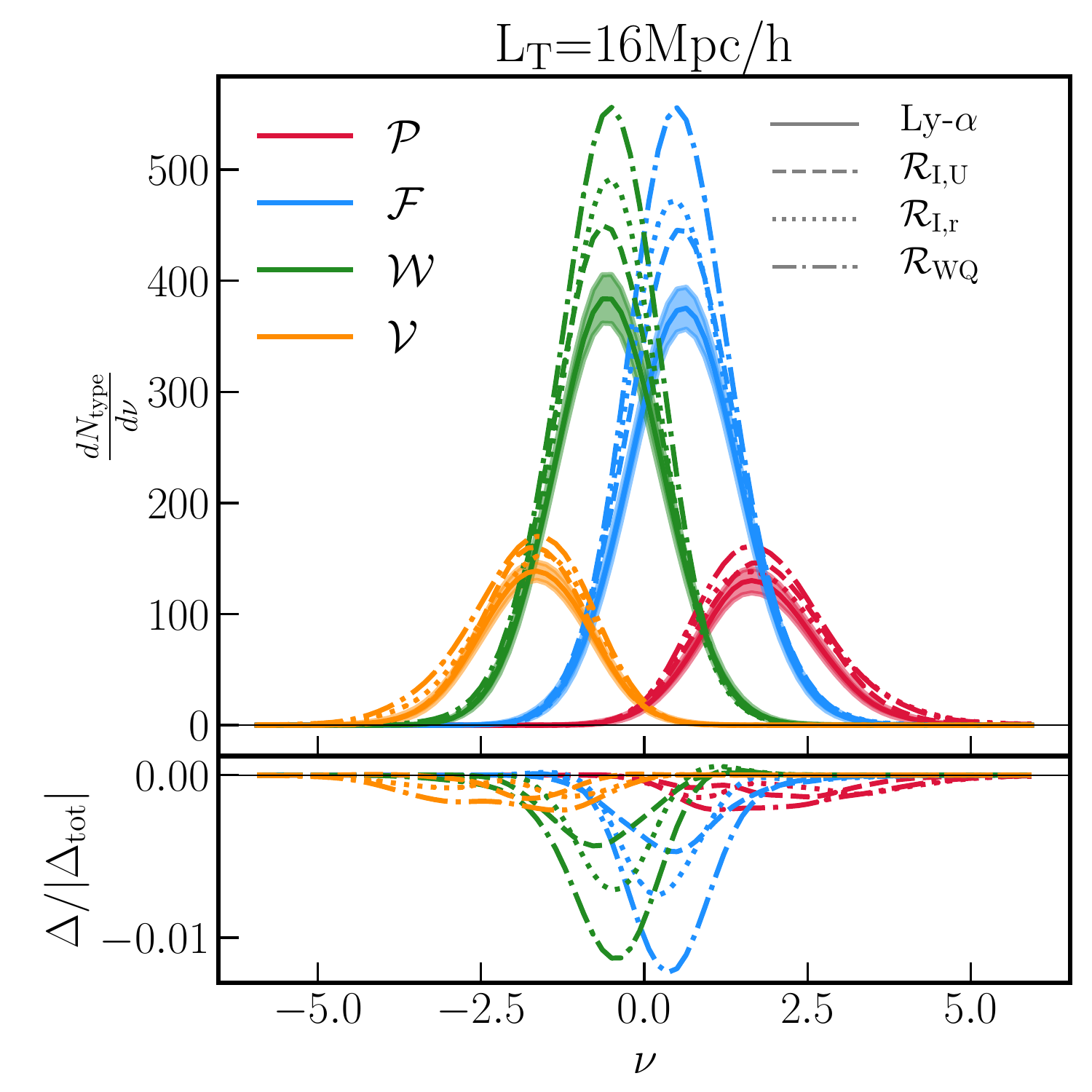}
\caption{\textit{Top:} Absolute number counts of critical points ($\mathcal{P}$: peaks, $\mathcal{F}$: filaments, $\mathcal{W}$: walls, and $\mathcal{V}$: voids) as functions of their rarity $\nu$ for the \lya (solid lines) and reconstructed fields \siu (dashed lines), \sir (dotted lines) and \svar (dash-dotted lines). \textit{Bottom:} Difference of the number counts of critical points of a given type normalised by the absolute value of the difference of the total number of critical points in the original and reconstructed fields. The shaded area corresponds to the error on the mean across the five mocks, shown for clarity only for the \lya reference field.
}
\label{fig:1pt_stats_all}
\end{figure}

Let us now focus on the comparison between the \lya reference field and the three types of reconstruction.
Fig.~\ref{fig:1pt_stats_all} shows the number counts of critical points as a function of rarity for the \lya flux (solid lines) together with the three configurations adopted for the reconstruction, namely \siu (dashed lines), \sir (dotted lines) and \svar (dash-dotted lines). 
As already noted in Section~\ref{sec:counts}, the number of critical points in any reconstructed field is larger than in the \lya reference field, with smallest differences for \siu and largest for \svar. 
The differences show a dependence on the rarity of a critical point. For filaments and walls they are confined to regions of intermediate rarity (in the vicinity of the maxima of the number counts distributions), while for peaks and voids the differences are more uniformly distributed over a much larger range of rarities. While the largest differences between the \lya and reconstructed fields are measured for filaments and walls, cumulative differences (for all rarities) are larger for peaks and voids (see Section~\ref{sec:counts}).

\subsection{Auto-correlation of critical points}
\label{sec:auto}
\begin{figure*}
\includegraphics[width=0.45\textwidth]{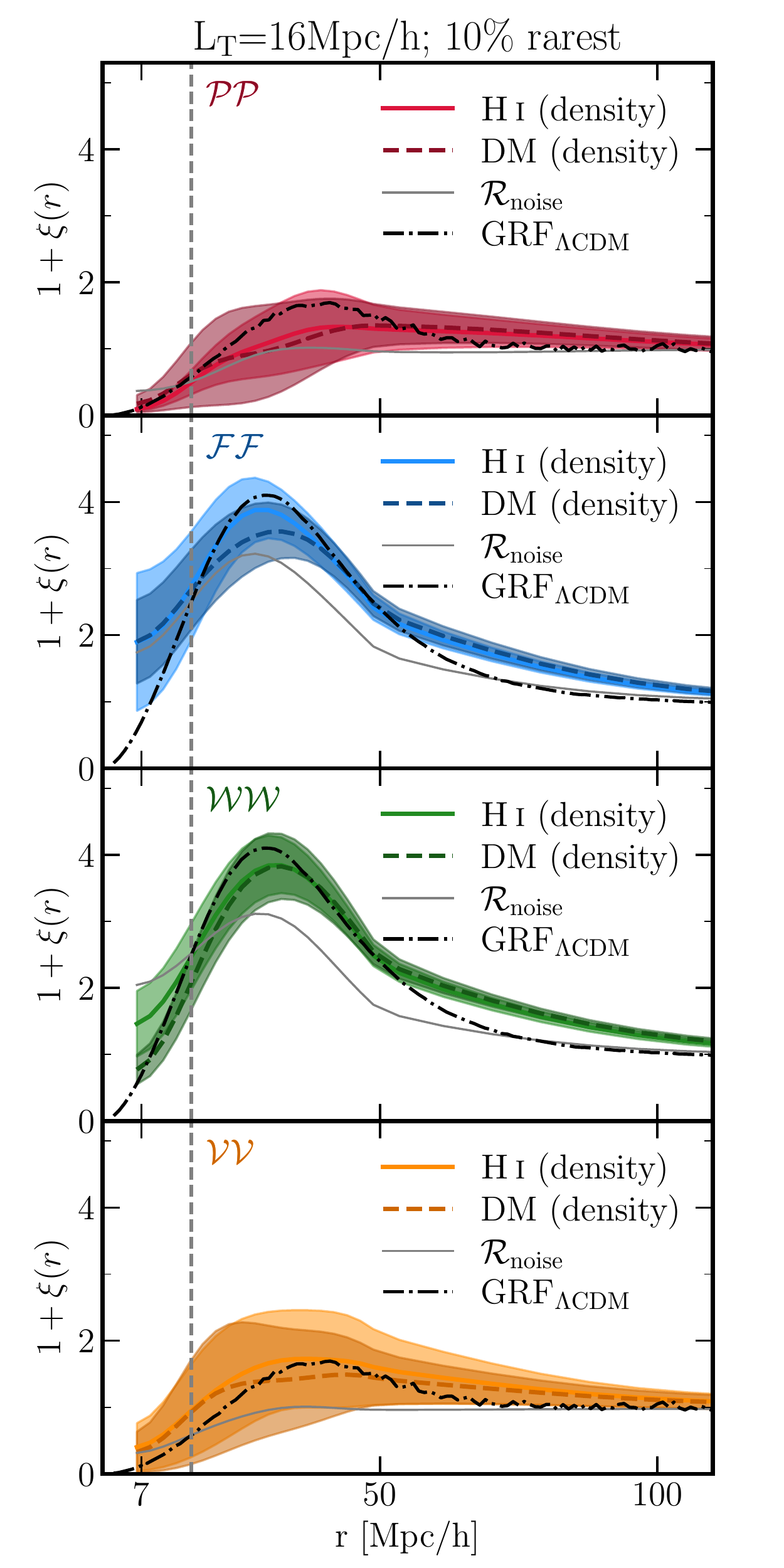}
\includegraphics[width=0.45\textwidth]{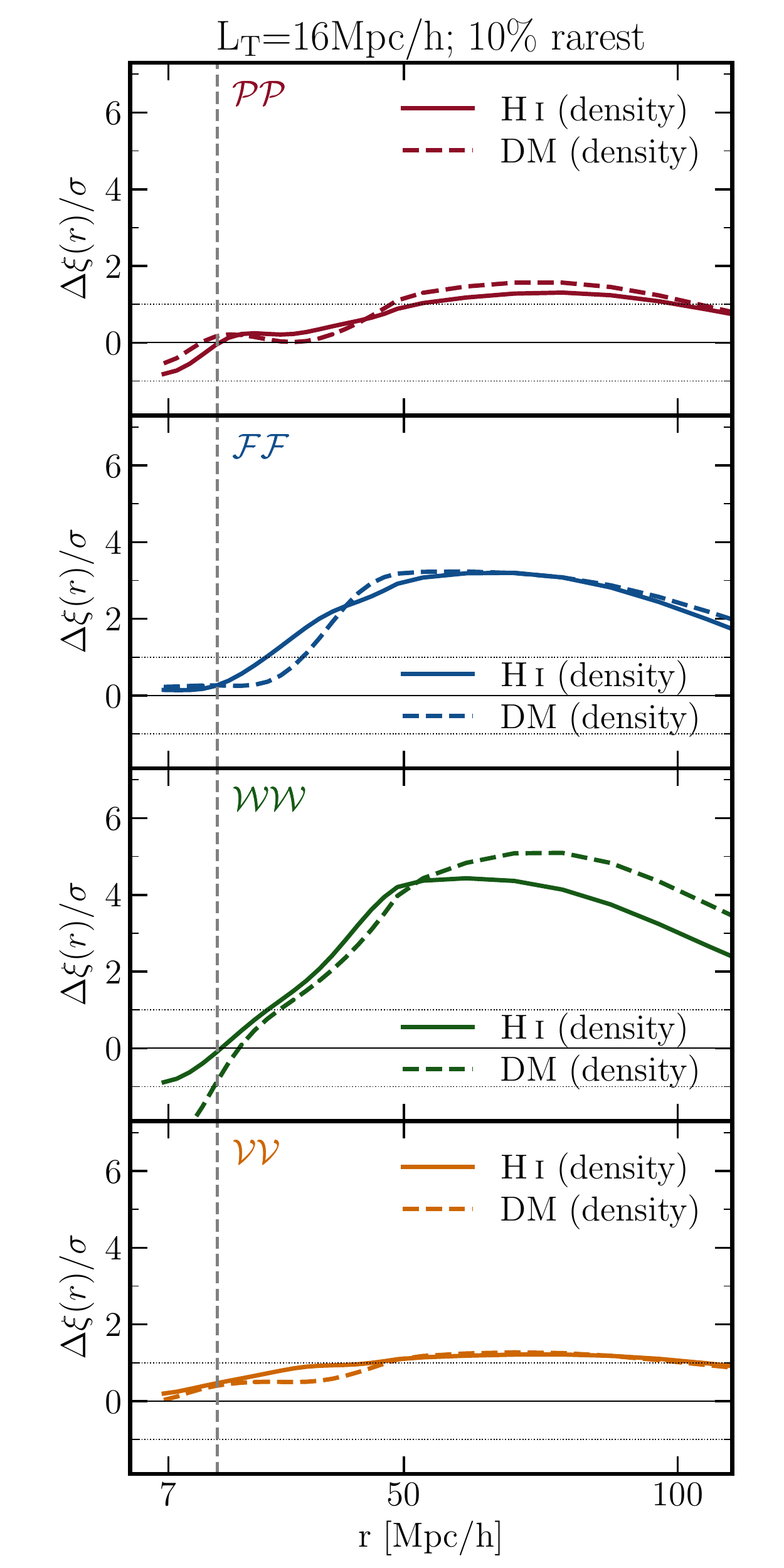}
\caption{\textit{Left:} Auto-correlations of critical points with 10\% abundance. ${\mathcal{PP}}$ (peak-peak), ${\mathcal{FF}}$ (filament-filament), ${\mathcal{WW}}$ (wall-wall), and ${\mathcal{VV}}$ (void-void) correlations are shown for \hi density field (coloured solid lines), DM field (coloured dashed lines) and field containing noise only (grey solid line).
Predicted auto-correlations of critical points for $\Lambda$CDM spectrum  are shown for comparison (black dash-dotted lines).
Smoothing of the fields is 16 Mpc/$h$. Shaded area corresponds to the error on the mean across five mocks. 
\textit{Right:} Differences of auto-correlations of critical points with respect to the noise in the units of $\sigma$ of a given field ($i$) and noise (\noise), i.e. $\sigma=\sqrt{\sigma_i^2+\sigma_{\mathcal{N}}^2}$, that we use to assess the significance of the measured signal.
}
\label{fig:auto_dm_rarity10_16}
\end{figure*}
%
Let us now move to the two-point statistics, starting with auto-correlation functions. Throughout the paper, we show results for the 10\% rarest critical points selected as follows. For peaks and filaments (resp. voids and walls), the critical points are extracted above (resp. below) the rarity yielding a given abundance. The choice of the 10\% rarity is a compromise between very rare outstanding events and therefore noisy measurements on the one hand, and less rare events with less enhanced characteristic features on the other hand \citep[see also][for a discussion regarding this choice]{Shim2021}.  The relative impact of this specific choice of rarity is addressed in Appendix~\ref{appendix:rarity}.
%
\begin{figure*}
\centering\includegraphics[width=0.45\textwidth]{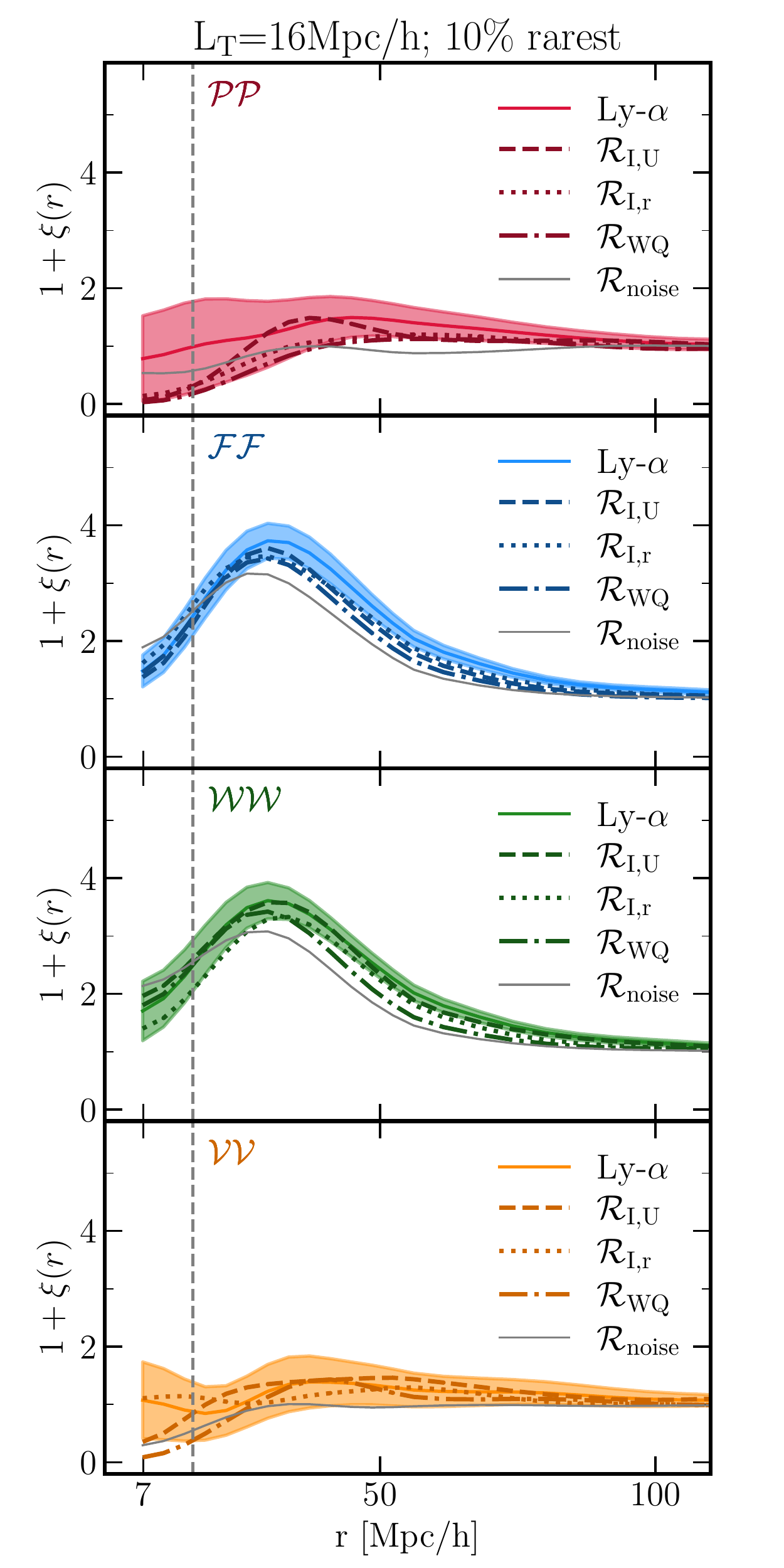}
\centering\includegraphics[width=0.45\textwidth]{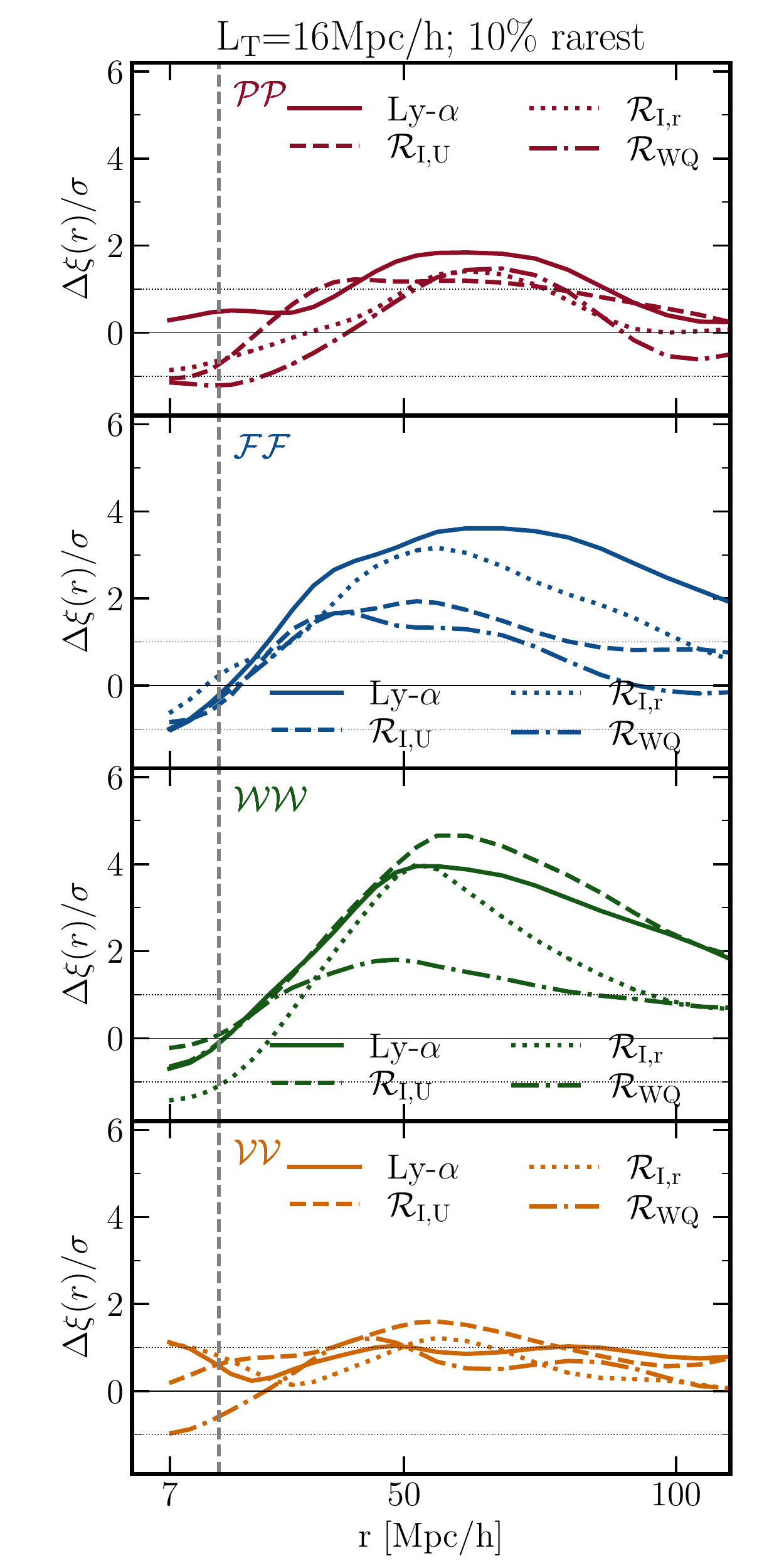}
\caption{Auto-correlations of critical points with 10\% abundance for the \lya flux and reconstructed fields (\textit{left}), and their relative difference with respect to the noise (\textit{right}). $\mathcal{PP}$ (peak-peak), $\mathcal{FF}$ (filament-filament), $\mathcal{WW}$ (wall-wall), and $\mathcal{VV}$ (void-void) correlations are shown for \lya field (solid lines) and reconstructed fields \sir (dashed lines), \siu (dotted lines) and \svar (dash-dotted lines).
The fields is smoothed over 16 Mpc/$h$.
Shaded area corresponds to the error on the mean across five mocks. For the sake of clarity, only errors for \lya are shown, those of \siu, \sir and \svar are comparable.  
As expected, the $\mathcal{FF}$ and $\mathcal{WW}$ correlations are best recovered. 
}
\label{fig:auto_all_rarity10_16}
\end{figure*}

Fig.~\ref{fig:auto_dm_rarity10_16} (left panels) shows the auto-correlations of the 10\% rarest critical points in the \hi density (color solid lines) and DM density (color dashed lines) fields. The corresponding auto-correlations in the Gaussian random field with $\Lambda$CDM power spectrum are shown for a comparison (black dash-dotted lines), as well as the auto-correlations for noise-only field (\noise;  thin gray solid line).
As expected, given the smoothing scale, there is a very good agreement between the \hi and DM density fields. 
The auto-correlations of all critical points follow qualitatively similar trends.
At small separations, they are negative ($\xi(r) < 0$), representing a region of anti-clustering or exclusion. They then increase,  and reach a positive maximum at $\approx 2$\lt for filaments and walls and $\approx 2.5-3$\lt for peaks and voids (see Table~\ref{tab:summary_numbers_max}),
before decreasing towards zero in the regime of large separations.
Filaments and walls show enhanced clustering at small separations, the maximum of their auto-correlation function occurs earlier, and their exclusion region is narrower compared to peaks and voids. {These differences between the auto-correlations of saddles and extrema potentially manifest different density and curvature conditions around extrema- and saddle-points \citep{Shim2021}.}

To quantify the significance of the  signal contained in these auto-correlations, we compute the difference between the density (\hi or DM) and noise-only fields auto-correlation, in units of the standard deviation ($\sigma=\sqrt{\sigma_i^2+\sigma_{\mathcal{N}}^2}$, with $\sigma_i$ and $\sigma_{\mathcal{N}}$ the standard deviations of the fields $i$, \hi or DM density, and noise-only, respectively) of the corresponding density field. As shown on Fig.~\ref{fig:auto_dm_rarity10_16} (right panels) these differences are  highest for filaments and walls, at $\approx $ 3-4$\sigma$, while for peaks and voids, the significance of the measured signal is at the level of $\approx 1\sigma$.

Fig.~\ref{fig:auto_all_rarity10_16} shows the auto-correlations of the critical points with abundance of 10\% in the \lya reference field, together with the three reconstructions (left panels) at \lt=16 Mpc/h. The corresponding differences between the auto-correlations of a given field and noise-only field are shown on the right panels.
These auto-correlations show qualitative  behaviour similar to the \hi and DM density correlations, with an anti-clustering at small separations, a maximum at 
$\approx 2$\lt for filaments and walls and at $\approx 2.5-3$\lt for peak and voids (see Table~\ref{tab:summary_numbers_max}), and a decrease towards zero at large separations. 
The quality of the reconstruction is typically the best for the regular distribution of lines of sight (\siu), it decreases for the reconstruction with the random distribution  (\sir) and degrades further when the noise on the spectra is added (\svar).  
The auto-correlations of filaments and walls are better recovered  than those of peaks and voids. To quantify  the quality of the signal contained in the \lya flux, we once again compute the differences of auto-correlations with respect to the noise-only field, shown on the right-hand side of Fig.~\ref{fig:auto_all_rarity10_16}. 
Remarkably, for filaments and walls, the most striking features in the auto-correlation functions can be measured with up to 5$\sigma$ of significance for \siu and up to 2$\sigma$ for \svar. 
Similar level of significance of the measured features in the auto-correlations is found for the critical points with 20\% rarity (see Fig.~\ref{fig:auto_rarity5_20_16}).
This is in line with the conclusions that saddle point statistics are more advantageous to use for extracting cosmological information \citep{Gay2012,Shim2021} because the cosmic evolution of saddle-points is less non-linear than that of extrema-points \citep[see Fig.~10 of][]{Gay2012}. Conversely, the constraining power for peaks and voids is limited at best to 1$\sigma$.

\subsection{Cross-correlation of critical points}
\label{sec:cross}

Let us start by considering the cross-correlation functions of over- and under-dense critical points, i.e. cross-correlations $\mathcal{PW}$, $\mathcal{PV}$, $\mathcal{FW}$ and $\mathcal{FV}$. The cross-correlations of critical points of the same over-density sign, i.e. $\mathcal{PF}$ and $\mathcal{WV}$ are addressed in \S\ref{sec:other-cross}. 

\subsubsection{Over and under-dense critical points}

\begin{figure*}
\centering\includegraphics[width=0.45\textwidth]{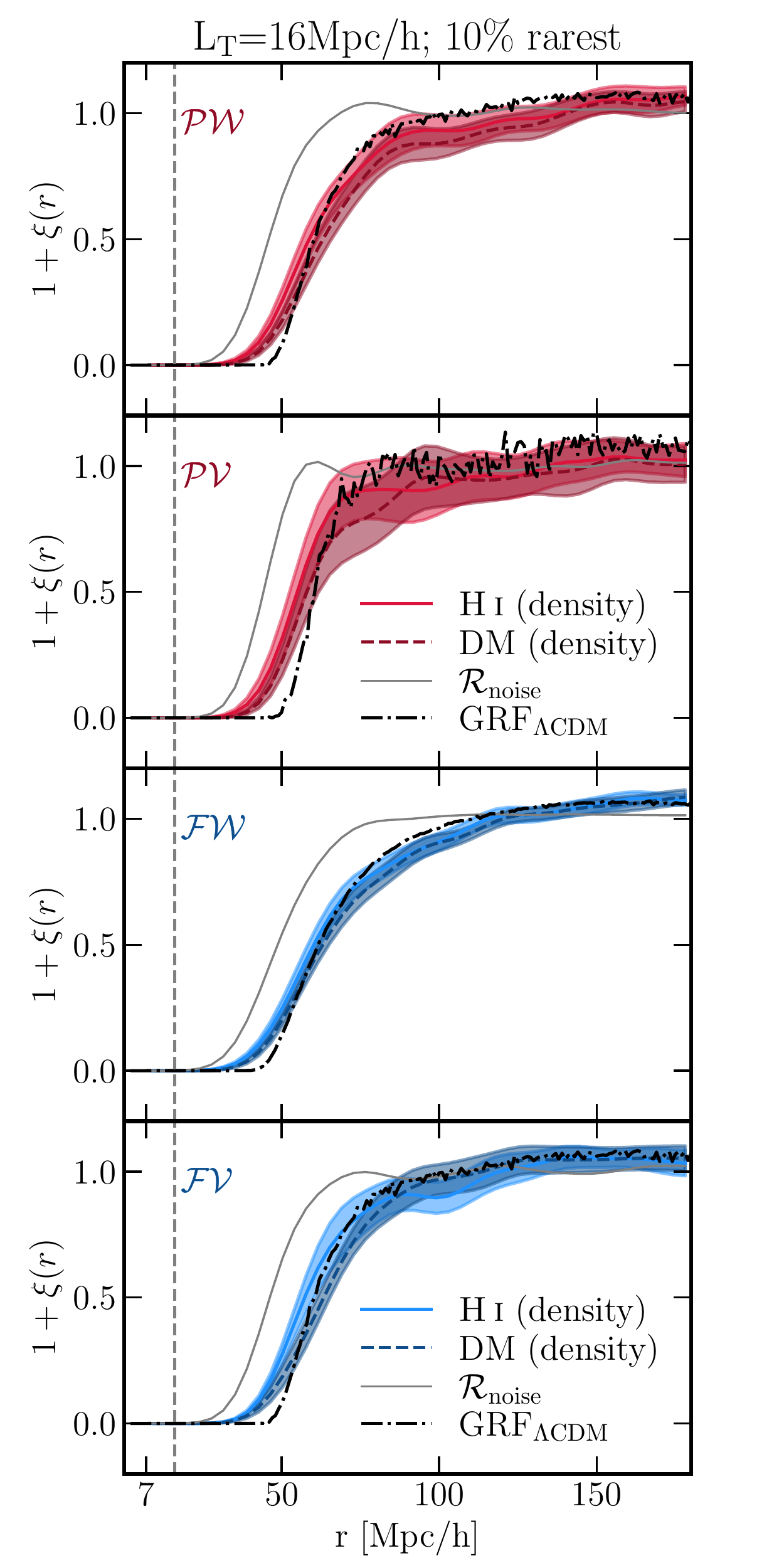}
\includegraphics[width=0.45\textwidth]{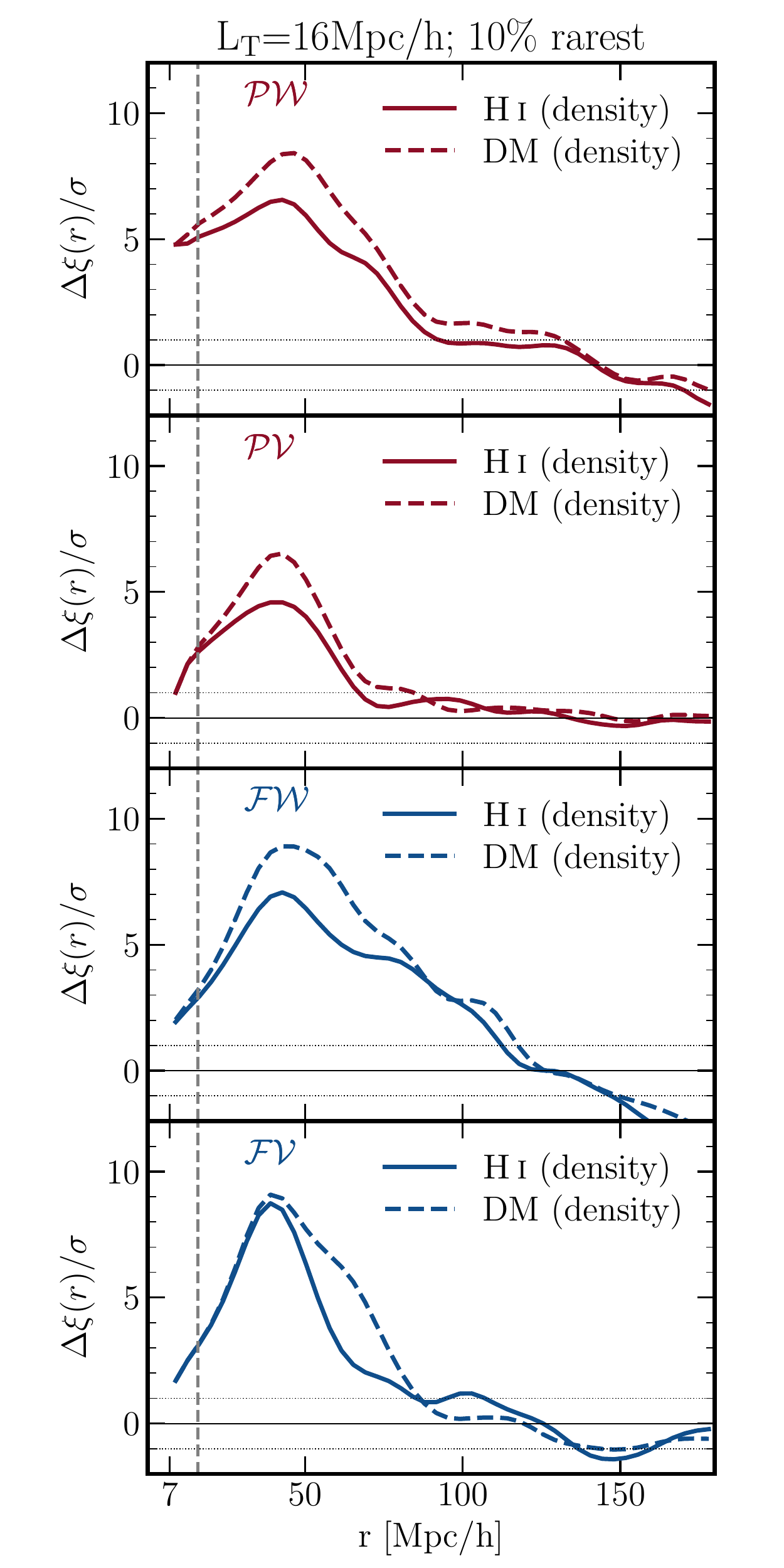}
\caption{Cross-correlations of critical points with 10\% abundance (\textit{left}) and their relative difference with respect to the noise in the units of the total sigma (multiplied by -1 for convenience, \textit{right}). $\mathcal{PW}$ (peak-wall), $\mathcal{PV}$ (peak-void), $\mathcal{FW}$ (filament-wall) and $\mathcal{FV}$ (filament-void) correlations. All these correlation's main features are  well detected, since they involve $\mathcal{F}$ and $\mathcal{W}$. As expected, the GRF correlations, which do not include peculiar velocities have wider exclusion zones.
} 
\label{fig:cc_dm_rarity10_4in1_16}
\end{figure*}

Fig.~\ref{fig:cc_dm_rarity10_4in1_16} shows such cross-correlation functions with 10\% abundance for the \hi and DM density fields (coloured solid and dashed lines, respectively). 
As in the case of auto-correlations, there is a very good agreement between the \hi and DM density fields and the cross-correlations of over- and under-dense critical points follow qualitatively similar trends. However, the cross-correlations of over- and under-dense critical points are very different from the auto-correlations. At small separations there is again an exclusion zone where $\xi(r)$ is close to -1, but then these cross-correlations monotonically increase, 
reaching zero at large separations. The size of the exclusion zone for \hi and DM density field is significantly larger ($>4\sigma$) compared to the the noise-only field for all cross-correlations of under- and over-dense critical points (see Table~\ref{tab:summary_numbers_rexc}). 
Note that the cross-correlations are negative at all separations, meaning that the over- and under-dense critical points are always anti-correlated. The appearance of the exclusion zone and  anti-clustering are consequences of the fact that these critical point pairs are oppositely biased tracers of the underlying dark matter density field. This is due to both curvature and density continuity constraints that force the positively and negatively biased critical points to  strongly separate, as mentioned in \citet{Shim2021}.

The significance of the outstanding features contained in these cross-correlations is again quantified. 
As shown on the right-hand side of Fig.~\ref{fig:cc_dm_rarity10_4in1_16}, the exclusion zone is very well constrained, with a significance level of up to $\approx 8 \sigma$.

\begin{figure*}
\centering\includegraphics[width=0.45\textwidth]{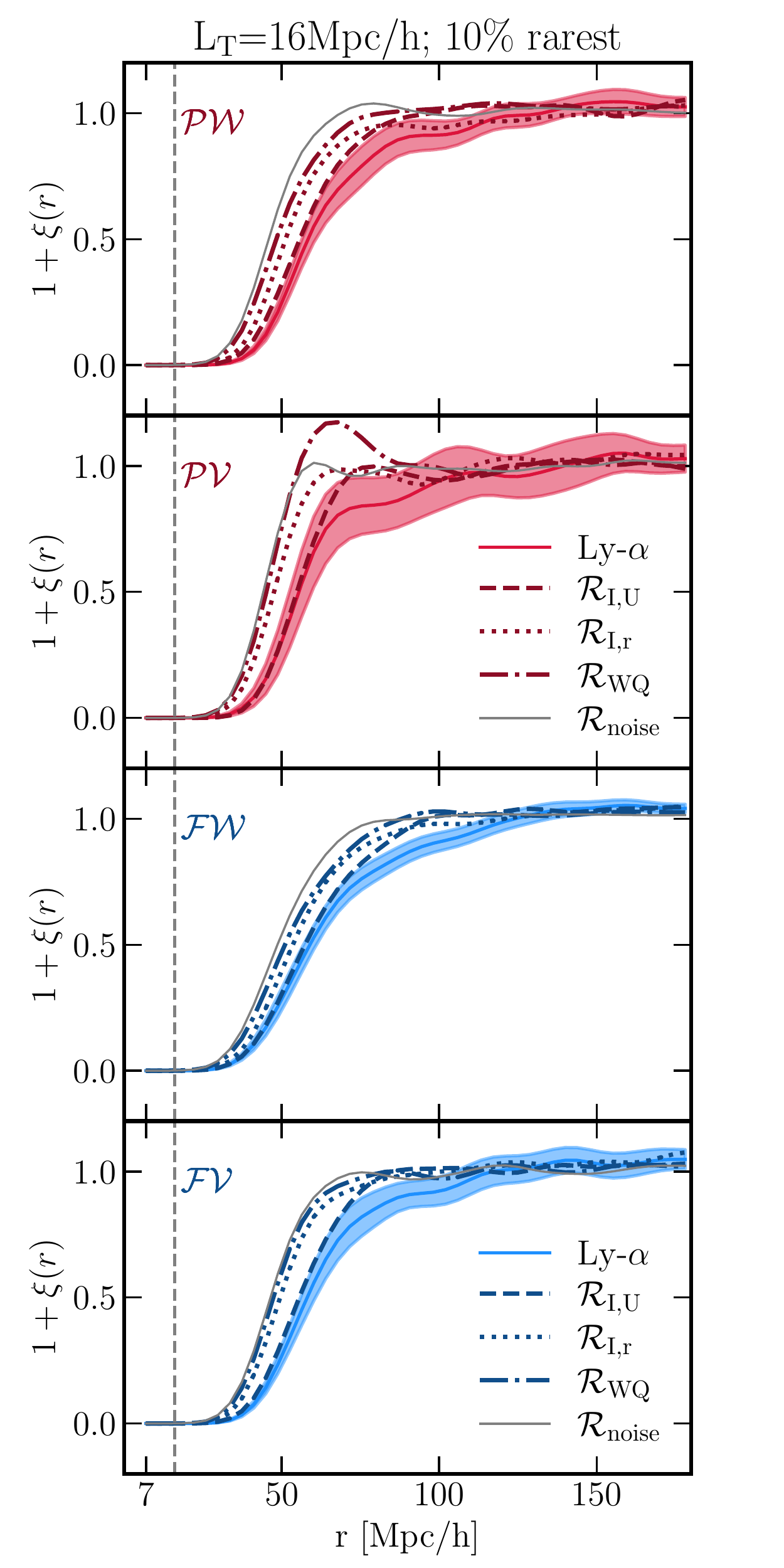}
\includegraphics[width=0.45\textwidth]{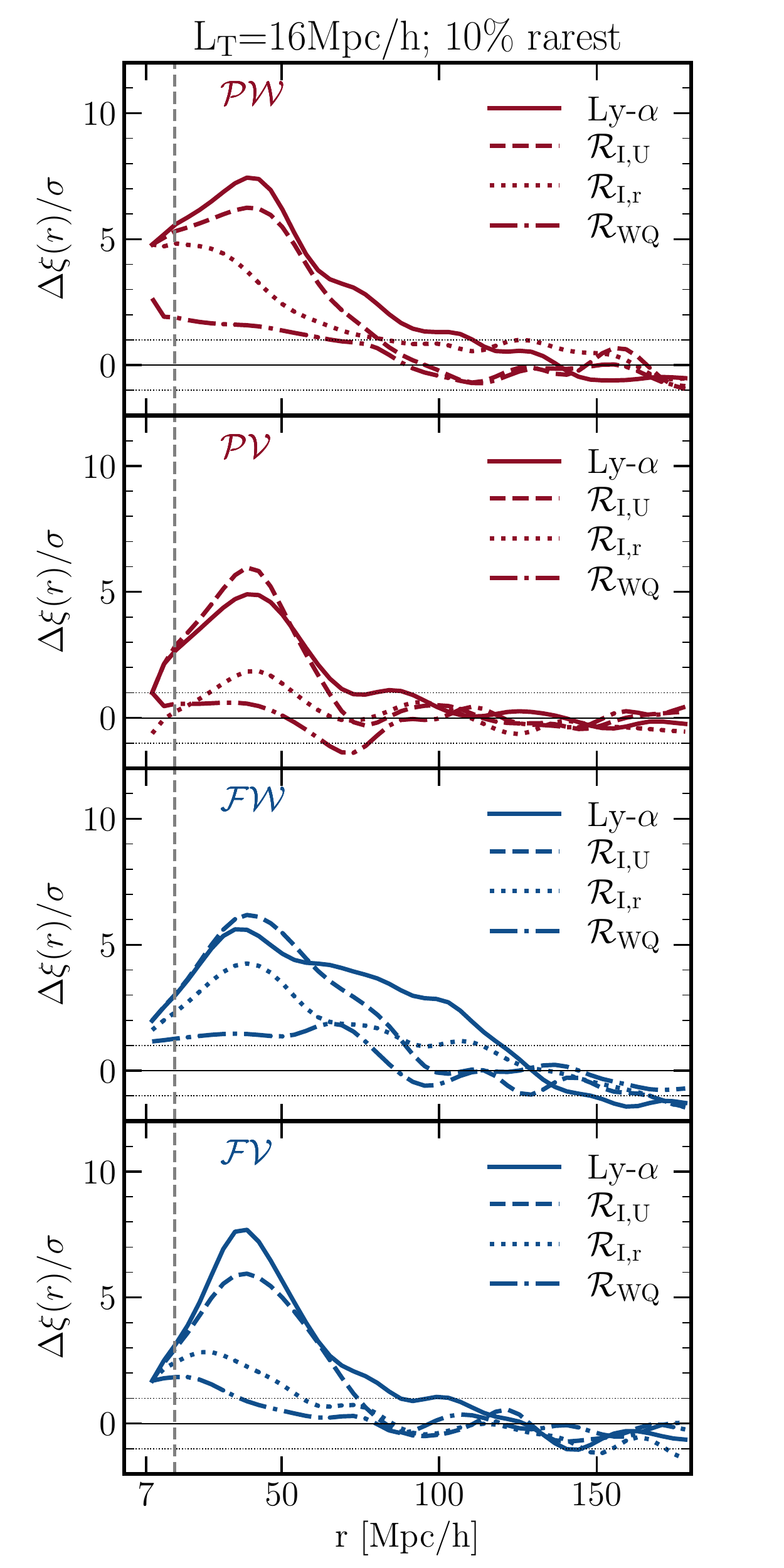}
\caption{\textit{Left:} Cross-correlations of critical points with 10\% abundance. \textit{Left:} $\mathcal{PW}$ (peak-wall), $\mathcal{PV}$ (peak-void), $\mathcal{FW}$ (filament-wall) and $\mathcal{FV}$ (filament-void) correlations for the original field (solid) and three reconstructions, \siu (dotted), \sir (dotted) and \svar (dash dotted).
\textit{Right:} Difference of cross-correlations between the noise and other fields.
}
\label{fig:cc_all_rarity10_4in1_16}
\end{figure*}

\medskip

Moving on to the comparison between the \lya and the reconstructed fields, Fig.~\ref{fig:cc_all_rarity10_4in1_16} shows their cross-correlations (left) together with the normalised differences with respect to the noise-only field (right).
For all fields, all four types of cross-correlation functions exhibit the same features, i) an exclusion zone at small separations, ii) a monotonic increase towards zero at large separations, and iii) anti-correlation ($\xi(r)<0$) at all scales. The quality of the reconstruction is again highest for the regular distribution of the sightlines (\siu), decreases for the random distribution of the sightlines (\sir) and is lowest with added noise on the spectra (\svar). The quality of the reconstruction also shows a variation with the type of the cross-correlation, in particular for the realistic configuration \svar.
While for \siu, the exclusion zone is constrained up to $\approx 6\sigma$ for all cross-correlations, 
for \svar it is only at the level of $\approx 1\sigma$ for peak-wall ($\mathcal{PW}$), filament-wall ($\mathcal{FW}$) and filament-void ($\mathcal{FV}$) and it decreases well below $1\sigma$ for peak-void ($\mathcal{PV}$) correlations (see Table~\ref{tab:summary_numbers_rexc} for the size of the exclusion zone for all fields). 
Contrarily to auto-correlations, the significance of the cross-correlations of over- and under-dense critical points increases with decreased rarity, in particular for $\mathcal{FW}$, where the level of significance reaches $4\sigma$ at 20\% abundance (see Fig.~\ref{fig:cc_orig_all_rarity5_20_4in1_16}).

\subsubsection{Same over-density sign critical points}\label{sec:other-cross}

\begin{figure*}
\centering\includegraphics[width=0.45\textwidth]{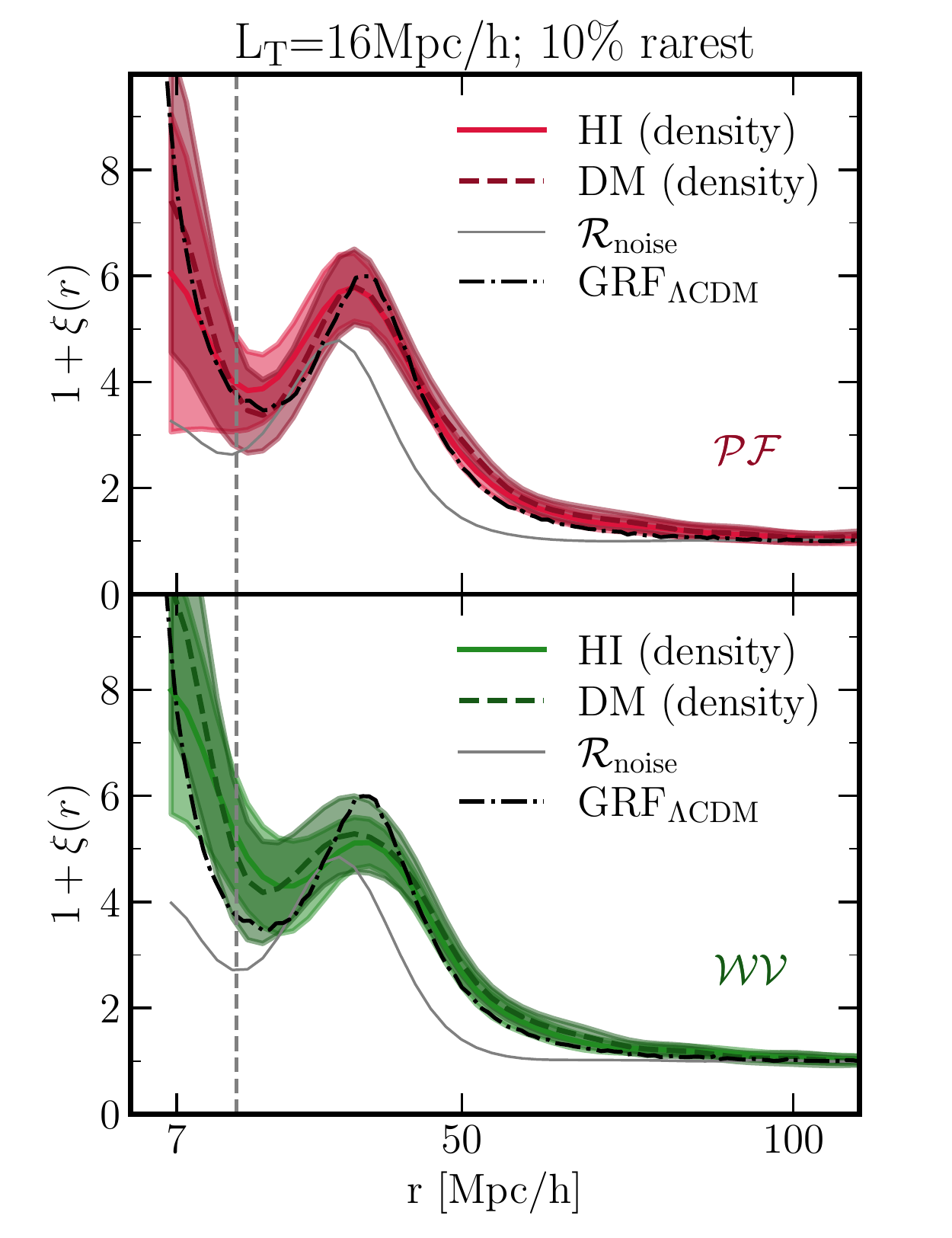}
\includegraphics[width=0.45\textwidth]{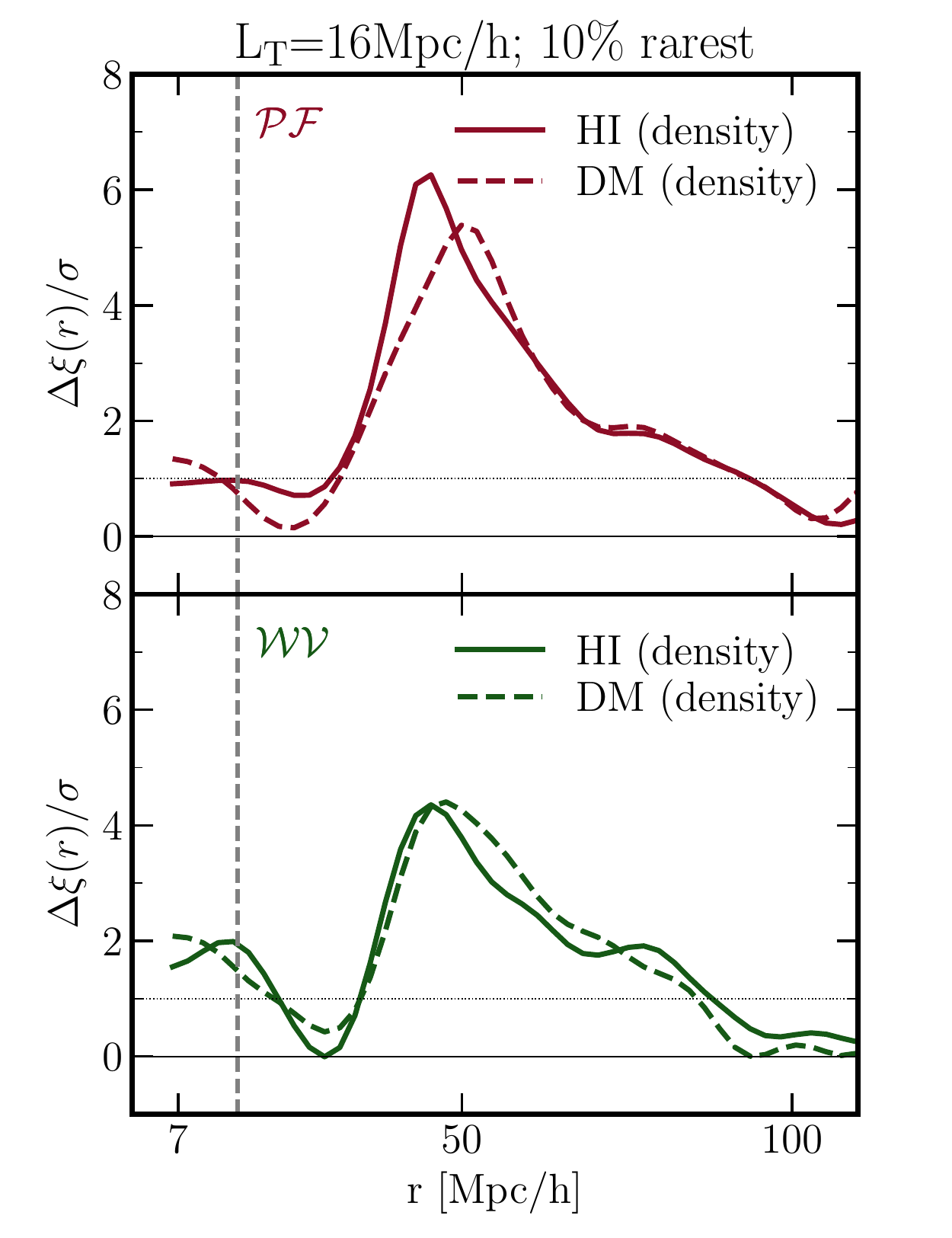}
\caption{Cross-correlations of critical points with 10\% abundance (\textit{left}) at the smoothing scale \lt~=16 Mpc/$h$ for \hi (solid coloured lines) and DM density (dashed coloured lines) fields and their differences relative to the noise in units of sigma (\textit{right}). $\mathcal{PF}$ (peak-filament) and $\mathcal{WV}$ (wall-void) correlations are shown on the top and bottom panels, respectively.
Vertical dashed gray line indicates the smoothing scale.
}
\label{fig:cc_dm_rarity10_2in1_16}
\end{figure*}

\begin{figure*}
\centering\includegraphics[width=0.45\textwidth]{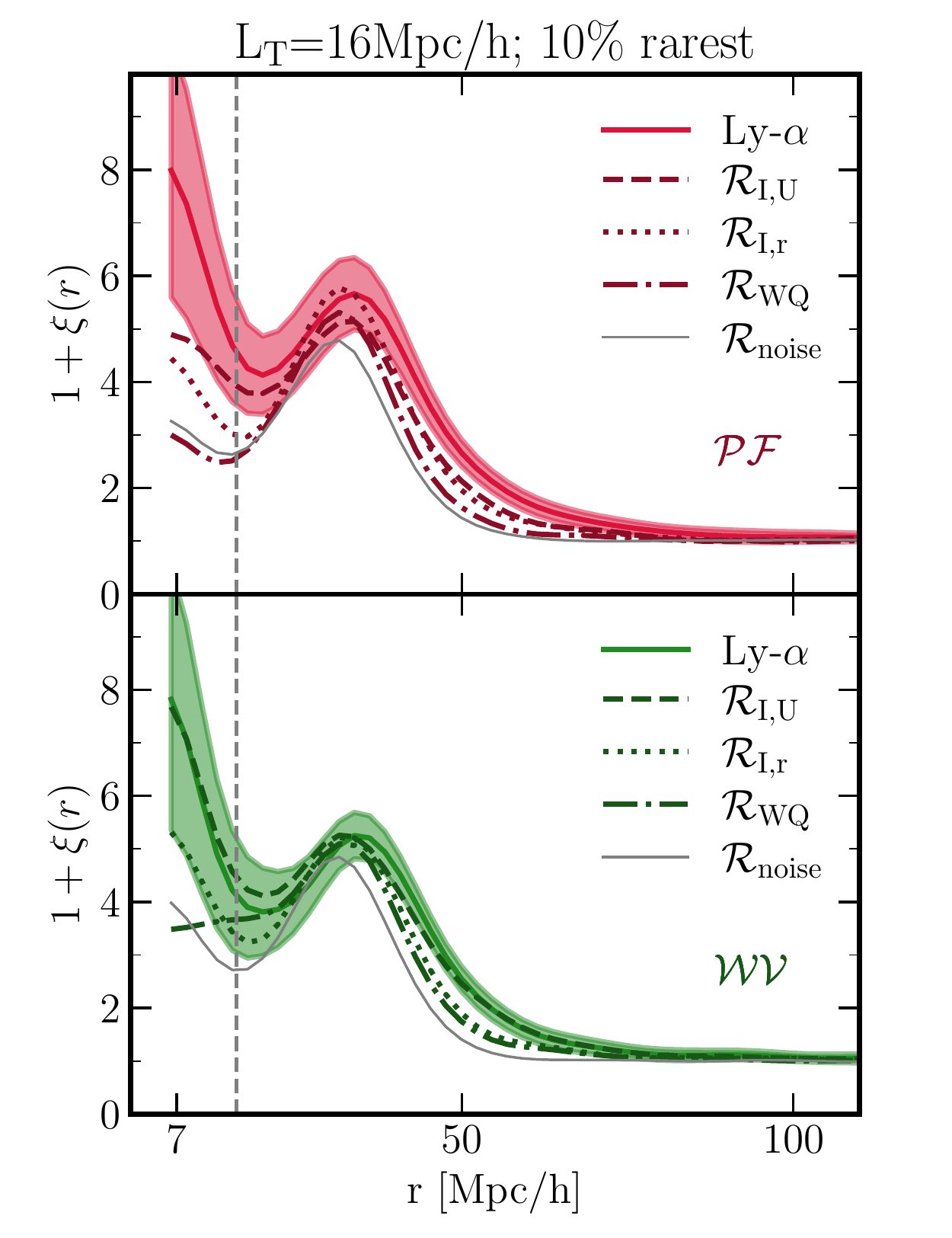}
\includegraphics[width=0.45\textwidth]{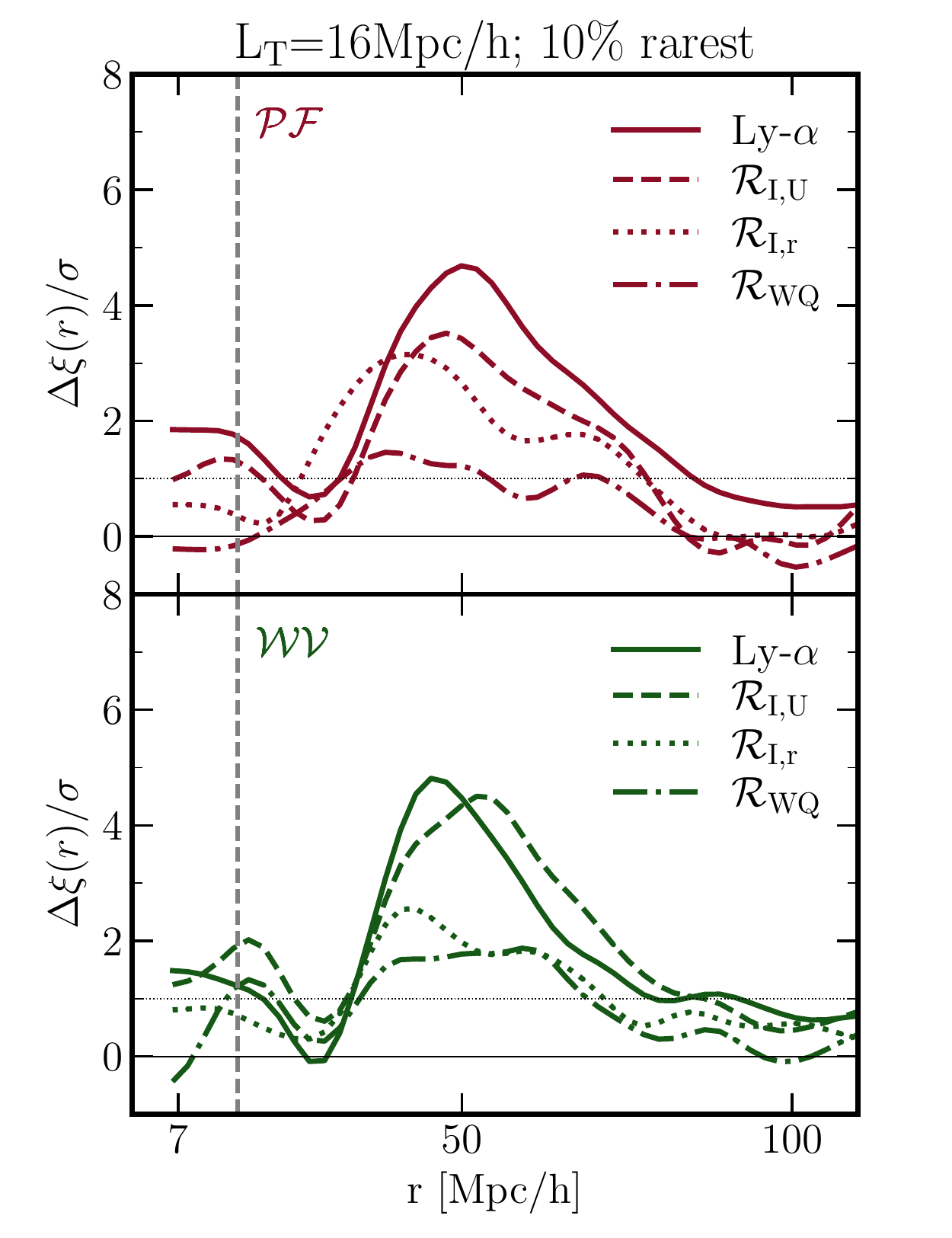}
\caption{\textit{Left:} Cross-correlations $\mathcal{PF}$ (peak-filament) and $\mathcal{WV}$ (wall-void)  with 10\% abundance at the smoothing scale \lt~=16 Mpc/$h$ for the \lya, the reconstructed fields and noise-only field (left). 
\textit{Right:} The differences of the cross-correlation functions with respect to the noise in the units of total sigma ($\sqrt{\sigma_i^2+\sigma_{\mathcal{N}}^2}$, with $i$ corresponding to $\sigma$ of \lya, \siu, \sir and \svar). 
Vertical dashed grey line indicates the smoothing scale.
}
\label{fig:cc_all_rarity10_2in1_16}
\end{figure*}

The two remaining cross-correlation functions are between the critical points of the same over-density sign, i.e. peak-filament ($\mathcal{PF}$) and wall-void ($\mathcal{WV}$) correlations.

Fig.~\ref{fig:cc_dm_rarity10_2in1_16} shows these cross-correlations for \hi and DM density fields (coloured solid and dashed lines, respectively) and 10\% abundance, together with the noise-only (thin grey line) and GRF with $\Lambda$CDM power spectrum (black dash-dotted line) for a comparison. The cross-correlations of the critical points of same over-density sign have a fundamentally different behaviour compared to the cross-correlations of over- and under-dense critical points (Fig.~\ref{fig:cc_dm_rarity10_4in1_16}). They diverge at zero separation without exhibiting any exclusion zone, nor anti-clustering at small separations; they have a local maximum at intermediate separations and finally, they approach zero at large separations.  
As discussed in \cite{Shim2021}, the divergence at zero separation is expected for two critical points with signatures difference of one and with overlapping ranges of density values. This behaviour is connected to the merging rate of these critical points at $r\rightarrow0$ when the field is smoothed on increasing scales \citep{Cadiou2020}. 
The position of the local maximum $r_{\rm max}$ is an expected geometrical feature of the cosmic web. This enhanced probability to find high peaks near high filament-type saddle points in the peak-filament cross-correlations (which  defines a statistically preferred distance between peaks- and filament-type saddles) is a measure of the typical length of filaments between two peaks, estimated as twice $r_{\rm max}$. 
For \hi and DM density fields, $r_{\rm max}$ is $\approx 2 \lt$ ($\approx$ 33.8 Mpc/$h$ and $\approx$ 31.5 Mpc/$h$, respectively),
therefore the typical length of filaments in both field is $\approx$ 64 Mpc/$h$ (see Table~\ref{tab:summary_numbers_max}).

In the cosmic web framework, the local maximum in wall-void cross-correlations can also be interpreted as the typical  radius of voids,
corresponding to $\approx$ 29.6 Mpc/$h$ and $\approx$ 33.8 Mpc/$h$ ($\approx 2 \lt$)  for \hi and DM density field, respectively (see also Table~\ref{tab:summary_numbers_max}). 

\medskip

Finally, Fig.~\ref{fig:cc_all_rarity10_2in1_16} shows the cross-correlation functions of the critical points with the same over-density sign and with 10\% abundance for \lya reference field, and is  compared to three reconstruction configurations, along with the noise-only field (left panels). The right panels show the relative differences in units of standard deviation with respect to the noise.
All salient features of these cross-correlations are recovered by all reconstructions considered in this work.    
For the peak-filament cross-correlation, the position of the local maximum (${r_{\rm max}}$) is best captured by the \siu reconstruction. However, for all fields, it is at $\approx 2\lt$ as in the case of \hi and DM density fields. For the height of the local maximum ($h_{\rm max}$), it is \sir that is closest to the reference field \lya (see also Table~\ref{tab:summary_numbers_max}). 
For the wall-void cross-correlation, the $r_{\rm max}$ is equally captured by all types of reconstruction, again at ${\approx 2\lt}$, while for the ${h_{\rm max}}$, \siu and \sir show a better match compared to \svar, however, the measured values are comparable, within the error bars. 

The existence of a local maximum is constrained with a significance up to ${4\sigma}$ for \siu and up to 1.5-2$\sigma$ for \svar for both peak-filament and wall-void cross-correlations, in contrast with the significance of  the cross-correlations of over- and under-dense critical points.

\subsection{Cosmic connectivity of critical points}\label{sub:connect}
\begin{figure}
\centering\includegraphics[width=0.9\columnwidth]{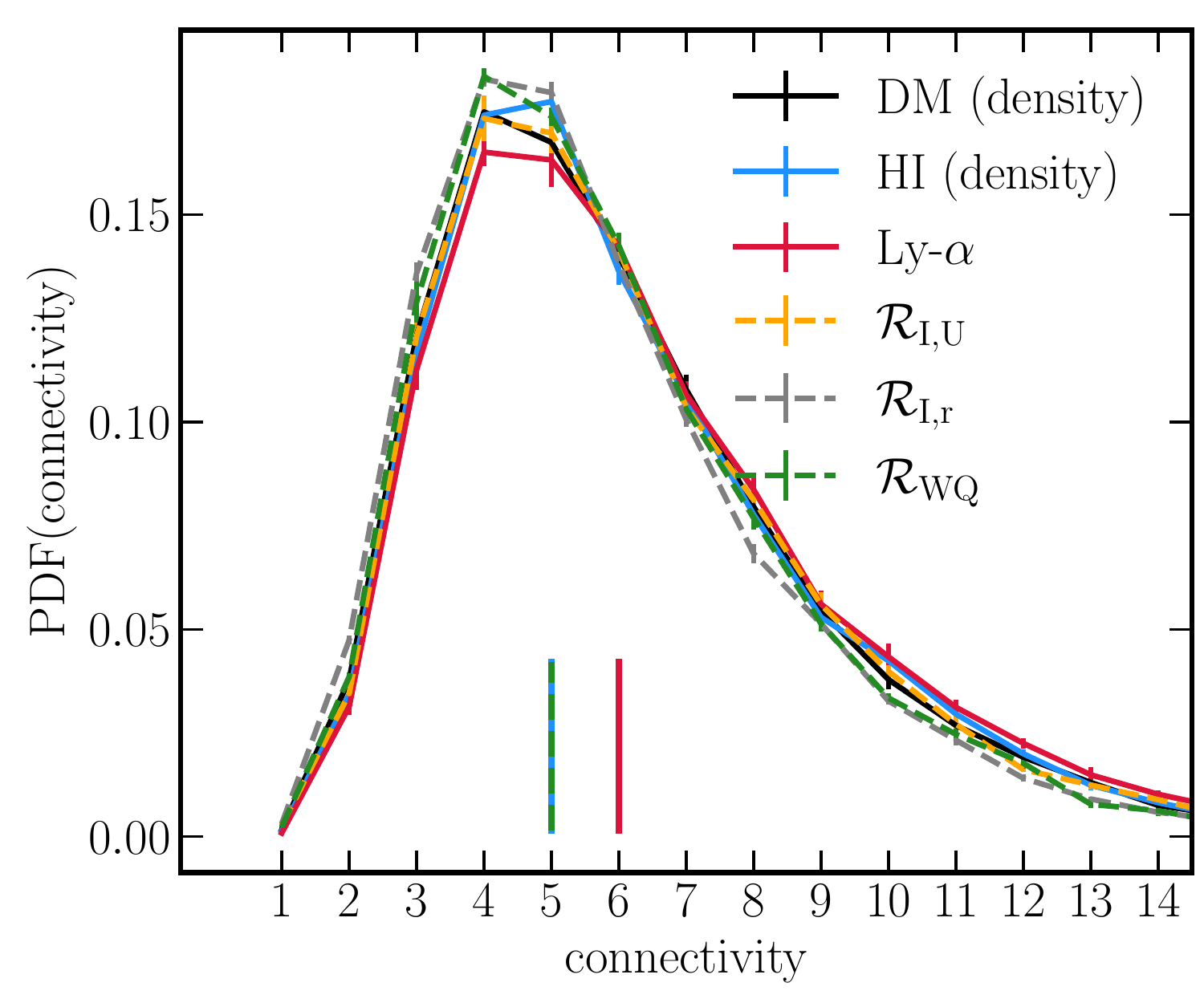}
\centering\includegraphics[width=0.9\columnwidth]{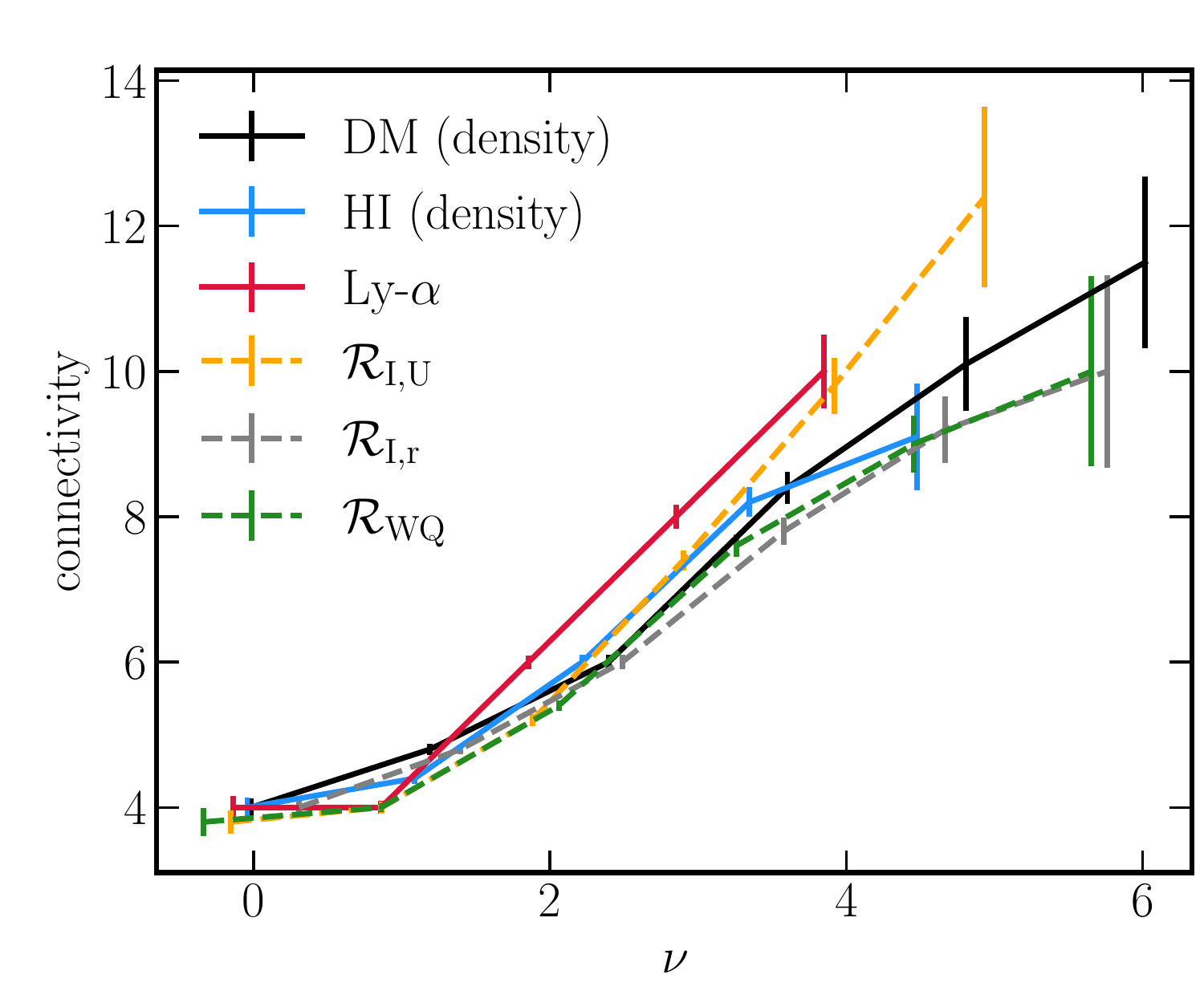}
\centering\includegraphics[width=0.9\columnwidth]{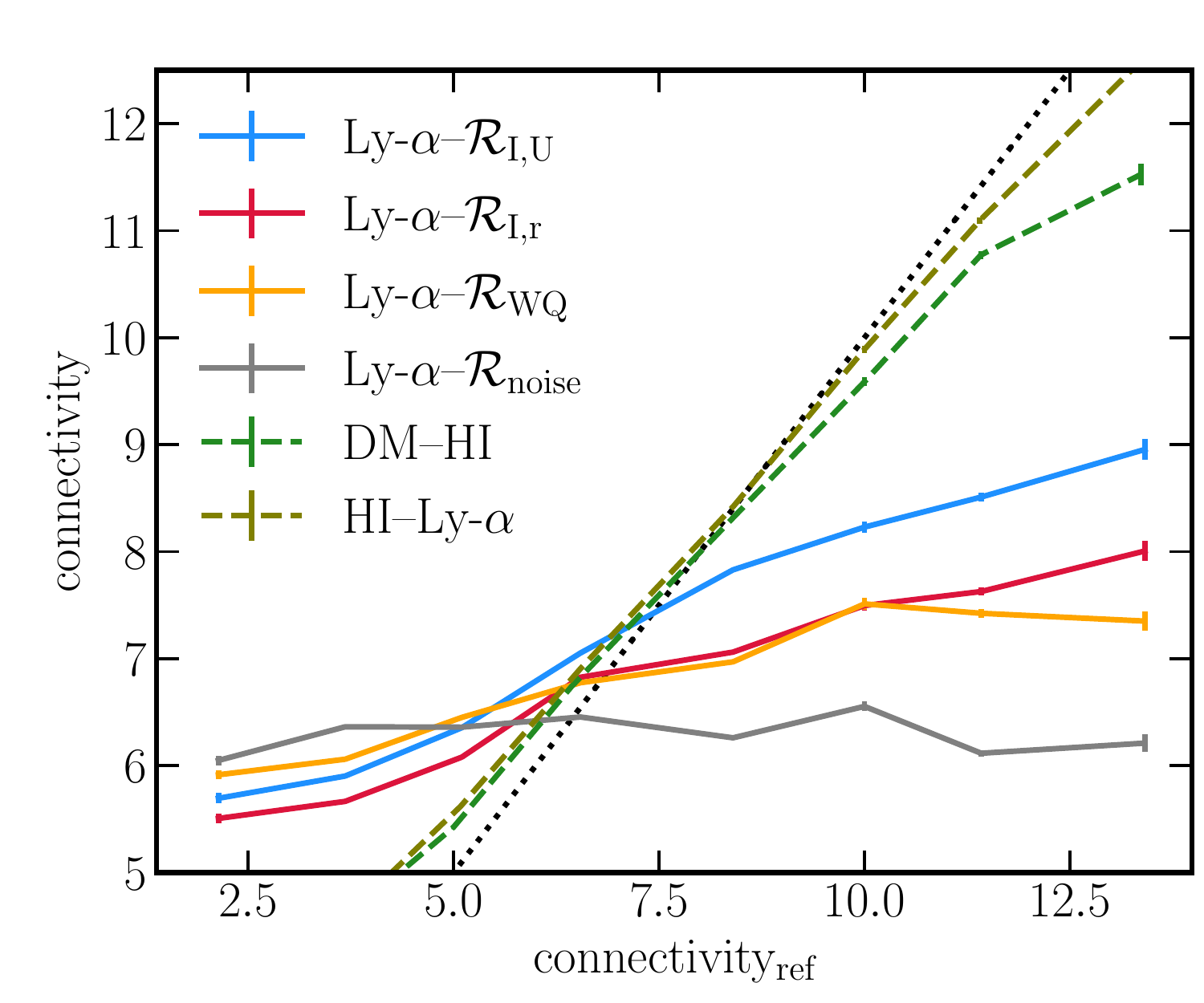}
\caption{\textit{Top:} PDF of the connectivity for DM and \hi density fields, \lya flux and the three reconstructed fields. The vertical lines represent the medians of the  connectivity across 5 mocks. This median is 5 for all fields but \lya for which it is 6. The mean values are close to 6 for all fields. \textit{Middle:} Connectivity as a function of rarity $\nu$ for the fields considered in this work. The mean connectivity of the peaks increases with  rarity for all fields. 
\textit{Bottom:} The connectivity of matched peaks across the range of fields considered. For each match, the reference field 
is the first in the label.  The closer the field to the reference one, the better the match. Note that both peak mismatch and change in the local geometry impact the connectivity. The black dotted line is the diagonal. 
The error bars are computed as the mean error across 5 mocks.  
}
\label{fig:connectivity_nu}
\end{figure}
Let us finally revisit our clustering results from the slightly different angle of topology \citep[see][for a first investigation  with tomographic reconstruction]{Caucci2008}. Indeed the relative positions of saddles and peaks impact the filamentary structure emerging from peaks, while the geometry of tunnels of given iso-contours is  set by the positions of wall- and filament-saddles. 
Morse theory \citep{Milnor1963} establishes a close  relationship 
between the distribution of critical points of the field on the one hand,  and the topology of its excursion sets (the iso-contours of the field) on the other hand.
The number of connected components within the excursion is one such quantity, and it is controlled by the connectivity of the field, defined as the number of ridges branching out of a given peak 
towards a given saddle point 
\citep{codisetal2018}\footnote{For instance, two distinct sets of iso-contours connect when reaching the height of the saddle point in between.}. Let us therefore measure this connectivity, in all the fields analysed in this study, using the ridge tracer algorithm \textsc{DisPerse} \citep{sousbie112}\footnote{For our purpose, \textsc{DisPerse} was run on the regular grid of density or flux contrast, depending on the field used, with a persistence threshold of 0.08 so as to obtain a total number of critical points comparable with the numbers given in Table~\protect\ref{tab:crits_pts}.} and assess our ability to recover it via a WEAVE-QSO like survey.

Fig.~\ref{fig:connectivity_nu} shows the PDF of the connectivity (top panel), the dependence of the median  connectivity on the rarity  $\nu$ (middle panel) 
for DM and \hi density fields, \lya flux and the three reconstructed fields at \lt=16~Mpc/$h$. While the median connectivity is~5 for all fields, except for \lya for which it is~6, the mean connectivity is close to~6 for all fields. This is completely expected from Table~\ref{tab:crits_ratio_pts} since it should match twice the ratio of the number of saddles to peaks \citep{2010AIPC.1241.1108P}.
Overall, there is a good agreement for both the PDF of the connectivity and the dependence of the connectivity on $\nu$ across all considered fields. As expected, the mean connectivity increases for peaks of higher rarity \citep{codisetal2018}, a feature which is recovered via the mock survey. This property seems robust, which is expected since connectivity reflects the underlying topology, hence does not discriminate well small features changes
across the various fields. 

Going one step further, we measured the cross match of the connectivity based on closest peak identification. 
In practice, for each peak in the reference field, we associate the closest peak in the matched field, 
without imposing any other condition on the match. We then compare the connectivities of these matched peaks.  
The result is presented in the bottom panel of Fig.~\ref{fig:connectivity_nu}\footnote{Note that the matching in the reverse order gives almost identical results.}. 
It reflects our ability to reconstruct the 
precise geometry of the field not only statistically, but also locally: the less noisy the reconstruction, the better the agreement. 
Conversely, the noise-only field falls back to the expected mean connectivity of 6.1 (see \S~\ref{sec:counts}).
Overall,  there is a good agreement between the recovered connectivity and the original one on a peak-to-peak basis. 
This is not unexpected given the consistency required by Morse theory. It does highlight that not only the critical points' relative distances are well preserved on average (as shown above from their clustering), but more generally their relative positions within the cosmic web (which controls the topology of the excursion set) is also recovered at various degrees, depending on the level of noise and the distribution of sightlines.

Measuring the connectivity of density peaks on a cluster-by-cluster basis could prove useful beyond cosmology, e.g. in the context of understanding the formation of galaxy groups and clusters \citep[][]{2019MNRAS.489.5695D,2020MNRAS.491.4294K,2021arXiv210705523L}. 
The size of the exclusion zone on these scales could be used to constrain the geometry of the warm-hot inter-galactic medium bubbles.
Beyond the connectivity, one could quantify the orientation and strength of filaments around reconstructed peaks,
as was investigated in 2D through stacking of lensing maps \citep[][]{2017A&A...605A..27G,2017A&A...605A..80C}.
Note finally that tomography would also allow us to compute  the dual connectivity of voids, which may prove more 
robust than that of peaks.

\section{Discussion}\label{sec:discussion-conclusions}
\label{sec:discussion}

 Let us discuss globally our main findings in terms of summary statistics, error budgets and  upcoming surveys.


\subsection{Summary statistics}

We start by computing physically-motivated summary statistics capturing the relative evolution of the outstanding  features in the two-point function of the critical points as a function of SNR and sampling strategy. Specifically, for the auto-correlations and cross-correlations of critical points of the same over-density sign we use the radius of their maximum \rmax and the corresponding height \hmax, while for the cross-correlations of over- and under-dense critical points we use the size of the exclusion zone \rexc.  

We recall that \rmax is defined as the separation at which the auto-correlation ${\cal C}_i{\cal C}_i$ or cross-correlation ${\cal C}_i{\cal C}_j$ peak (see Table~\ref{tab:summary_numbers_max}), 
\hmax is the height of the maximum of auto-correlation 1 + ${\cal C}_i{\cal C}_i$ or cross-correlation 1 + ${\cal C}_i{\cal C}_j$ (see Table~\ref{tab:summary_numbers_max}), and 
\rexc corresponds to the radius at which the cross correlation ${\cal C}_i{\cal C}_j $ correlation departs from -1. In practice, due to the noise on the measurement, we allow for a departure from this value by 0.01 (see Table~\ref{tab:summary_numbers_rexc}). 

From these numbers we extract ratios for the three reconstructed fields and the noise-only field w.r.t. the reference field \lya, and study how these ratios vary with the parameters of the reconstruction method. These ratios are shown in Tables~\ref{tab:summary_stats_max} and \ref{tab:summary_stats_rexc}. 
All of these ratios confirm the overall conclusion based on the detailed analysis of the two-point correlation functions, that is, globally, the quality of the reconstruction is the highest for the regular distribution of sightlines (\siu), it degrades for their random distribution (\sir) and is further reduced when the realistic noise on the spectra is added (\svar). Such a trend is best captured by the ratios of \rexc (Table~\ref{tab:summary_numbers_rexc}). The ratios of \rmax and \hmax (Table~\ref{tab:summary_stats_max}) do not allow us to well discriminate between the reconstructions and typically do not capture all the details contained in two point statistics. 

Note that even though our choice of summary statistics is physically motivated,  it is not  ideal  to highlight the differences between the reconstruction at the (relatively high) level of noise in the WEAVE-QSO mocks and the reference \lya flux, w.r.t. to the noise-only field, and therefore to constrain cosmology. In particular, the slopes and widths of the maximum auto- and cross-correlations might be more powerful for discriminating between different underlying power spectra. Beyond summary statistics derived from single auto and cross-correlations, one could also consider measuring the ratio between complementary auto and cross-correlations e.g. $\mathcal{FF/WW}$, $\mathcal{PV/FW}$, etc., as a way to get rid of some systematics inherited from the noise introduced in the reconstruction.

\subsection{How to improve the error budgets in future surveys?}
\label{sec:errorbudget}

\paragraph*{Decreasing rarity:} Part of the discrepancy between the original and reconstructed field comes from the non-regular sampling of the density field, which in turn depends on the spatial distribution of background sources. When sightlines are randomly distributed, in some regions of the reconstructed volume they will be less clustered than the correlation length set at the reconstruction stage (taken here as the mean inter-sightline distance), which will inevitably degrade the count and correlation of critical points. This can be seen for example on Fig.~\ref{fig:cc_all_rarity10_4in1_16}, where the discrepancy is the largest between the random distribution of sightlines ({dotted} lines) and the regular distribution of sightlines ({dashed} lines). Adding extra-noise on the spectra ({dashed-dotted } line) has on overall a smaller impact. To minimize the consequence of this random sampling (without changing the survey design), one could increase the correlation length in the reconstruction and carry out the study at larger scale. Obviously, when considering a fixed area on the sky, increasing the smoothing scale would also increase the statistical uncertainties since the number of volume elements would decrease. 
To mitigate this effect, we could choose to include in the statistics less rare critical points (e.g. taking all the 20\% rarest rather than only the 10\% rarest). Appendix~\ref{appendix:rarity} shows indeed that the signal is more significant while decreasing rarity.

\paragraph*{Combining galaxies and quasars background sources:} Another straightforward but costly way to mitigate the impact of shot noise due to the random sightline distribution could be to increase the number of sightlines. In the HIGHDENS footprint, we assumed that all quasars brighter than 23.5 in the $r$-band will be observed. Given that the galaxy number counts dominate over the quasar number counts for $r>22.5$, one could complement the survey with bright star-forming galaxies in order to efficiently increase the number of sightlines in the reconstruction. This is the strategy adopted in other surveys \citep[e.g. in PFS:][]{PFS2014}, but at the price of observing a smaller area. 

\paragraph*{Increasing the volume of the survey:} Recall that in this study we explored a configuration compatible with the HIGHDENS footprint in terms of sightline density, but, as mentionned above, our simulation set covers  $\sim 6.8\times$ the expected volume in this footprint. As a consequence, one should expect the size of the errorbars to be about $\sqrt{6.8} \simeq 2.6$ times larger when restricting ourselves to a volume comparable to the HIGHDENS footprint. 
 
 On the other hand, to improve the overall significance of the detection, one could consider carrying out the analysis on a larger volume by exploiting the WIDE footprint, which  will cover a much larger area (6000 deg$^2$). However the limited redshift window ($2.5<z<3$) will be higher than the one studied here. For the same magnitude limit in the background quasar distribution, this will translate into a lower sightline density. More precisely, in the WIDE footprint, the reconstruction could be performed at a scale $L_{\rm T}\simeq 19\,{\rm Mpc}/h$ over 13.6 (Gpc$/h$)$^3$, resulting into $\sim 2.0\times 10^{6}$ volume elements, while in the HIGDENS footprint, the reconstruction could be performed at a scale of $L_{\rm T}\simeq 16\,{\rm Mpc}/h$ over 0.7 (Gpc$/h$)$^3$, resulting into $\sim 1.8\times 10^{5}$ volume elements. Therefore, the size of the errorbars is expected to be divided by $\sqrt{11.1} \simeq 3.3$ when using the WIDE instead of the HIGDENS footprint. In other words, the errorbars in the WIDE footprint are expected to be $3.3/2.6 \simeq 1.3$ times smaller than those displayed in this paper, turning a $4\sigma$ detection into a $5\sigma$ one. 
 
 It remains to be seen though whether the scientific gain (from the point of view of constraining cosmology through the statistics of critical points) is higher in the HIGHDENS footprint (high density of quasars, small volume, lower redshift) or in the WIDE one (lower density of quasars, much larger volume, higher redshift). This question can be fully addressed only after having explored which scale/rarity is the most effective for cosmology (Shim et al. in prep).

\subsection{Prospects}\label{subsec:prospect}

From a cosmological perspective, given that the radii of exclusion zone and maximum correlation negligibly evolve with time \citep{Shim2021}, one can make use of these particular scales as standard rulers to measure the expansion of the Universe  because they are analytically predictable from first principles and nearly redshift independent. Because the correspondence between redshift and distance of an object depends on the underlying cosmology, the characteristic clustering scales of critical points will remain constant and match the theoretical prediction only when the correct cosmological parameters are adopted. Requesting such a  match yields an estimator for the corresponding parameters.
It then  becomes  crucial  to assess the ability  to recover these characteristic clustering scales from observations. 
Our forecasts (Figs.~\ref{fig:auto_all_rarity10_16}, ~\ref{fig:cc_all_rarity10_4in1_16} and ~\ref{fig:cc_all_rarity10_2in1_16} for 10\% rarity, and the corresponding Figs.~\ref{fig:auto_rarity5_20_16}, \ref{fig:cc_orig_all_rarity5_20_4in1_16} and \ref{fig:cc_orig_all_rarity5_20_2in1_16} for 20\% rarity) show that 
the cross-correlations
$\mathcal{PF}$ and $\mathcal{WV}$ are constrained at the $2\sigma$ level, the auto-correlations $\mathcal{FF}$, $\mathcal{WW}$ at up to $\sim2\sigma$ level, the cross-correlations $\mathcal{PW}$, $\mathcal{PV}$, $\mathcal{FV}$ 
up to $3\sigma$, $1\sigma$, $3\sigma$ level, respectively 
and $\mathcal{FW}$ up to the $\sim4\sigma$ level,
which represents the strongest significance for the WEAVE-QSO-like configuration. 
Thus, to provide tighter constraints on cosmological parameters it is more advantageous to use 
the characteristic features of the two-point functions involving filaments and/or walls (e.g. the exclusion zone in the $\mathcal{FW}$, $\mathcal{PW}$, $\mathcal{FV}$ cross-correlations, or the radius at a local maximum in the $\mathcal{FF}$, $\mathcal{WW}$ auto-correlations and in the $\mathcal{PF}$, $\mathcal{WV}$ cross-correlations)
as standard rulers because they can be measured with a higher significance.
Using these features as cosmic rulers  complements current approaches that rely on the BAO scale to measure cosmological parameters.
Interestingly, these scales associated to the two-point correlation functions  are smaller than BAO's and thus probe different part of the power spectrum, with more modes available within a given survey geometry.
Shim et al. (in prep.) investigates the cosmology dependence  of the clustering of critical points, exploring alternative cosmology models.  
Eventually, connecting their results and ours will allow us to make efficient cosmic forecasts from Ly-$\alpha$ tomography,
relying on both one \citep{Gay2010,Codis2013} and two-point  \citep[][]{Shim2021} statistics predictions.

While the present paper was focused  on the technical specification of WEAVE-QSO,
other upcoming tomographic surveys such as PFS or DESI  could help to improve  error budgets. It would be of interest to 
calibrate the best compromise one should make in terms of surface area, depth, tracers (Lyman-break galaxies versus QSOs) and expected SNR. 
It would clearly be an asset to complement  spectroscopic surveys  with photometric  redshift ones, as they could straightforwardly be 
integrated into the reconstruction. It would of course also be of interest to  quantify the clustering  of  critical point  on intensity maps in two dimensions, and our ability to extract such points from the corresponding surveys.

Moving beyond critical points this could be further completed by investigating the cosmic evolution of critical lines e.g.
connecting saddles together \citep{Pogosyan2009}, through  the statistics of their (differential) length,
as has been attempted for galactic catalogues \citep[][]{Sousbie2008} and in ELT mocks \citep{2019A&A...632A..94J}.

\section{Conclusions}
\label{sec:conclusions}

Mocks were used to asses our ability to recover the connectivity and clustering properties
of critical points of the reconstructed large-scale structure from 
\lya tomography in the context of a realistic quasar survey configuration (WEAVE-QSO). The mocks were produced with the \lya
Mass Association Scheme \citep[][]{peirani14,peirani21}.

Our main findings are the following:
\begin{itemize}
   \item \textbf{General:} As expected,  the quality of reconstruction decreases with randomness in the  distribution of  lines-of-sight and with the inclusion of noise on the spectra. Conversely, the measured signal increases with decreasing rarity of the critical points and with increasing smoothing scale, but at the expense of less marked features. 
   \item \textbf{Critical points number counts:} The total number of the critical points is larger in the reconstructed field compared to the original (reference) field by about 15\% for \siu, 19\% for \sir and 35\% for \svar.
    This fraction is slightly higher for peaks than for voids. However, as expected, reconstructed filaments and walls are about 3-times more abundant than peaks and voids, while the ratio between the number of peaks and walls over filaments and voids is close to one for all the reconstructions.
    \item \textbf{Auto-correlations of critical points:} The reconstruction captures the main expected features of the auto-correlation functions: exclusion  zones at small separations, maxima at $\approx 2-3$ \lt and convergence towards zero at large separations, in particular for saddles  (even for high rarity).
    \item \textbf{Cross-correlations of over- and under-dense critical points:}  
    The large exclusion zone at small separations and monotonic increase toward zero at large separations are well recovered.
    The amplitude of these cross-correlations is however systematically higher compared to the original \lya field for all explored reconstruction configurations.
    \item \textbf{Cross-correlations of the same over-density sign critical points:}
    Again, the reconstruction recovers the correlation's main features:  divergence at zero separation, lack of negative correlations, and exclusion zone at small scales, presence of a local maximum at similar separation, when compared to the auto-correlations. 
    \item  \textbf{Resilience of saddles:} Fortunately, the (cross-) correlations involving the least non-linear critical points (walls, filaments), which display the least amount of  variation with redshift are also those which are best reconstructed from WEAVE-like \lya tomography. This validates a posteriori using the clustering of  saddle  points  as a novel cosmic probe. 
    The significance of auto-correlations reaches 2$\sigma$ (1$\sigma$) for walls (filaments). 
    It is up to 4$\sigma$ for the cross-correlations of filaments and walls of 20\% abundance and up to 2$\sigma$ for the cross-correlations peak-filament and wall-void.
    \item \textbf{Connectivity:} the topology of the recovered field, as traced by its connectivity, is in good agreement with the initial one both statistically and in the vicinity of given peaks. This is consistent with the persistence of the clustering properties of critical points with respect to tomographic reconstruction.
\end{itemize}

Our conclusions highlight that the main features of the two-point correlation functions  of critical points
can be recovered with a good-degree of confidence in a WEAVE-QSO-like 
tomographic surveys \citep{Pieri2016,PFS2014,2016arXiv161100036D}. As they show little evolution with redshift \citep{Shim2021}, their clustering
should provide useful complementary estimators for  dark energy experiments.

\section*{Acknowledgements}
This work was supported by the \textit{Programme National Cosmology et Galaxies} (PNCG) of CNRS/INSU with INP and IN2P3, co-funded by CEA and CNES. 
We thank Prof. S.C. Trager for calculations regarding the throughput of the WEAVE instrument that underlie the SNR/\AA\, estimates in Fig. 3.
KK acknowledges support from the DEEPDIP project (ANR-19-CE31-0023). CL and CP thanks the LAM for hospitality. CP thanks P.~Petitjean for originally stimulating this line of research and S. Colombi for an introduction to Morse theory. The research of SC is partially supported by the chaire "Nouvelle Equipe" of Paris University, Fondation Merac and the French {\sl Agence Nationale de la Recherche} (grant ANR-18-CE31-0009). VI is supported by the Kavli Foundation. This work has made use of the Infinity cluster on which the simulation was post-processed, hosted by the Institut d'Astrophysique de Paris. We warmly thank S.~Rouberol for running it smoothly and T. Sousbie for distributing \textsc{DisPerSE}.  %
\section*{Data Availability}
The data underlying this article will be shared on reasonable request to the corresponding author.
%
%
\def\bibfont{\footnotesize}

\bibliographystyle{mnras}
\bibliography{author}


\appendix

\section{Quality versus SNR and ${L}_{\rm T}$}
\label{appendix:quality}

\begin{figure*}
\centering 
\includegraphics[width=0.95\textwidth]{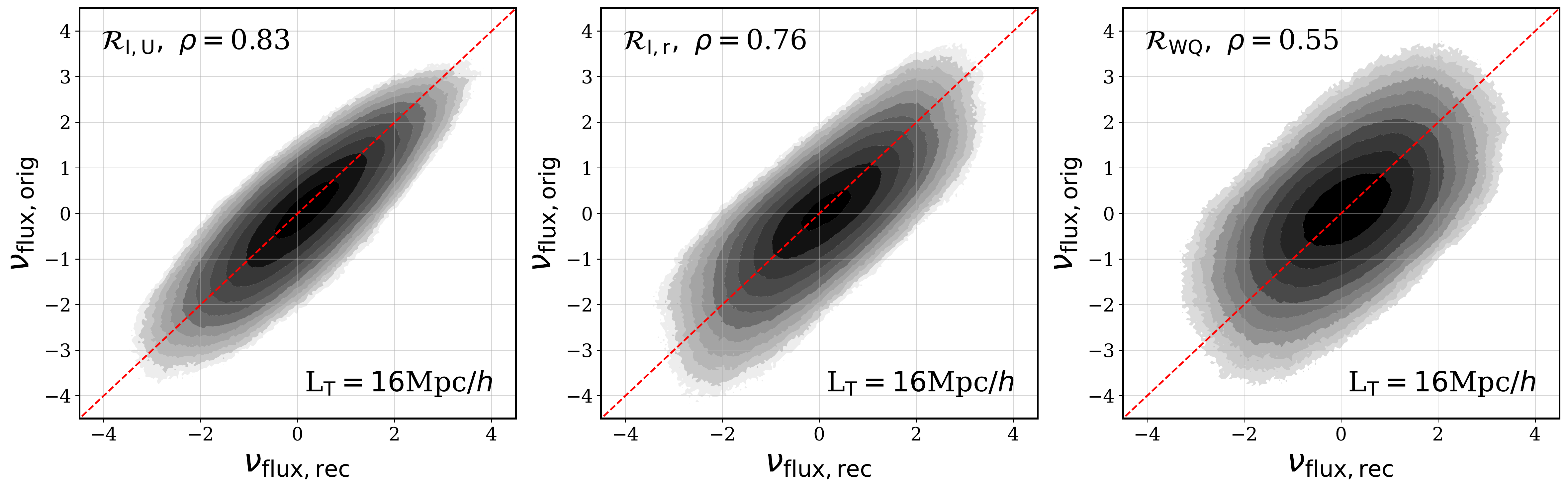}
\caption{The original flux density contrast (in units of rms fluctuations of the field) plotted against the reconstructed flux density contrast for the different noise configurations envisaged in this study, i.e. regular distribution of sightlines ($\mathcal{R}_{\rm I,U}$), random distribution of sightlines ($\mathcal{R}_{\rm I,r}$), and fiducial configuration ($\mathcal{R}_{\rm WQ}$). Also indicated is the Pearson correlation coefficient $\rho$ between the two fields.  }
\label{fig:densitycomparison}
\end{figure*}


Fig.~\ref{fig:densitycomparison} quantifies the pixel-to-pixel correlation between the 3D maps of the original and the reconstructed flux contrast in units of the rms fluctuations,  for the different configuration studied in this work (as presented in Section~\ref{sec:configurations}).  On each panel is displayed the pearson correlation coefficient, that we can use as a single metric to assess the overall agreement between both fields. 
%

\begin{figure}
\centering 
\includegraphics[width=0.45\textwidth]{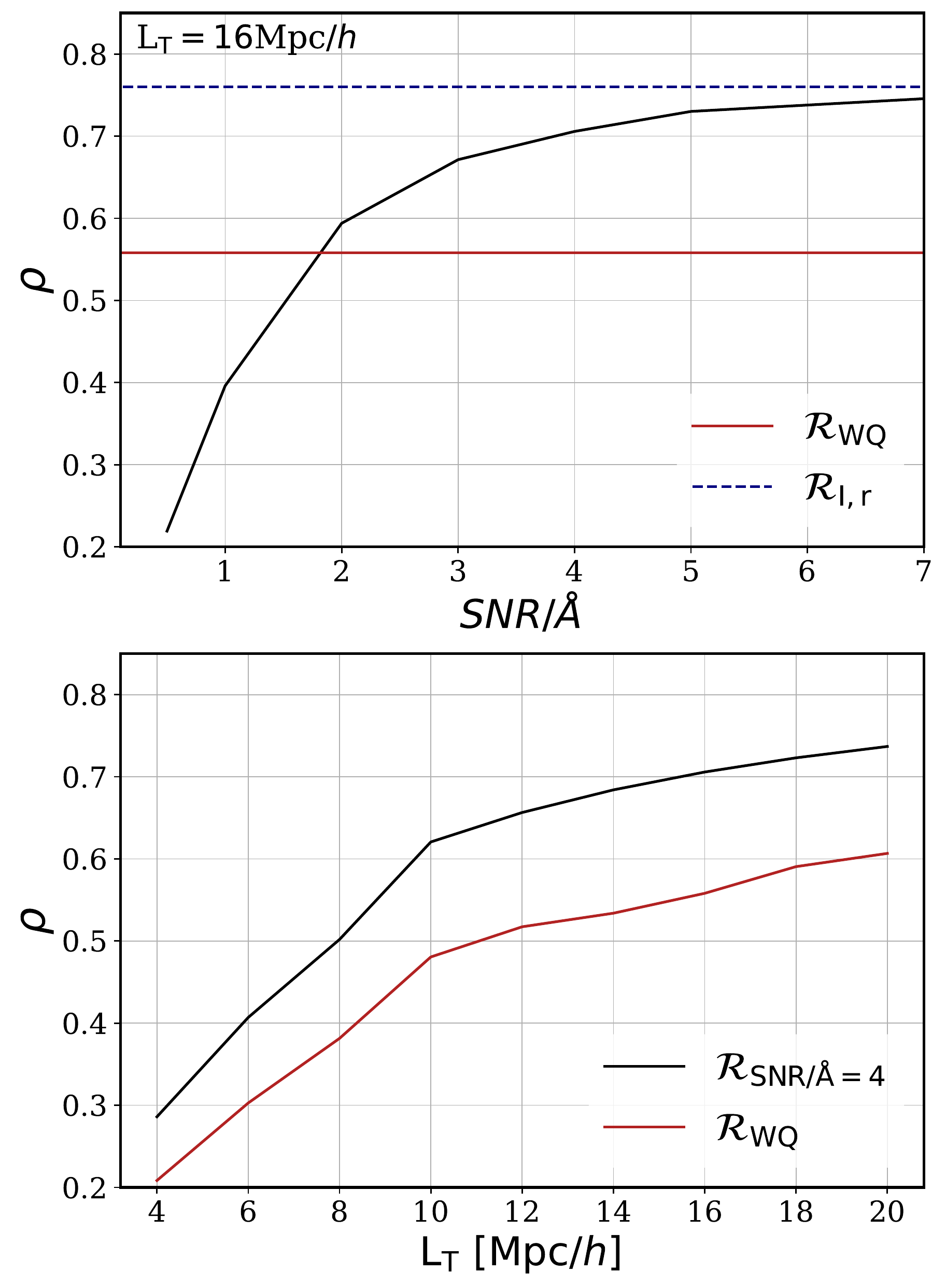}
\caption{\textit{Top}: Pearson correlation coefficient between the Ly-$\alpha$ reference field and the reconstructed fields (with ${L}_{\rm T}=16{\rm Mpc}/h$) for different realisations of the SNR on the sightlines. Also indicated as straight lines are the correlation coefficients for $\mathcal{R}_{\rm WQ}$ ({red}) and $\mathcal{R}_{\rm I,r}$ ({dashed blue}). \textit{Bottom}: Correlation coefficient between the original and the reconstructed fields when different correlation scales $L_{\rm T}$ have been adopted when performing the reconstruction, adopting a constant SNR$/$\AA$=4$ ({black}) or a realistic SNR distribution ({red}). }
\label{fig:correlation}
\end{figure}
Let us now use this metric to quantify the degradation of the correlation between the original and the reconstructed fields. We performed the reconstruction with a similar density of sightlines but different levels of Gaussian white noise added on the Ly-$\alpha$ forest of each sightline prior to the reconstruction. The corresponding pearson correlation coefficient is displayed on the {top} panel of Fig.~\ref{fig:correlation}. On overall, increasing the SNR$/$\AA~  brings the reconstructed and original fields in better agreement. However, one also note that the correlation coefficient reaches a plateau at  SNR$/$\AA$>4$. This suggests that, at high SNR$/$\AA~ (on spectra), the noise budget starts to be  dominated by shot noise due to the finite sampling and clustering of sightlines. 

On the {bottom} panel of Fig.~\ref{fig:correlation}, the impact of choosing different correlation lengths $L_{\rm T}$ when performing the reconstruction (see Sec.~\ref{sec:reconstruction}) is quantified. The reconstruction are performed with the same density of sightlines (as for $\cal{R_{\rm WQ}}$) and  with a realistic SNR distribution (similar to $\cal{R_{\rm WQ}}$) on the one hand, and with a constant SNR$/$\AA$=4$ on the other hand. As a reminder, $L_{\rm T}=16$ Mpc$/h$ was chosen  as it was reflecting the mean separation between sightlines.  Unsurprisingly, one notes that the agreement improves when a larger $L_{\rm T}$ is used in the reconstruction (several sightlines will contribute to the same volume element in the reconstructed map, which will enhance the signal over the noise). Interestingly, the agreement between the original and reconstructed fields is relatively well preserved even when decreasing $L_{\rm T}$ below its fiducial value, down to $\sim 10$ Mpc$/h$, from which it dropped brutally. One should note however that, when decreasing $L_{\rm T}$, the quality of the reconstruction will be spatially less and less homogeneous, due to the non-regular distribution of the sightlines. The pearson correlation coefficient, which is a global metric, is not very sensitive to this degradation, but this effect might dramatically impacts the count and correlation of the critical points .

\section{Predictions in the linear regime}
\label{appendix:GRF}
Throughout the main text, we have compared measurements for the critical points counts and cross-correlations with Gaussian random fields predictions. To compute those predictions, we rely on the formalism extensively described in \citep{Shim2021} (we refer the reader to Appendix A of this paper for more details) with a few modifications. We first compute the covariance matrix of a Gaussian random field and its first and second derivatives at two spatial positions separated by a distance $r$ and characterized by a power spectrum given by the linear power spectrum used in the simulation smoothed with a Gaussian kernel of length $L_{\rm T}=16$ Mpc$/h$. From this $20\times 20$ covariance matrix, we compute the corresponding joint probability distribution function (PDF) of the field and its first and second derivatives and then evaluate through a MCMC integration scheme the probability of finding two critical points with specified signatures separated by $r$ in order to get their cross-correlation functions. Each point of the function is evaluated by generating 10 millions random numbers satisfying the joint PDF with a zero gradient constraint and is kept only if the critical points conditions are fulfilled (signatures and density thresholds). The result is displayed with dash-dotted black lines on the figures of the main text.

\section{Summary statistics}
\label{appendix:tables}
This Appendix provides a summary of measured positions and heights of the maxima of auto-correlations and cross-correlations $\mathcal{PF}$, $\mathcal{WV}$, sizes of the exclusion zones of cross-correlations $\mathcal{PW}$, $\mathcal{PV}$, $\mathcal{FW}$, $\mathcal{FV}$ and summary statistics presented in the main text. 

Table~\ref{tab:summary_numbers_max} summarises the position (\rmax) and height (\hmax) of the maximum of auto-correlations and cross-correlations $\mathcal{PF}$, $\mathcal{WV}$ for all the fields used in this work at the smoothing scale \lt=16 Mpc/$h$.

Table~\ref{tab:summary_numbers_rexc} shows the size of the exclusion zone (\rexc) of cross-correlations $\mathcal{PW}$, $\mathcal{PV}$, $\mathcal{FW}$, $\mathcal{FV}$ for all the fields used in this work at the smoothing scale \lt=16 Mpc/$h$.

Tables~\ref{tab:summary_stats_max} and \ref{tab:summary_stats_rexc} reports the ratios of \rmax, \hmax and \rexc with respect to the \lya reference field for the reconstructed and noise-only fields at the smoothing scale \lt=16 Mpc/$h$.

\begin{table*}
\centering
\caption{Position (\rmax [Mpc/$h$]) and height (\hmax) of the maximum of auto-correlations ($\mathcal{PP}$, $\mathcal{FF}$, $\mathcal{WW}$, $\mathcal{VV}$) and cross-correlations $\mathcal{PF}$, $\mathcal{WV}$ for 10\% rarity, for all the fields used in this work at the smoothing scale \lt=16 Mpc/$h$. The errors are the standard deviations of the mean across all mocks.} 
\label{tab:summary_numbers_max}
\begin{tabular*}{\textwidth}{@{\extracolsep{\fill}}lccccccc}
\hline
\hline
&  & $\mathcal{PP}$ & $\mathcal{FF}$ & $\mathcal{WW}$ & $\mathcal{VV}$ & $\mathcal{PF}$ & $\mathcal{WV}$ \\ 
\hline
\multirow{7}{*}{\rmax} & \hi & 30.82$\pm{3.45}$ & 25.13$\pm{0.67}$ & 28.92$\pm{1.44}$ & 36.50$\pm{3.93}$  & 
33.78$\pm{1.72}$  & 29.64$\pm{1.99}$ \\
& DM & 37.92$\pm{2.07}$ & 30.34$\pm{1.82}$ & 27.97$\pm{1.82}$ & 35.55$\pm{3.18}$  & 
31.48$\pm{1.72}$  & 33.78$\pm{1.46}$\\
& \lya & 39.77$\pm{7.42}$ & 27.33$\pm{1.35}$ & 24.31$\pm{1.35}$ & 32.64$\pm{1.6}$  &
33.78$\pm{1.72}$ & 35.63$\pm{1.51}$ \\
& \siu & 35.66$\pm{3.14}$ & 29.61$\pm{2.14}$ & 30.36$\pm{3.10}$ & 34.91$\pm{4.86}$  &
32.86$\pm{1.05}$ & 31.94$\pm{0.41}$ \\
& \sir & 40.96$\pm{3.03}$ & 25.82$\pm{1.07}$ & 28.09$\pm{1.73}$  & 51.13$\pm{3.96}$  & 29.64$\pm{1.20}$ & 31.48$\pm{1.13}$ \\
& \svar & 51.83$\pm{7.09}$ & 27.34$\pm{0.83}$ & 26.58$\pm{1.66}$ & 43.24$\pm{3.65}$  & 31.48$\pm{0.65}$ & 34.24$\pm{0.41}$ \\
& \noise & 37.38$\pm{6.44}$ & 26.20$\pm{0.36}$ & 29.61$\pm{1.51}$ & 40.37$\pm{2.88}$ &  31.09$\pm{0.63}$ & 29.87$\pm{0.47}$  \\
\hline
\multirow{7}{*}{\hmax} & \hi &  3.85$\pm{0.45}$ & 5.66$\pm{0.43}$ & 5.76$\pm{0.42}$ & 5.15$\pm{0.44}$   
& 7.77$\pm{0.67}$  & 7.38$\pm{0.31}$  \\
& DM &  3.64$\pm{0.49}$ & 5.11$\pm{0.45}$ & 6.34$\pm{0.19}$ & 4.83$\pm{0.64}$   &
8.47$\pm{0.53}$  & 7.67$\pm{0.25}$ \\
& \lya &  3.35$\pm{0.38}$ & 4.90$\pm{0.12}$ & 4.76$\pm{0.09}$ & 3.34$\pm{0.19}$  &  
7.39$\pm{0.61}$ &  6.68$\pm{0.31}$  \\
& \siu &  2.86$\pm{0.31}$ & 4.64$\pm{0.30}$ & 4.31$\pm{0.27}$ & 2.59$\pm{0.17}$  &  
7.03$\pm{0.38}$ & 6.82$\pm{0.22}$ \\
& \sir &  2.12$\pm{0.21}$ & 4.49$\pm{0.15}$ & 4.40$\pm{0.09}$ & 2.39$\pm{0.21}$  &  
7.29$\pm{0.19}$ & 7.16$\pm{0.36}$ \\
& \svar &  1.75$\pm{0.07}$ & 4.34$\pm{0.13}$ & 4.53$\pm{0.23}$ & 2.54$\pm{0.23}$ &  
7.22$\pm{0.12}$ & 6.64$\pm{0.49}$  \\
& \noise & 2.09$\pm{0.22}$ & 4.17$\pm{0.12}$ & 3.88$\pm{0.13}$ & 1.96$\pm{0.10}$  & 
6.43$\pm{0.18}$  & 6.27$\pm{0.19}$  \\
\hline
\end{tabular*}
\end{table*}

\begin{table}
\centering
\caption{Size of the exclusion zone \rexc (in [Mpc/$h$]) of cross-correlations $\mathcal{PW}$, $\mathcal{PV}$, $\mathcal{FW}$, $\mathcal{FV}$ for 10\% rarity, for all the fields used in this work at the smoothing scale \lt=16 Mpc/$h$. The errors are the standard deviations of the mean across all mocks.} 
\label{tab:summary_numbers_rexc}
\begin{tabular*}{\columnwidth}{@{\extracolsep{\fill}}lcccc}
\hline
\hline
&  $\mathcal{PW}$ & $\mathcal{PV}$ & $\mathcal{FW}$ & $\mathcal{FV}$\\
\hline
\hi &  44.23$\pm{0.82}$ & 47.24$\pm{1.26}$ & 38.22$\pm{1.26}$ & 43.48$\pm{0.76}$  \\
DM & 45.74$\pm{0.67}$ & 49.49$\pm{0.67}$ & 42.73$\pm{1.06}$ & 44.98$\pm{1.34}$  \\
\lya &  44.95$\pm{1.52}$ & 49.51$\pm{1.27}$ & 38.09$\pm{0.68}$ & 43.42$\pm{0.83}$ \\
\siu &  40.38$\pm{0.68}$ & 46.47$\pm{0.83}$ & 36.57$\pm{0.68}$ & 41.14$\pm{1.08}$ \\
\sir &  41.14$\pm{1.08}$ & 41.90$\pm{1.27}$ & 36.57$\pm{0.68}$ & 41.14$\pm{1.08}$ \\
\svar & 36.58$\pm{0.68}$ & 43.42$\pm{1.74}$ & 32.77$\pm{0.68}$ & 38.86$\pm{0.83}$ \\
\noise & 36.19$\pm{0.55}$ & 40.00$\pm{0.55}$ & 32.77$\pm{0.48}$ & 36.58$\pm{0.48}$ \\
\hline
\end{tabular*}
\end{table}

\begin{table*}
\centering
\caption{Summary statistics. Ratios of \rmax and \hmax with respect to the \lya reference field for the reconstructed and noise-only fields at the smoothing scale \lt=16 Mpc/$h$ and for 10\% rarity.}
\label{tab:summary_stats_max}
\begin{tabular*}{\textwidth}{@{\extracolsep{\fill}}lccccccc}
\hline
\hline
 &  & $\mathcal{PP}$ & $\mathcal{FF}$ & $\mathcal{WW}$ & $\mathcal{VV}$ & $\mathcal{PF}$ & $\mathcal{WV}$ \\
\hline
\multirow{4}{*}{\rmax} & \siu & 0.89$\pm{0.19}$ & 1.08$\pm{0.09}$ & 1.25$\pm{0.13}$ & 1.07$\pm{0.15}$ & 0.97$\pm{0.06}$& 0.89$\pm{0.04}$ \\
& \sir & 1.03$\pm{0.20}$ & 0.95$\pm{0.06}$ & 1.15$\pm{0.09}$ & 1.57$\pm{0.11}$ & 0.88$\pm{0.06}$ & 0.88$\pm{0.05}$ \\
& \svar & 1.30$\pm{0.26}$ & 1.0$\pm{0.06}$ & 1.09$\pm{0.09}$ & 1.21$\pm{0.11}$ & 0.93$\pm{0.05}$ & 0.96$\pm{0.04}$ \\
& \noise & 0.94$\pm{0.25}$& 0.96$\pm{0.05}$ & 1.22$\pm{0.08}$ & 1.24$\pm{0.10}$ & 0.92$\pm{0.05}$ & 0.84$\pm{0.04}$ \\
\hline
\multirow{4}{*}{\hmax} & \siu & 0.85$\pm{0.14}$ & 0.95$\pm{0.07}$ & 0.91$\pm{0.06}$ & 0.78$\pm{0.08}$ & 0.95$\pm{0.10}$ & 1.02$\pm{0.06}$ \\
& \sir & 0.63$\pm{0.12}$ & 0.92$\pm{0.04}$ & 0.92$\pm{0.03}$ & 0.72$\pm{0.09}$ & 0.99$\pm{0.09}$ & 1.07$\pm{0.07}$ \\
& \svar & 0.52$\pm{0.09}$ & 0.89$\pm{0.04}$ & 0.95$\pm{0.05}$ & 0.76$\pm{0.09}$ & 0.98$\pm{0.08}$ & 0.99$\pm{0.09}$ \\
& \noise & 0.62$\pm{0.12}$ & 0.85$\pm{0.03}$ & 0.82$\pm{0.03}$ & 0.59$\pm{0.06}$ & 0.87$\pm{0.08}$ & 0.94$\pm{0.05}$ \\
\hline
\end{tabular*}
\end{table*}

\begin{table}
\centering
\caption{Summary statistics. Ratios of \rexc with respect to the \lya reference field for the reconstructed and noise-only fields at the smoothing scale \lt=16 Mpc/$h$ and for 10\% rarity.} 
\label{tab:summary_stats_rexc}
\begin{tabular*}{\columnwidth}{@{\extracolsep{\fill}}lcccc}
\hline
\hline
 & $\mathcal{PW}$ & $\mathcal{PV}$ & $\mathcal{FW}$ & $\mathcal{FV}$\\
\hline
\siu & 0.89$\pm{0.04}$ & 0.94$\pm{0.03}$ & 0.96$\pm{0.03}$ & 0.95$\pm{0.03}$ \\
\sir & 0.92$\pm{0.04}$ & 0.85$\pm{0.04}$ & 0.96$\pm{0.03}$ & 0.95$\pm{0.03}$\\
\svar & 0.81$\pm{0.03}$ & 0.88$\pm{0.04}$ & 0.86$\pm{0.03}$ & 0.89$\pm{0.03}$ \\
\noise & 0.81$\pm{0.03}$ & 0.81$\pm{0.03}$ & 0.86$\pm{0.02}$ & 0.84$\pm{0.02}$ \\
\hline
\end{tabular*}
\end{table}

\section{Impact of rarity and smoothing}
\label{appendix:rarity}
Let us explore the impact of rarity and smoothing on the two-point correlation functions. 

\begin{figure*}
\centering\includegraphics[width=0.45\textwidth]{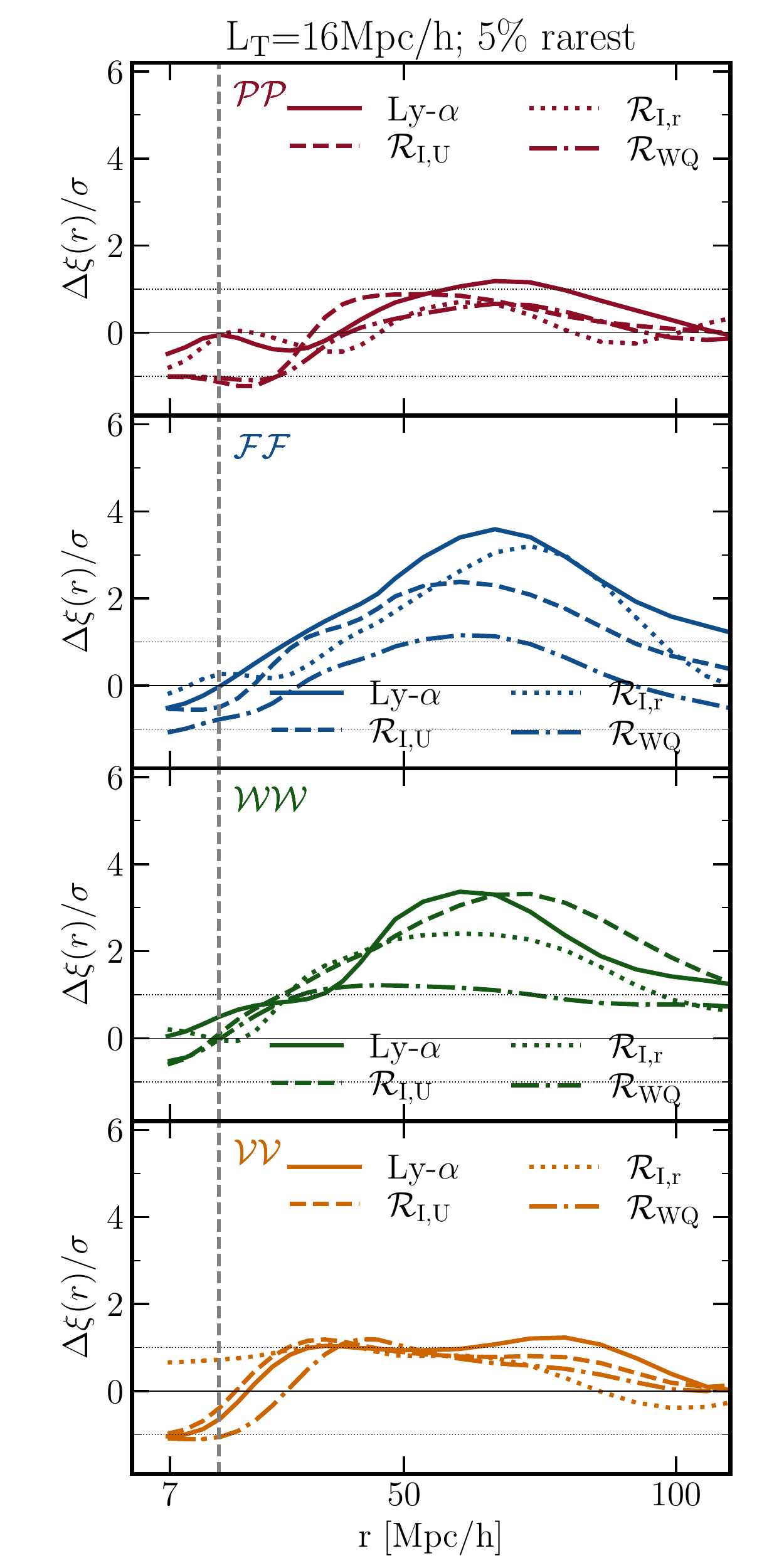}
\includegraphics[width=0.45\textwidth]{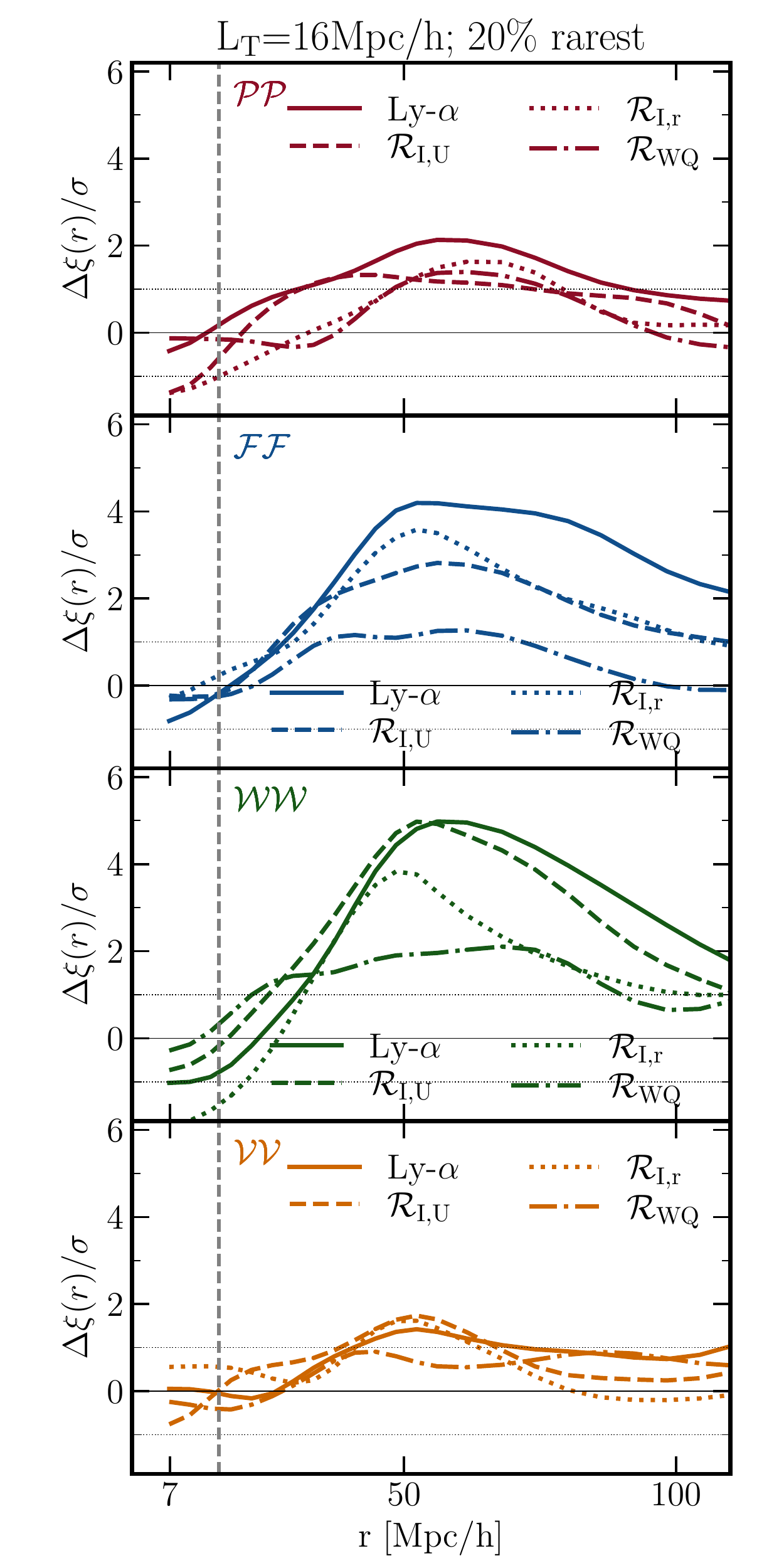}
\caption{Differences of auto-correlations of critical points of \lya and three reconstructed fields with respect to the noise for 5\% (left) and 20\% rarity (right) at the smoothing scale \lt=16~Mpc/h. 
Vertical dashed gray lines indicate the smoothing scale. While with decreasing rarity the significance of the differences does not change for $\mathcal{VV}$ auto-correlations and it only slightly increases for  $\mathcal{PP}$ and $\mathcal{FF}$, it is enhanced by a factor of about 1.5-2 for $\mathcal{WW}$. The differences obtained for 20\% rarity (right) are comparable to those of 10\% (see Fig.~\ref{fig:auto_all_rarity10_16}). 
}
\label{fig:auto_rarity5_20_16}
\end{figure*}

Starting with rarity, we complement the  10\% rarity results presented in the main text with rarities of 5\% and 20\%.
Fig.~\ref{fig:auto_rarity5_20_16} shows the differences of the auto-correlations of the \lya reference field and the three reconstructions, \siu, \sir, and \svar, w.r.t. the noise-only field $\mathcal{R}_{\rm noise}$ for the critical points with abundance of 5\% (left panels) and 20\% (right panels) at \lt=16 Mpc/h. 
As was the case for 10\% rarity (see Fig.~\ref{fig:auto_all_rarity10_16}), the most striking features in the auto-correlation functions are measured for filaments and walls. The increased significance with decreased rarity is notable for the $\mathcal{WW}$ auto-correlations, where the significance increases from about 2.5$\sigma$ (1$\sigma$) for \siu (\svar) at 5\% abundance to up to 5$\sigma$ (2$\sigma$) at 20\% abundance. For the $\mathcal{FF}$  auto-correlations this significance  is mush less striking. The significance of the outstanding features in the auto-correlations of the critical points at 10\% abundance is comparable to that of 20\% abundance.

\begin{figure*}
\centering\includegraphics[width=0.45\textwidth]{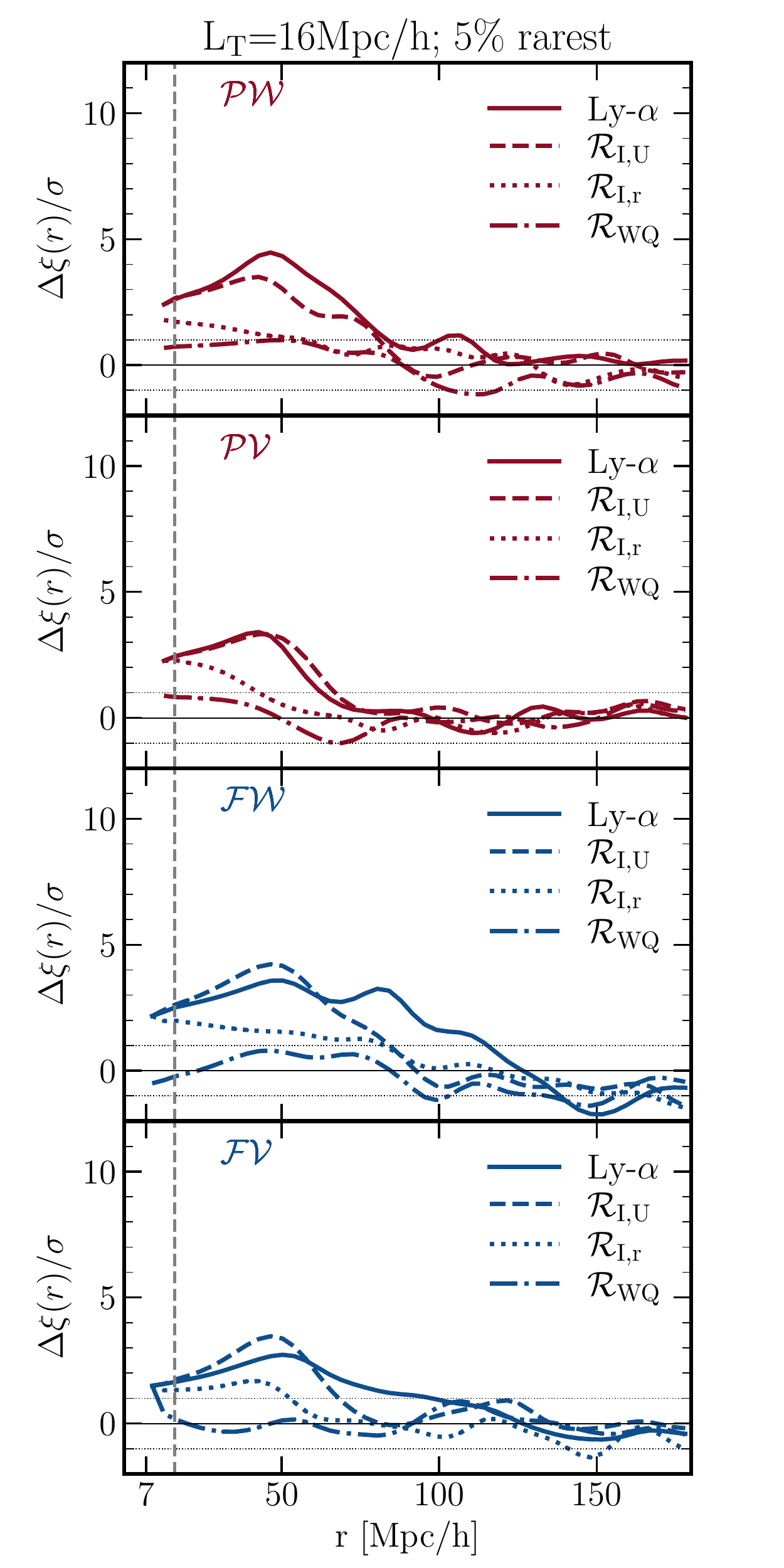}
\includegraphics[width=0.45\textwidth]{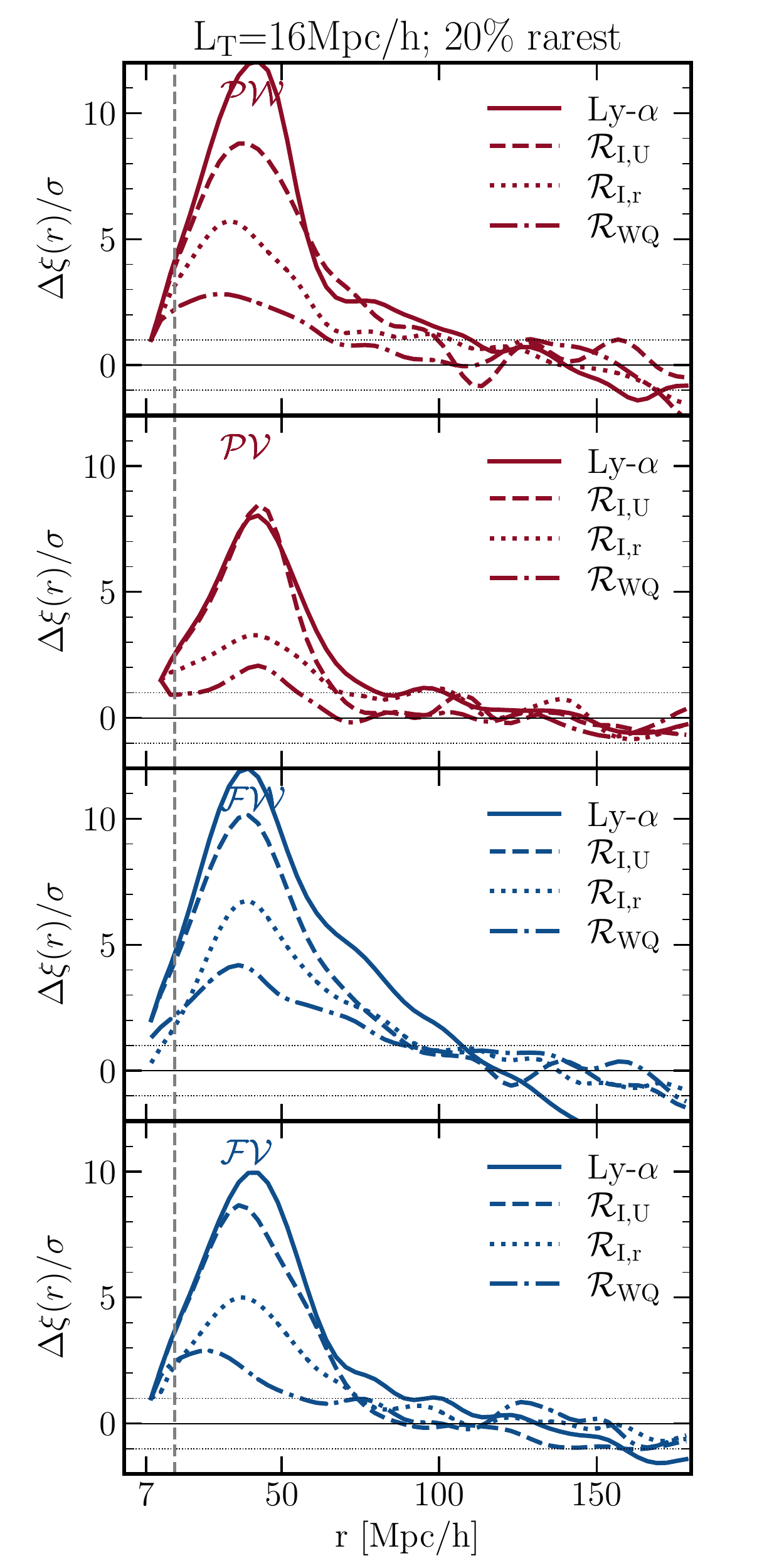}
\caption{Differences of cross-correlations of over- and under-dense critical points of \lya and three reconstructed fields with respect to the noise with 5\% (left) and 20\% rarity (right) at the smoothing scale \lt=16~Mpc/h. Vertical dashed gray line indicates the smoothing scale.
For all cross-correlations, 
the significance of their differences between all fields and the noise only field increases by about a factor of two with decreased rarity (from 5\% to 20\%).
}
\label{fig:cc_orig_all_rarity5_20_4in1_16}
\end{figure*}

Fig.~\ref{fig:cc_orig_all_rarity5_20_4in1_16} compares the differences of the cross-correlations of \lya reference field and the three reconstructions, \siu, \sir, and \svar, w.r.t. the noise-only field $\mathcal{R}_{\rm noise}$ for the over- and under-dense critical points with abundance of 5\% (left panels) and those of 20\% (right panels) at \lt=16~Mpc/h. 
In contrast to the overall mild increase of the significance of the outstanding features contained in the auto-correlations, their significance increases strikingly for all cross-correlations of under and overdense critical points, i.e. $\mathcal{PW}$, $\mathcal{PV}$, $\mathcal{FW}$ and $\mathcal{FV}$. 
While for $\mathcal{PW}$ and $\mathcal{PV}$ the significance increases by about the factor of two between 5\% and 20\% abundance (up to 8$\sigma$ at 20\% for \siu), for $\mathcal{FW}$ this factor is even higher (between factor of 2.5 for \siu and factor of 4 for \svar). The most striking increase of the significance of outstanding features with decreasing rarity is seen for $\mathcal{FV}$, where for \siu the significance increases from 3$\sigma$ to 9$\sigma$ and for \svar it is from 0$\sigma$ to up to 3$\sigma$.

\begin{figure*}
\centering\includegraphics[width=0.45\textwidth]{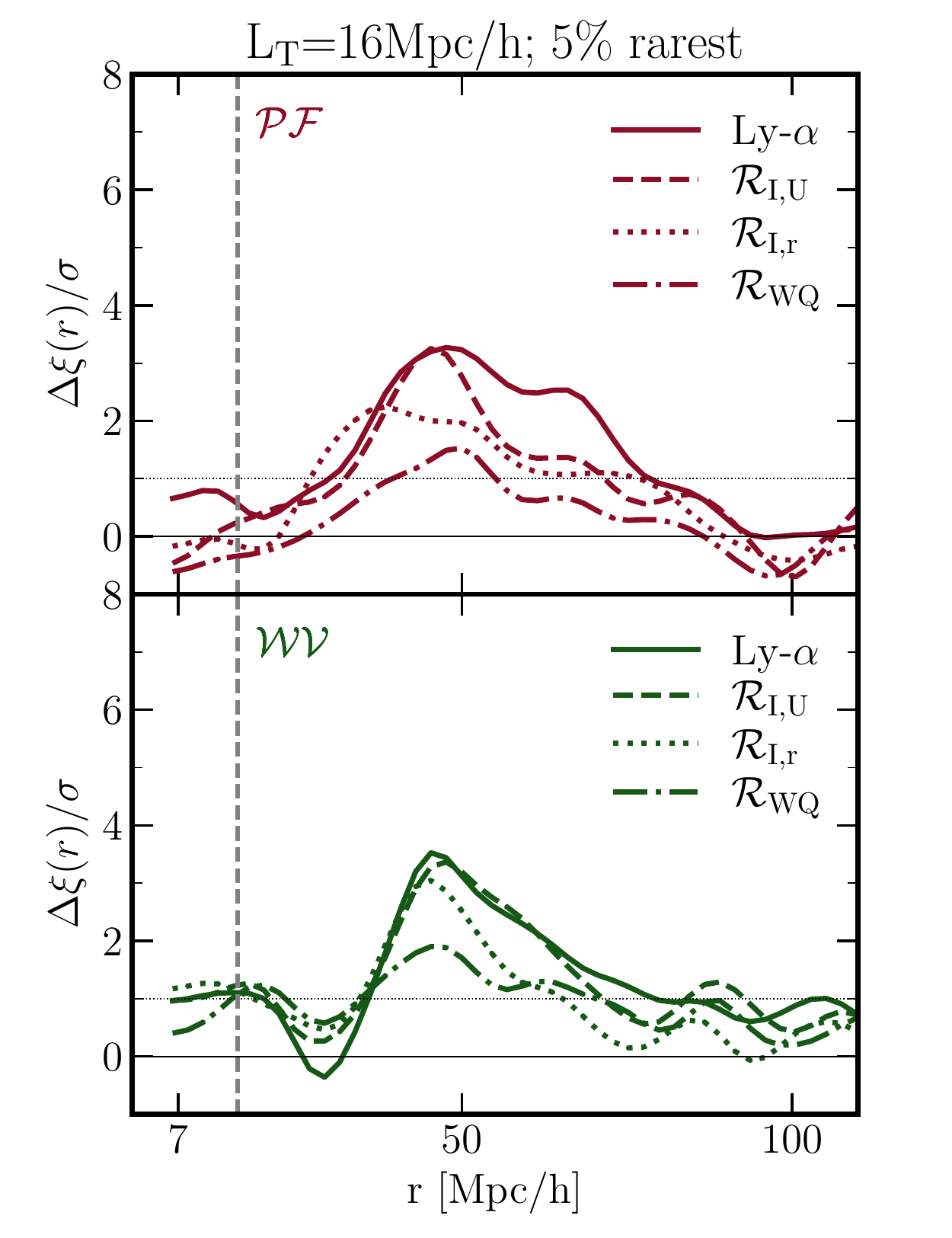}
\includegraphics[width=0.45\textwidth]{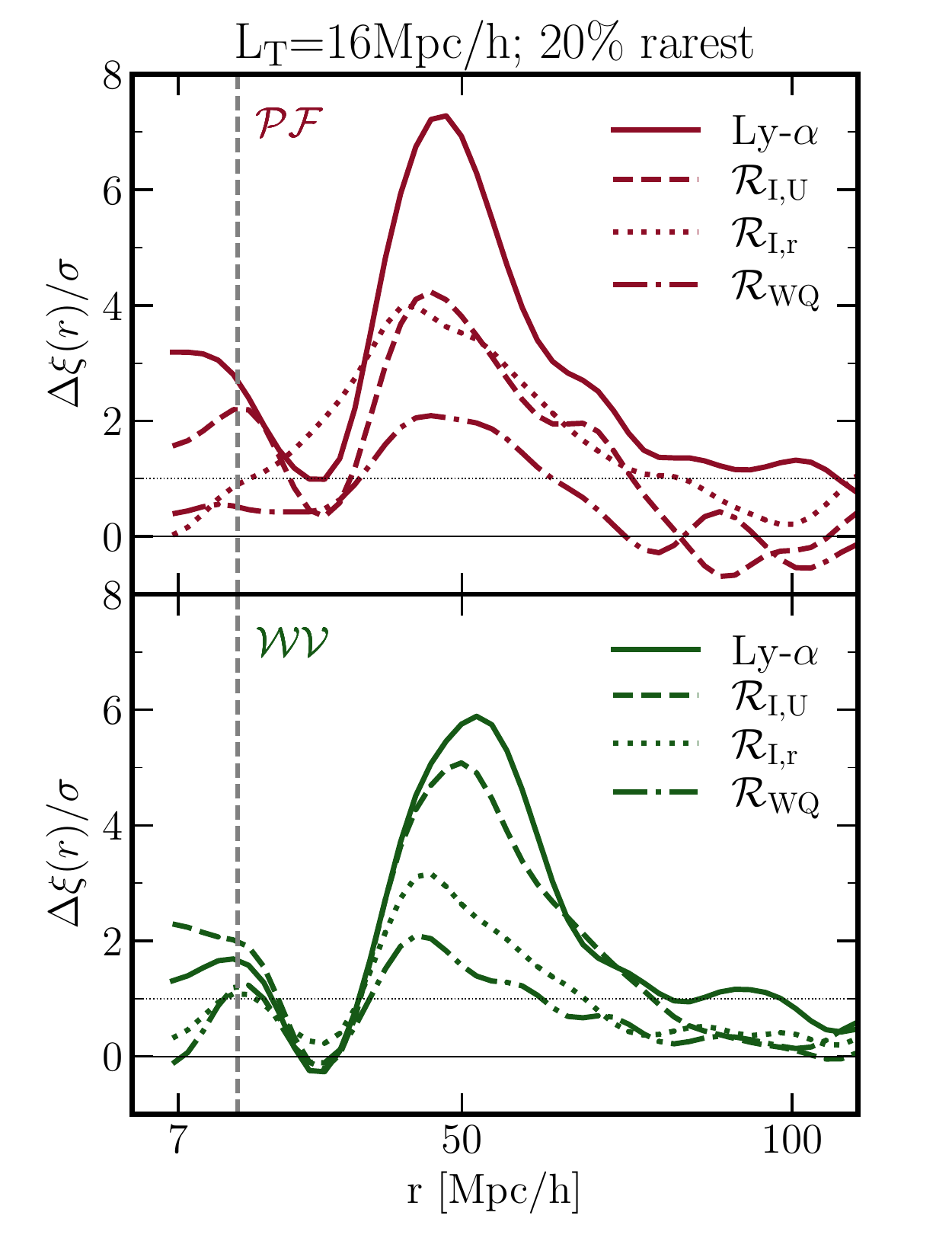}
\caption{Differences of cross-correlations of critical points with the same sign of overdensity of \lya and three reconstructed fields with respect to the noise with 5\% (left) and 20\% rarity (right) at the smoothing scale \lt=16~Mpc/h. Vertical dashed gray line indicates the smoothing scale.
As in the case of under and over-dense critical points, for all cross-correlations, there is about a factor of up to two increase of the significance of differences 
with decreased rarity. 
}
\label{fig:cc_orig_all_rarity5_20_2in1_16}
\end{figure*}

Similarly, but to a lesser extent, the significance increases with decreasing rarity for the cross-correlations of the critical points with the same overdensity sign, i.e. $\mathcal{PF}$ and $\mathcal{WV}$, as was shown in Fig.~\ref{fig:cc_orig_all_rarity5_20_2in1_16}.

\begin{figure}
\includegraphics[width=0.45\textwidth]{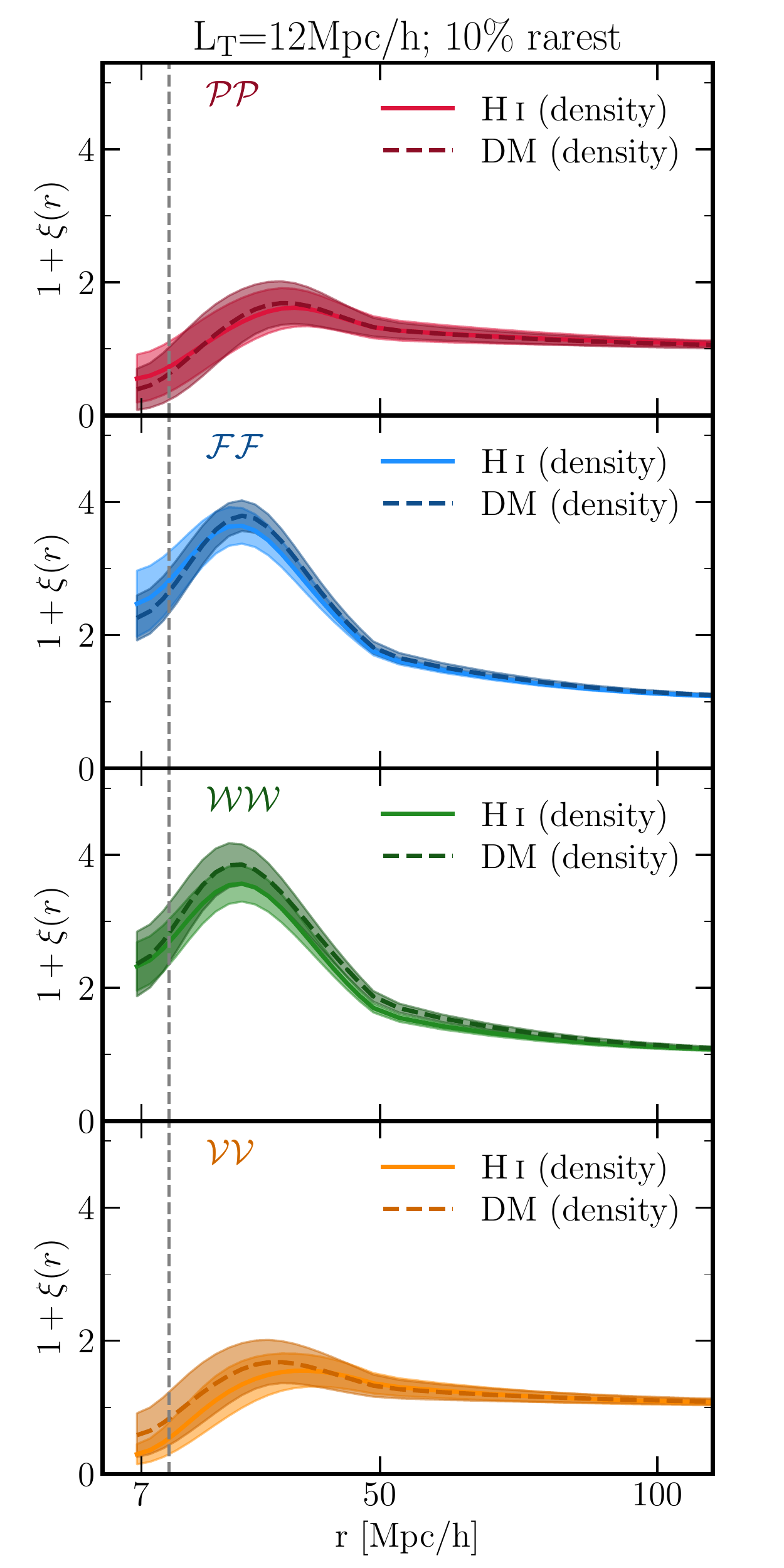}
\caption{Auto-correlations of critical points with 10\% abundance. ${\mathcal{PP}}$ (peak-peak), ${\mathcal{FF}}$ (filament-filament), ${\mathcal{WW}}$ (wall-wall), and ${\mathcal{VV}}$ (void-void) correlations (from top to bottom panels) are shown for \hi density field (coloured solid lines) and DM field (coloured dashed lines) at the smoothing scale of 12~Mpc/$h$. Shaded area corresponds to the error on the mean across five mocks. Vertical dashed gray line indicates the smoothing scale.
}
\label{fig:auto_dm_rarity10_12}
\end{figure}

Let us now examine the impact of  smoothing on the two point correlation functions. We will focus on the comparison between the \hi and DM density fields.
Fig.~\ref{fig:auto_dm_rarity10_12} shows the auto-correlations of the 10\% rarest critical points in the original (\hi density; color solid lines) and DM density (color dashed lines) fields, as in Fig.~\ref{fig:auto_dm_rarity10_16}, but smoothed at the scale \lt=12~Mpc/$h$. The auto-correlations of all critical points follow qualitatively similar trends, as in the case of 16~Mpc/$h$ smoothing scale, with a good
agreement between the \hi and DM density fields. As expected, the maximum of the auto-correlations is again reached at $\approx 2$\lt for filaments and walls and $\approx 2.5-3$\lt for peaks and voids (see Table~\ref{tab:summary_numbers_max_12}). The positions and heights of the maxima are much better constrained at \lt=12~Mpc/$h$ smoothing, compared to  16 Mpc/$h$.


\begin{table}
\centering
\caption{Size of the exclusion zone \rexc (in [Mpc/$h$]) of cross-correlations $\mathcal{PW}$, $\mathcal{PV}$, $\mathcal{FW}$, $\mathcal{FV}$ for 10\% rarity, for \hi and DM density fields at the smoothing scale \lt=12 Mpc/$h$. The errors are the standard deviations of the mean across all mocks.} 
\label{tab:summary_numbers_rexc_12}
\begin{tabular*}{\columnwidth}{@{\extracolsep{\fill}}lcccc}
\hline
\hline
&  $\mathcal{PW}$ & $\mathcal{PV}$ & $\mathcal{FW}$ & $\mathcal{FV}$\\
\hline
\hi &  30.69$\pm{1.26}$ & 33.70$\pm{0.82}$ & 28.44$\pm{0.67}$ & 31.45$\pm{1.06}$  \\
DM & 32.95$\pm{0.82}$ & 35.96$\pm{1.25}$ & 31.45$\pm{0.0}$ & 33.70$\pm{0.82}$  \\
\hline
\end{tabular*}
\end{table}

\begin{figure}
\centering\includegraphics[width=0.45\textwidth]{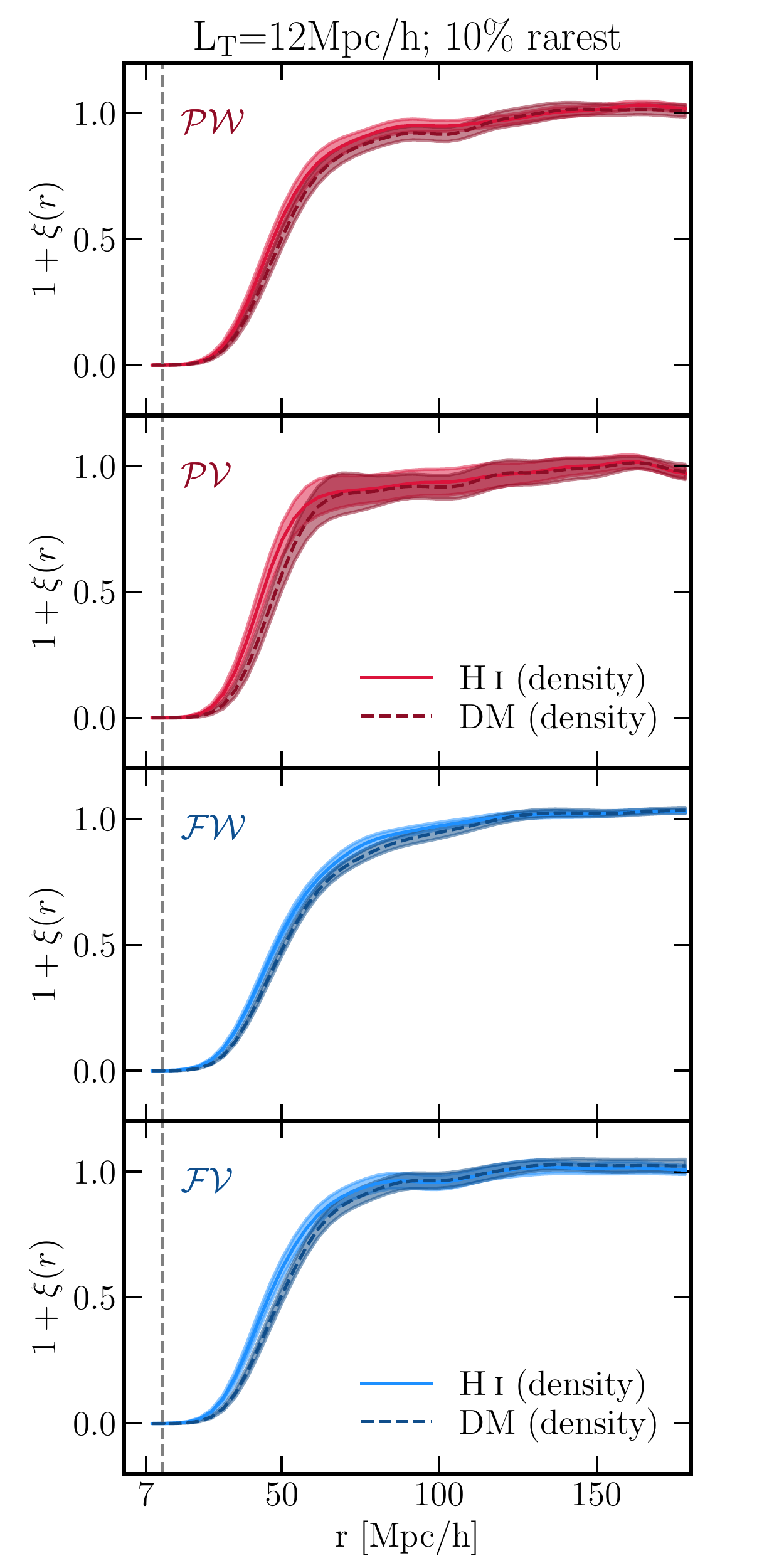}
\caption{Cross-correlations of critical points with 10\% abundance. $\mathcal{PW}$ (peak-wall), $\mathcal{PV}$ (peak-void), $\mathcal{FW}$ (filament-wall) and $\mathcal{FV}$ (filament-void) correlations (from top to bottom panels) are shown for the \hi\! density field (coloured solid lines) and DM field (coloured dashed lines) at the smoothing scale of 12~Mpc/$h$. The shaded area corresponds to the error on the mean across five mocks. Vertical dashed gray line indicates the smoothing scale. 
} 
\label{fig:cc_dm_rarity10_4in1_12}
\end{figure}

Fig.~\ref{fig:cc_dm_rarity10_4in1_12} shows the cross-correlation function of under and overdense critical points with 10\% abundance for the \hi and DM density fields (coloured solid and dashed lines, respectively), following Fig.~\ref{fig:cc_dm_rarity10_4in1_16}, but smoothed at the scale \lt=12~Mpc/$h$. As for the auto-correlations, there is a better agreement between the \hi and DM density fields compared to the smoothing \lt=16~Mpc/$h$. The size of the exclusion zone is identical for  the \hi and DM density fields and it is smaller compared to the fields smoothed at \lt~=~16~Mpc/$h$ (see Table~\ref{tab:summary_numbers_rexc_12}).

\begin{figure}
\centering\includegraphics[width=0.45\textwidth]{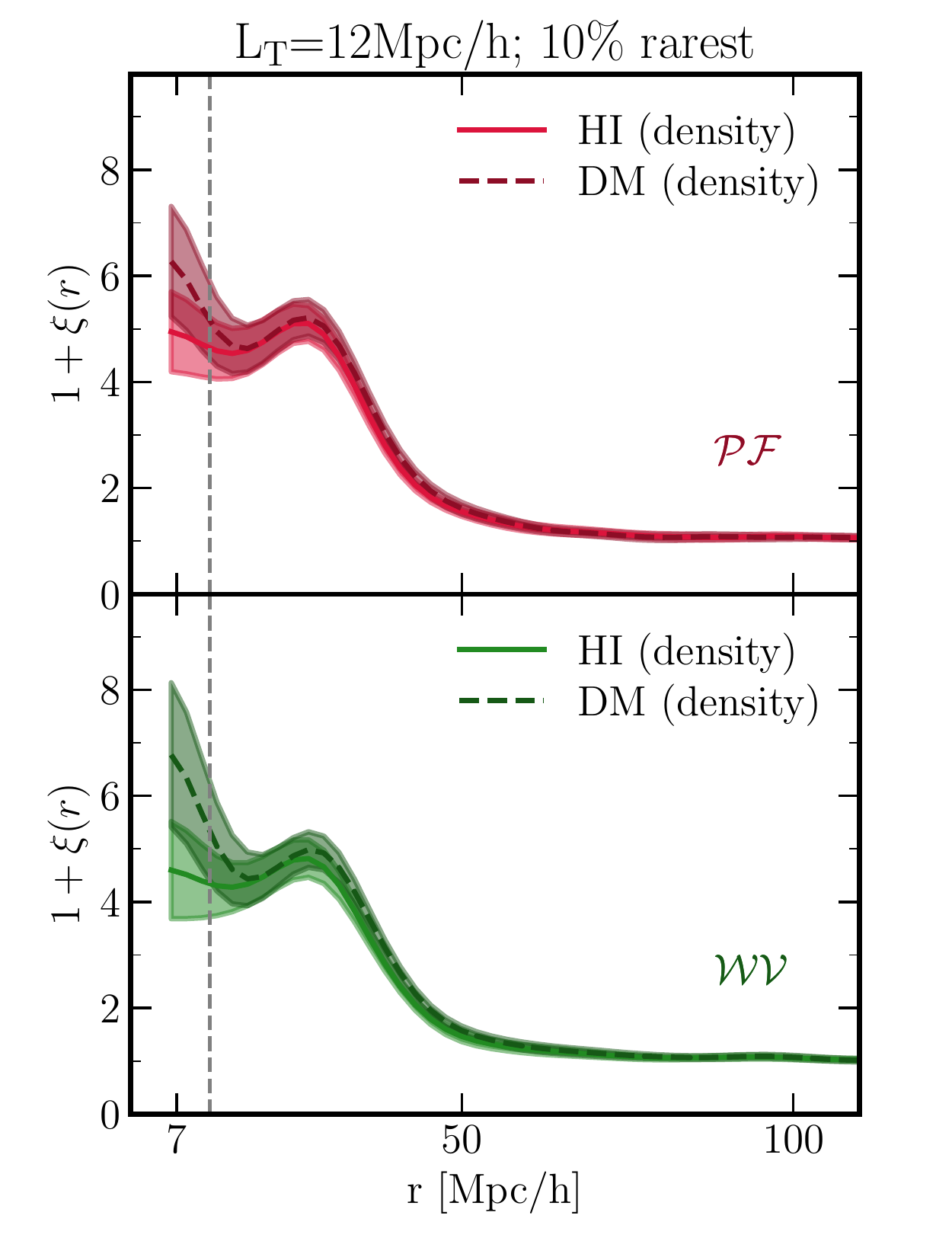}
\caption{Cross-correlations of critical points with 10\% abundance at the smoothing scale \lt=12~Mpc/$h$ for \hi\! (solid coloured lines) and DM  (dashed coloured lines) density fields. The $\mathcal{PF}$ (peak-filament) and $\mathcal{WV}$ (wall-void) correlations are shown on the top and bottom panels, respectively.
Vertical dashed gray line indicates the smoothing scale, while vertical colored lines mark the position of the maximum of each curve.
}
\label{fig:cc_dm_rarity10_2in1_12}
\end{figure}

Fig.~\ref{fig:cc_dm_rarity10_2in1_12} shows the cross-correlation function of the critical points of the same overdensity sign for \hi and DM density fields (coloured solid and dashed lines, respectively) and 10\% abundance, following Fig.~\ref{fig:cc_dm_rarity10_2in1_16}, but at the smoothing scale of 12 Mpc/$h$. The agreement between the two fields with decreased smoothing scale is again confirmed. For both \hi and DM density fields, the position of maxima $r_{\rm max,i}$ is $\approx$ 26.9 Mpc/$h$ ($\approx 2 \lt$). The height of the maxima $h_{\rm max,i}$ are also in a good agreement between the two fields (see Table~\ref{tab:summary_numbers_max_12}).

\begin{table*}
\centering
\caption{Position (\rmax [Mpc/$h$]) and height (\hmax) of the maximum of auto-correlations and cross-correlations $\mathcal{PF}$, $\mathcal{WV}$ for 10\% rarity, for \hi and DM density fields at the smoothing scale \lt=12 Mpc/$h$. The errors are the standard deviations of the mean across all mocks.} 
\label{tab:summary_numbers_max_12}
\begin{tabular*}{\textwidth}{@{\extracolsep{\fill}}lccccccc}
\hline
\hline
&  & $\mathcal{PP}$ & $\mathcal{FF}$ & $\mathcal{WW}$ & $\mathcal{VV}$ & $\mathcal{PF}$ & $\mathcal{WV}$ \\ 
\hline
\multirow{2}{*}{\rmax} & \hi & 34.13$\pm{0.42}$ & 25.61$\pm{1.24}$ & 24.66$\pm{0.79}$ & 27.50$\pm{3.85}$  & 26.88$\pm{0.65}$  & 25.49$\pm{1.05}$ \\
& DM & 30.34$\pm{3.10}$ & 23.24$\pm{1.03}$ & 20.39$\pm{0.67}$ & 29.87$\pm{2.42}$  & 25.95$\pm{0.82}$  & 29.18$\pm{0.65}$\\
\hline
\multirow{2}{*}{\hmax} & \hi &  2.86$\pm{0.21}$ & 4.93$\pm{0.28}$ & 4.94$\pm{0.24}$ & 3.15$\pm{0.38}$   & 5.10$\pm{0.34}$  & 4.81$\pm{0.35}$  \\
& DM &  3.15$\pm{0.18}$ & 4.83$\pm{0.11}$ & 4.96$\pm{0.26}$ & 3.03$\pm{0.23}$   & 6.34$\pm{0.19}$  & 6.15$\pm{0.29}$ \\
\hline
\end{tabular*}
\end{table*}


\bsp	
\label{lastpage}
\end{document}